\documentclass[fleqn,usenatbib]{mnras}

\usepackage{longtable}

\usepackage[T1]{fontenc}

\DeclareRobustCommand{\VAN}[3]{#2}
\let\VANthebibliography\thebibliography
\def\thebibliography{\DeclareRobustCommand{\VAN}[3]{##3}\VANthebibliography}

\usepackage{graphicx}	
\usepackage{amsmath}	
\usepackage{amssymb}
\usepackage{caption}
\usepackage{subcaption}

\title[MIGHTEE-H{\sc i} detected galaxies at $z>0.25$]{MIGHTEE-H{\sc i}: The direct detection of neutral hydrogen in galaxies at $z>0.25$}

\author[M. J. Jarvis et al.]{Matt J. Jarvis$^{1,2}$\thanks{E-mail: matt.jarvis@physics.ox.ac.uk},
Madalina N. Tudorache$^{1}$,
I. Heywood$^{1,3,4}$,
Anastasia A. Ponomareva$^{1,5}$,
\and
M. Baes$^{6}$,
Natasha Maddox$^{7}$, 
Kristine Spekkens$^{8}$,
Andreea V\u{a}r\u{a}\c{s}teanu$^{1}$,
C. L. Hale$^{1}$, 
Mario G. Santos$^{2}$,
\and
R. G. Varadaraj$^{1}$, 
Elizabeth A. K. Adams$^{9,10}$,
Alessandro Bianchetti$^{11,12}$,
Barbara Catinella$^{13,14}$,
\and
Jacinta Delhaize$^{15}$,
M. Maksymowicz-Maciata$^{7}$,
Pavel E. Mancera Pi\~{n}a$^{16}$,
Hengxing Pan$^{1}$,
\and
Am\'elie Saintonge$^{17,18}$,
Gauri Sharma$^{19,2,20,21,22}$,
O.~Ivy~Wong$^{23,13}$
\\
\noindent
$^{1}$Astrophysics, Department of Physics, University of Oxford, Keble Road, Oxford, OX1 3RH, UK \\
$^{2}$Department of Physics and Astronomy, University of the Western Cape, Robert Sobukwe Road, 7535 Bellville, Cape Town, South Africa \\
$^{3}$Centre for Radio Astronomy Techniques and Technologies, Department of Physics and Electronics, Rhodes University, PO Box 94, \\Makhanda, 6140, South Africa. \\
$^{4}$South African Radio Astronomy Observatory, 2 Fir Street, Black River Park, Observatory, Cape Town, 7925, South Africa.\\
$^{5}$Centre for Astrophysics Research, University of Hertfordshire, College Lane, Hatfield AL10 9AB, UK \\
$^{6}$Sterrenkundig Observatorium, Universiteit Gent, Krijgslaan 281 S9, B-9000 Gent, Belgium\\
$^{7}$School of Physics, H.H. Wills Physics Laboratory, Tyndall Avenue, University of Bristol, Bristol, BS8 1TL, UK\\
$^{8}$Department of Physics, Engineering Physics and Astronomy, Queen's University, Kingston, Ontario, K7L 3N6, Canada\\
$^{9}$ASTRON, the Netherlands Institute for Radio Astronomy, Oude Hoogeveesedijk 4,7991 PD Dwingeloo, The Netherlands\\
$^{10}$Kapteyn Astronomical Institute, University of Groningen, PO Box 800, 9700 AV Groningen, The Netherlands\\
$^{11}$Department of Physics and Astronomy, Università degli Studi di Padova, Vicolo dell’Osservatorio 3, I-35122, Padova, Italy\\
$^{12}$INAF - Osservatorio Astronomico di Padova, Vicolo dell’Osservatorio 5, I-35122, Padova, Italy\\
$^{13}$International Centre for Radio Astronomy Research, The University of Western Australia, Crawley, WA 6009, Australia\\
$^{14}$ARC Centre of Excellence for All Sky Astrophysics in 3 Dimensions (ASTRO 3D), Australia\\
$^{15}$Department of Astronomy, University of Cape Town, Private Bag X3, Rondebosch 7701, South Africa\\
$^{16}$Leiden Observatory, Leiden University, P.O. Box 9513, 2300 RA, Leiden, The Netherlands\\
$^{17}$Department of Physics and Astronomy, University College London, London, WC1E 6BT, UK\\
$^{18}$Max-Planck Institute for Radio Astronomy, Auf dem H\"ugel 69, 53121 Bonn, Germany\\
$^{19}$University of Strasbourg, CNRS UMR 7550, Observatoire astronomique de Strasbourg, F-67000 Strasbourg, France      \\
$^{20}$SISSA International School for Advanced Studies, Via Bonomea 265, I-34136 Trieste, Italy\\
$^{21}$INFN-Sezione di Trieste, via Valerio 2, I-34127 Trieste, Italy\\
$^{22}$IFPU Institute for Fundamental Physics of the Universe, Via Beirut, 2, 34151 Trieste, Italy\\
$^{23}$CSIRO Space \& Astronomy, PO Box 1130, Bentley, Western Australia 6102, Australia 
}

\date{Accepted XXX. Received YYY; in original form ZZZ}

\pubyear{\the\year{}}

\begin{document}
\label{firstpage}
\pagerange{\pageref{firstpage}--\pageref{lastpage}}
\maketitle

\begin{abstract}
Atomic hydrogen constitutes the gas reservoir from which molecular gas and star formation in galaxies emerges. However, the weakness of the line means it has been difficult to directly detect in all but the very local Universe. Here we present results from the first search using the MeerKAT International Tiered Extragalactic Exploration (MIGHTEE) Survey for high-redshift ($z>0.25$) H{\sc i} emission from individual galaxies. By searching for 21-cm emission centered on the position and redshift of optically-selected emission-line galaxies we overcome difficulties that hinder untargeted searches. We detect 11 galaxies at $z>0.25$, forming the first sample of $z>0.25$ detections with an interferometer, with the highest redshift detection at $z = 0.3841$. We find they have much larger H{\sc i} masses than their low-redshift H{\sc i}-selected counterparts for a given stellar mass. This can be explained by the much larger cosmological volume probed at these high redshifts, and does not require any evolution of the H{\sc i} mass function. We make the first-ever measurement of the baryonic Tully-Fisher relation (bTFr) with H{\sc i} at $z>0.25$ and find consistency with the local bTFr, but with tentative evidence of a flattening in the relation at these redshifts for higher-mass objects. This may signify evolution, in line with predictions from hydrodynamic simulations, or that the molecular gas mass in these high-mass galaxies could be significant. This study paves the way for future studies of H{\sc i} beyond the local Universe, using both searches targeted at known objects and via pure H{\sc i} selection.
\end{abstract}




\section{Introduction}
Neutral hydrogen is the pristine reservoir of gas that underpins the formation of all the galaxies in the Universe. This gas accretes on to dark matter haloes, and subsequently forms molecular gas from which stars emerge. Therefore, understanding its relation to galaxies and the large-scale structure of the Universe over cosmic time is a critical piece of information that is currently missing from our picture of galaxy evolution. This is largely because direct observations of the neutral atomic hydrogen are limited to either absorption line studies of the Lyman-$\alpha$ transition, which require space-based ultraviolet-sensitive spectrographs for studies at low redshift  ($z \lesssim 2$), or deep observations using radio telescopes to detect the 21-cm emission from the spin-flip transition of the hydrogen electron.

Although the 21~cm emission is accessible to ground-based radio telescopes, the transition probability is many orders of magnitude lower than that for Lyman-$\alpha$, and the 21~cm emission line is much fainter. For this reason the detection of H{\sc i} emission from individual galaxies at 21-cm has, until recently, been restricted to the relatively low-redshift Universe, with large-area untargeted surveys only sensitive to $z \lesssim 0.1$ \citep[e.g.][]{ALFALFA,HIPASS}. In recent years, there has been effort to extend this redshift range using deep targeted observations of galaxies \citep[e.g.][]{CatinellaCortese2015}, galaxy clusters \citep[][]{BUDHIES2,BUDHIES3} and  surveys of small regions of the sky with extensive multi-wavelength data. An example of the latter is the COSMOS H{\sc i} Large Extragalactic Survey (CHILES), which reported the detection of what was the highest unlensed H{\sc i} detection \citep{Fernandez2016} until recently, although the validity of this has recently been called into question \citep{Heywood2024}. Such targeted approaches have also been complemented by spectral stacking of samples selected from optical spectroscopic surveys. These provide statistical measurements of the average H{\sc i} mass for various subsamples of galaxies based on star-formation rate, colour, stellar mass or environment \citep[e.g.][]{Sinigaglia2022,Bera2023,Sinigaglia2024,Bianchetti2025}.

The new Five-hundred-meter Aperture Spherical Telescope (FAST) deep survey has recently reported the detection of six H{\sc i}-detected galaxies at $0.38<z<0.5$, utilising the vast collecting area of the single dish \citep{Xi2024}. However, the poor angular resolution of single-dish telescopes can lead to overestimates of the H{\sc i} mass in individual systems due to possible confusion with neighbouring gas-rich galaxies.
Deep spectral-line observations with interferometers can therefore play a crucial role in enhancing our understanding of neutral hydrogen in galaxies. The MeerKAT telescope is ideally suited to this goal, combining extremely high sensitivity with a baseline distribution that is sensitive to diffuse emission, characteristic of low-density H{\sc i} gas in galaxies. Indeed, several of the large survey programmes using the MeerKAT telescope are designed to target this diffuse H{\sc i} emission with a variety of science goals. The most relevant to the search for high-redshift H{\sc i} emission are the deep  extragalactic surveys, namely the MeerKAT International Tiered Extragalactic Exploration (MIGHTEE) Survey \citep{Jarvis2016} and the Looking at the Distant Universe with the MeerKAT Array \citep[LADUMA; ][]{Blyth2016}.  These surveys have also recently found the two highest redshift hydroxyl megamasers to date \citep{Glowacki2022, Jarvis2024}, demonstrating that they also have the potential to detect spectral lines from  high-redshift galaxies.

In this paper we report on a targeted search for H{\sc i} emission from individual galaxies at $0.25<z<0.5$ using optical spectroscopic redshifts from the Dark Energy Spectroscopic Instrument Early Data Release \citep[DESI; ][]{DESI_ES}, using the recent release of the full MIGHTEE spectral cube covering COSMOS field \citep{Heywood2024}.

In Section~\ref{sec:data} we describe the MIGHTEE-H{\sc i} along with the DESI data that we use to identify potential H{\sc i} galaxies at $0.25<z<0.5$. Section~\ref{sec:search} details how we search for the H{\sc i} galaxies using the optical spectroscopic information and also how we mitigate against false positives. In Section~\ref{sec:sampleproperties} we investigate the sample properties through various scaling relations and the baryonic Tully-Fisher relation (bTFr), using the wealth of ancillary data over the field to carry out spectral energy distribution fitting to derive the host galaxy properties. Our conclusions are presented in Section~\ref{sec:conclusions}.
Throughout this paper we assume $\Lambda$CDM cosmology with $H_0 = 70$~km\,s$^{-1}$\,Mpc$^{-1}$ and $\Omega_{\rm M} = 0.3$ and $\Omega_\Lambda = 0.7$.

\section{Data}\label{sec:data}

The MeerKAT International GigaHertz Tiered Extragalactic Exploration survey \citep{Jarvis2016}, is a medium-deep, medium-wide survey providing simultaneous radio continuum \citep{Heywood2022,Hale2024}, spectral line \citep{Maddox2021, Heywood2024} and polarisation observations \citep{Taylor2024}. It will eventually cover the four well-known extragalactic deep fields: COSMOS, XMM-LSS, Extended Chandra Deep Field South (ECDFS) and ELAIS-S1. The MIGHTEE Data Release 1 L1 spectral line cube over the COSMOS field, described in \citet[][]{Heywood2024}, extends over $\sim5$\,deg$^2$, with a spectral range 960--1150\,MHz, channel width of 104.5\,kHz and angular resolution FWHM$\sim 15$\,arcsec. We use this cube to search for neutral hydrogen emission at the position and redshift of galaxies with measured optical spectroscopic redshifts from the DESI public catalogue \citep{DESI_ES}. 

DESI provides the ideal database for targeting potential H{\sc i}-rich galaxies over the redshift range accessible with the MIGHTEE spectral cube, due to the targeting of emission line galaxies that are likely actively star forming with large gas reservoirs. We downloaded the DESI catalogue covering the footprint of the MIGHTEE-H{\sc i} Data Release 1 and limited the redshift range to $0.23 <z< 0.5$, i.e. corresponding to the  redshifts that are accessible with the L1 spectral window (960--1150\,MHz) defined in \cite{Heywood2024}, and where the radio data has relatively low noise. The redshift distribution for the optical spectroscopic redshifts is shown in Figure~\ref{fig:DESIzdist}.
We then extracted $80\times 80$~arcsec$^2$ cubelets covering the full spectral window of the L1 data, at the position of each optical spectroscopic source that fell within the redshift range $0.25 <z <0.5$. In this initial step there were 915 galaxies with data in the MIGHTEE-L1 spectral cube.

\section{The search for H{\sc i} galaxies}\label{sec:search}
We are searching for what may be faint line emission, likely with a broad range of line widths and profiles. We therefore choose to initially identify potential candidates by eye. To do this we first collapse 3-D cubelets to 2-D images covering $80\times80$\,arcsec$^2$ centred on the optical galaxy position, summing the emission from a spectral region of $\pm 1$\,MHz ($\pm 285$\,km~s$^{-1}$ at a central frequency of $\sim 1050$\,MHz) at the frequency of H{\sc i} that corresponds to the optical spectroscopic redshift.
The signal-to-noise in the collapsed image depends on both the width {\em and} the flux within any putative emission line. We therefore also extracted, in parallel, the 1-D spectrum spanning a much larger spectral region of $\sim 13$\,MHz (see Figure~\ref{fig:M0Spec}) for all the 915 spectroscopically detected galaxies using a $10\times 10$~arcsec$^2$ aperture, centred on the optical galaxy position, to also check if there was a prominent signal in the spectral dimension. We then visually assessed whether there was likely to be an emission line at the redshift of the optical galaxy using both the collapsed image and the 1-D spectrum with the following: (i) a prominent signal in the 2-D collapsed image centered within 4~arcsec of the spatial position (well within the positional accuracy of an unresolved source but accounting for any possible emission that may be offset from the centre, e.g. in an extended disk), and which appeared similar in extent to the main lobe of the synthesised beam, and; (ii) have significant emission above the noise in the spectral domain, at the frequency corresponding to H{\sc i} at the optical redshift, and which spans at least five spectral channels ($\sim 150\,$km~s$^{-1}$ at $z\sim 0.35$).

At this stage no statistical measurement is made to quantify the robustness of the candidates. As such, this stage should purely be seen as an initial filtering of the most likely H{\sc i}-emission line candidates, removing obviously noisy regions and noise spikes below the resolution of the data. We also note that given this initial selection then one should be judicious about what science the final sample is used for, for  example one should not attempt to perform statistical studies that rely on accurate estimates of the sample completeness. 
This initial selection led to 43 candidates to be investigated in more detail.

Using the extracted 1-D spectrum for each of the objects we measured the integrated flux across the plausible H{\sc i} emission line by estimating where the flux of the identified line appears to fall to a mean of zero and remains around zero. This allows for the potential of real line structures that may, for example, exhibit troughs between two peaks caused by a rotating disk. We also account for the synthesised beam area by scaling the flux by the ratio of the peak flux to the $10\times 10$ ~arcsec$^2$ aperture flux in the synthesised beam, and assuming that the H{\sc i} emission region is unresolved  at the resolution ($\sim 15$\,arcsec FWHM) of the MIGHTEE L1, robust = 0.0 spectral cube \citep[see ][ for further details]{Heywood2024}. The $\sim 15$ arcsec resolution of the data corresponds to $\sim 75$\,kpc at $z\sim 0.35$, which if we assume the local relationship between H{\sc i} mass and the diameter of H{\sc i} disks \citep[e.g.][]{BroeilsRhee1997,VerheijenSancisi2001, Rajohnson2022} holds at $z\sim 0.35$, corresponds to H{\sc i} masses of $M_{\rm HI} \sim 10^{10}$\,M$_{\odot}$. Thus, although it is possible that high H{\sc i}-mass galaxies may be marginally resolved in our data, the amount of missing flux is likely to be small with our measurement using an aperture of $10\times 10$\,arcsec$^2$. 

To determine the signal-to-noise (SNR) of the integrated line flux we measured the integrated flux across $\sim 500$ similar-sized regions within 2~arcmin of the central source of interest, using the same number of channels over which we determine the emission line flux, as indicated by the blue part of the spectra in Figure~\ref{fig:M0Spec}, varying both the central position and the central frequency (within 10~MHz of the H{\sc i} emission line candidate). This method for estimating the noise accounts for the intrinsic noise variation across the mosaicked COSMOS cube, both spatially and spectrally, whilst also accounting for the correlated noise within the size of the extracted region. Using these $\sim 500$ positions we measure the 16th and 84th percentiles of the flux-density histogram to determine the noise for the 3-D voxel used to extract the H{\sc i} emission line. As we make no a-priori assumption of Gaussianity or symmetric noise, we determine the mean of the 50th - 16th and 84th - 50th percentiles, and calculate the SNR as the emission line flux divided by this value.
We note that the 16th and 84th percentiles are broadly similar around the median 50th percentiles, suggesting that the noise is largely Gaussian. To confirm this we use the Anderson-Darling test \citep{AndersonDarling} and find that the distribution is consistent with being Gaussian at $>95$ per cent confidence. The noise distribution around IDs 2 and 3 are marginally rejected at this level, and this is reflected in the estimate of the SNR ratio for these objects.
We use these individual noise estimates for each source to calculate the SNR of the integrated line flux and retained all galaxies in our sample with an integrated line flux SNR $> 4$. This resulted in a final sample of 11 H{\sc i} detected galaxies.  

The H{\sc i} masses for these 11 galaxies were determined by summing the spectrum across the estimated spectral extent of the H{\sc i} emission line using Equation~45 in \cite{Meyer2017}, under the assumption of the line being optically thin. We use the noise extracted from the $\sim 500$ independent voxels to determine the 1$\sigma$ uncertainty on the H{\sc i} mass.
The spectra for the final high-confidence (SNR $>4$) sample are presented in Figure~\ref{fig:M0Spec}, alongside the moment-0 maps spanning $80 \times 80$~arcsec$^{2}$ integrated over the emission line (shown in blue in Figure~\ref{fig:M0Spec}). H{\sc i} properties of the sources are provided in Table~\ref{tab:sample}.

We note that a SNR$>4$ at the position and redshift of a known galaxy means that the actual probability of the source being real is in fact higher than 4$\sigma$, as we are estimating the noise at random positions but the candidate itself is already known to have a coincident optical counterpart at the redshift of the H{\sc i} emission line. However, a danger when searching for emission-line galaxies by eye is the possibility of confirmation bias, i.e. knowledge of where a line should be may lead to the "detection" of more spurious sources.

We therefore repeated the search for H{\sc i} emission at positions offset from the optical galaxy position by 50~arcmin, but at the same redshift. This resulted in 34 potential candidates. 
We carry out the same analysis of determining the integrated signal-to-noise ratio across the putative emission lines, but all were ruled out by the final cut requiring SNR$>4$ to be considered a significant detection. The fact that we find $\sim 20$ per cent fewer potential sources at the offset positions by the same search method used for optical spectroscopic redshift, supports the number of sources that we identify as robust candidates (i.e. 11 of our initial sample of 43 galaxies were retained). This check also accounts for any possible non-Gaussianity in the noise resulting in spurious positive detections, if the noise properties of the data cubes were skewed towards positive fluxes. Although we note that the noise across the spectral cubes has been shown to be consistent with being Gaussian, \citep{Heywood2024}.
We are therefore confident that the 11 sources presented are highly likely to be real H{\sc i} detections. 

Another possible source of false positives, particularly when selecting based on properties related to star formation (i.e. with our spectroscopic redshift selection), is the possible presence of radio continuum emission at the position of the candidate H{\sc i} detection. Such continuum emission may not only increase the noise at the position of the sources, but if the continuum subtraction leaves artifacts they may manifest as putative H{\sc i} lines. We use the MIGHTEE radio continuum Data Release 1 \citep{Hale2024} to determine the radio continuum flux density of our 11 H{\sc i} detections. We find nine of the 11 are detected in the radio continuum map, and their flux densities at $\sim 1.28$\,GHz are given in Table.~\ref{tab:seds}.
Although it would be very unlucky for an artifact from imperfect continuum subtraction to reside at the exact frequency corresponding to the optical redshift, we check whether there are any detections with SNR$>4$ at the spatial position of our sources but spanning the full frequency range of the L1 cube, excluding the region around the optical redshift. We find no lines with SNR$>4$ along the spectral frequency window, lending further confidence to the robustness of our H{\sc i} detections.

As a further check on the robustness of our H{\sc i} emission line candidates we also remeasure the SNR using the noise characteristics measured from random spatial positions around the source but fix the frequency range to that of the extracted emission lines. This allows us to assess whether there is increased noise at the specific frequency of the emission, e.g. from radio-frequency interference. We do this for both an aperture spanning the $10\times 10$\,arcsec$^2$ used in the selection, and also from the single pixel at the position of the optical galaxy. Both methods have pros and cons. The first assumes the correction from an aperture flux to a total flux  using the synthesised beam and an appropriate correction factor. This method potentially measures additional flux if the source is marginally resolved providing a more appropriate measurement of the total H{\sc i} content for such sources and also smooths out noise variations from single-pixel measurements of the peak flux. We note that this effect may be important given our initial selection of inspection by eye, which would favour including sources with "hot pixels" at the position of the optical galaxy.
The second method assumes that the H{\sc i} emission is completely unresolved and adopts the assumption that the peak H{\sc i} flux is representative of the total H{\sc i} flux of the galaxy. 

We list the SNRs measured using these different estimates in Table~\ref{tab:sample} alongside the SNR used to identify the H{\sc i} emission line in the first instance. One can see that the SNR can vary significantly dependent on the method used to estimate the noise. In particular, there are significant differences in the SNR between the aperture measurements and the single-pixel peak flux measurements. This suggests that some of the galaxies may be marginally resolved at the resolution of our data, indeed the 5 objects with low SNR from the single-pixel peak-flux measurement show extended contours in their emission (Figure~\ref{fig:M0Spec}). The optical data (Figure~\ref{fig:RGB_SED}) also suggest that this may be the case, with many of the sources showing extended morphologies over 5-15~arcsec. Thus, our estimates for the H{\sc i} mass are dependent on this. We therefore provide a systematic uncertainty on the H{\sc i} mass in Table~\ref{tab:sample}, for which we use the difference between the single pixel flux measurement and the aperture flux measurement, noting that the aperture flux will be closer to the true line flux for marginally resolved sources, although we still may miss more extended emission. Such extended flux could potentially be recovered by using a larger aperture, however increasing the aperture would also increase the noise and consequently decrease the SNR on the detection. Given that the sources are only likely to be marginally resolved, the missing H{\sc i} mass is likely to be low. We therefore adopt the systematic uncertainty derived above and assume it is symmetric around the mesaured H{\sc i} mass to account for uncharacterised extended emission.
As a check, we convolve a typical disk galaxy of 16~arcsec in extent with the beam of our data, and find that the missing flux is $<10$\ per cent for the total flux, well within the conservative systematic uncertainties that we adopt. After these additional checks, ID3 becomes a significantly less secure candidate, with a SNR=1.5 from its single-pixel measurement. ID9 has a SNR=2.7 from the single-pixel measurement, however we retain them both in our sample as they formally meets the original selection criteria, although we caution that they may be false positives. 

\begin{figure}
  \includegraphics[width=\columnwidth]{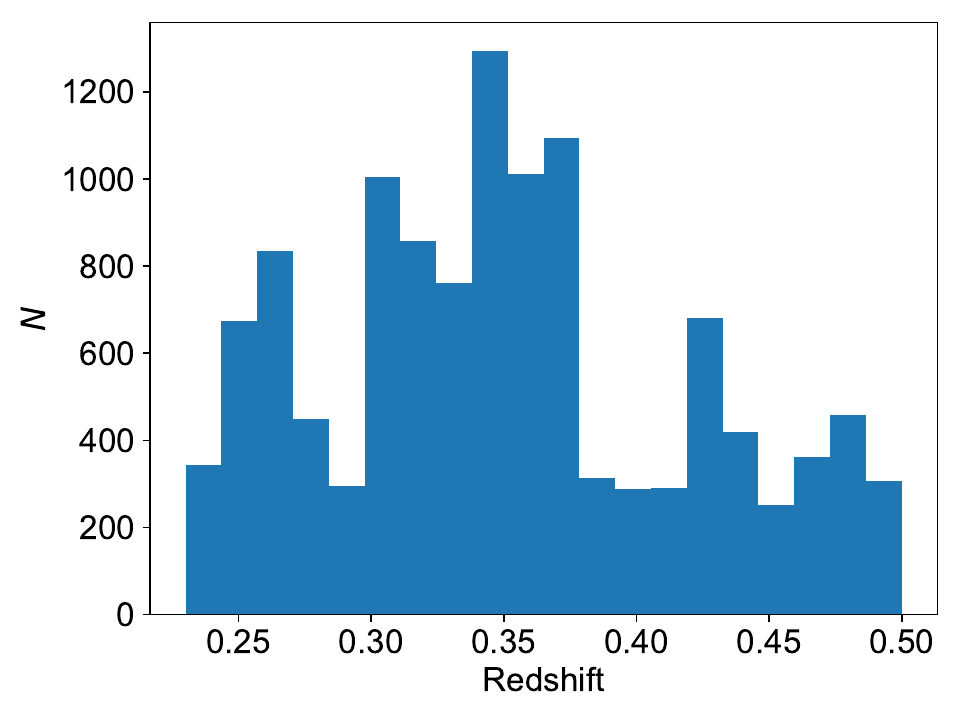}
   \caption{Histogram of the redshift distribution of the parent DESI redshift catalogue over the MIGHTEE-H{\sc i} DR1 area for the redshift range $0.23 < z< 0.5$, corresponding to the MIGHTEE L1 spectral window. }
    \label{fig:DESIzdist}
\end{figure}

Our high-confidence sources span the redshift range $0.25 \lesssim z \lesssim 0.39$ with no reliable sources at $0.39 < z <  0.5$, even though the L1 cube extends to frequencies corresponding to these redshifts. Given that we detect one source at a SNR $= 6.4$ with $z=0.3285$,  naively it may be surprising that we do not detect any sources at higher redshifts. However, the redshift distribution of the parent DESI galaxy sample, shown in Figure~\ref{fig:DESIzdist}, shows that $\sim 75$ per cent of the sample lies at $0.25<z<0.4$ compared to $\sim $25 per cent at $0.4<z<0.5$. Thus, it is unsurprising that we do not detect any H{\sc i} emission at $0.4<z<0.5$ given the redshift distribution of the parent sample, coupled with the depth of the MIGHTEE data.

Figure~\ref{fig:HIzdist} shows the redshift versus H{\sc i} mass plane for the eleven H{\sc i}-detected galaxies and full information for the sample is also provided in Table~\ref{tab:sample}.

\section{H{\sc i} galaxy sample properties}\label{sec:sampleproperties}

The MIGHTEE survey fields were chosen to overlap with surveys at other wavelengths, providing a wealth of ancillary data with which to determine the properties of host galaxies for both H{\sc i} and radio continuum sources. 
We cross-matched the high-confidence H{\sc i} galaxies with the Deep Extragalactic VIsible Legacy Survey \citep[DEVILS;][]{devils} photometric
catalogue \citep[][]{devils-photom}. 
This catalogue provides ultraviolet through to mid- and far-infrared data measured in a consistent way \citep[using ProFOUND; ][]{Robotham2018} on imaging data from the {\em Galaxy Evolution Explorer} \citep[GALEX; ][]{Zamojski2007}, the Canada-France-Hawaii Telescope \citep[CFHT; ][]{Ilbert2006,Capak2007} ($u$-band), HyperSuprimeCam \citep[HSC; ][]{Aihara2019} ($grizy$), Visible-Infrared Survey Telescope for Astronomy \cite[VISTA; ][]{McCracken2012, Jarvis2013} ($YJHK_{s}$), {\em Spitzer Space Telescope} \citep[][]{Lonsdale2003,Sanders2007,Mauduit2012} (mid-infrared), and the {\em Herschel Space Observatory} \citep[][]{Oliver2012}. We find that ten of the galaxies (IDs 1--10) are in the DEVILS catalogue and we show postage stamps of these data in Figure~\ref{fig:RGB_SED}.

ID11 lies beyond the area covered by the multi-wavelength data and we use the data from the {\sc legacy} survey\footnote{\url{viewer.legacysurvey.org}} for this source. There are two plausible counterparts for ID11 within the MIGHTEE synthesised beam, which lie 7.7~arcsec apart (Figure~\ref{fig:RGB11}) and separated by $\sim 180$\,km~s$^{-1}$. We therefore use the stellar mass and SFR given in the DESI catalogue, noting that these are based on a more limited set of imaging data and therefore more uncertain than the stellar masses and SFRs determined for IDs 1--10, and provide the information for both potential counterparts. However, it is feasible that the H{\sc i} emission we detect has a contribution from both of these galaxies given their close proximity both spatially and in redshift space. This may also explain the high H{\sc i} mass and very high velocity width of the line, although we note that the width of the line is highly dependent on whether the line is truly double-horned. We therefore denote these with red symbols in all subsequent relevant figures.

\begin{figure*}
        \includegraphics[width=0.95\columnwidth]{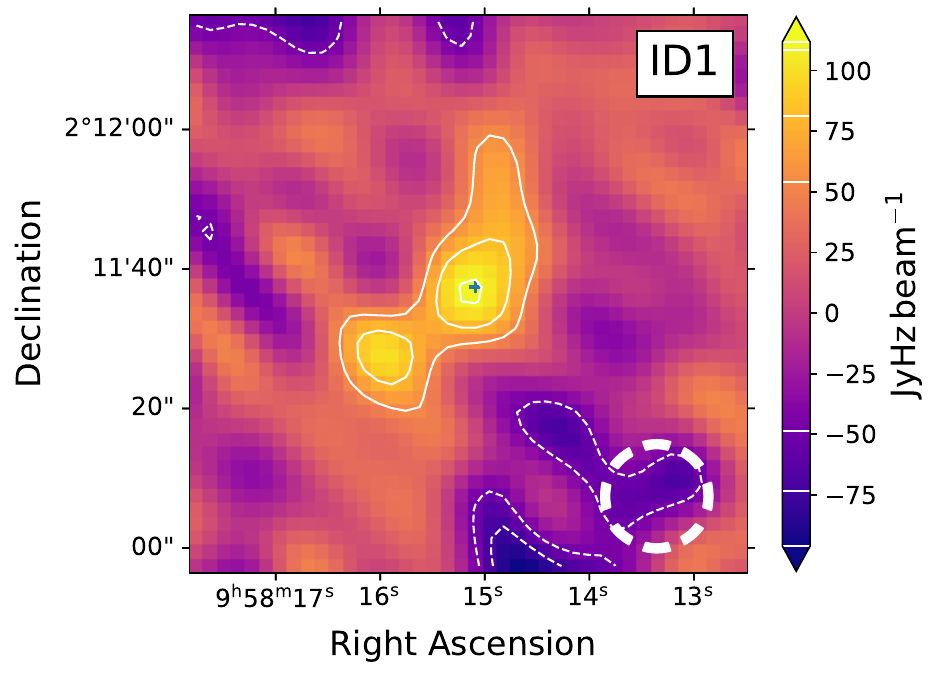}
        \includegraphics[width=0.95\columnwidth]{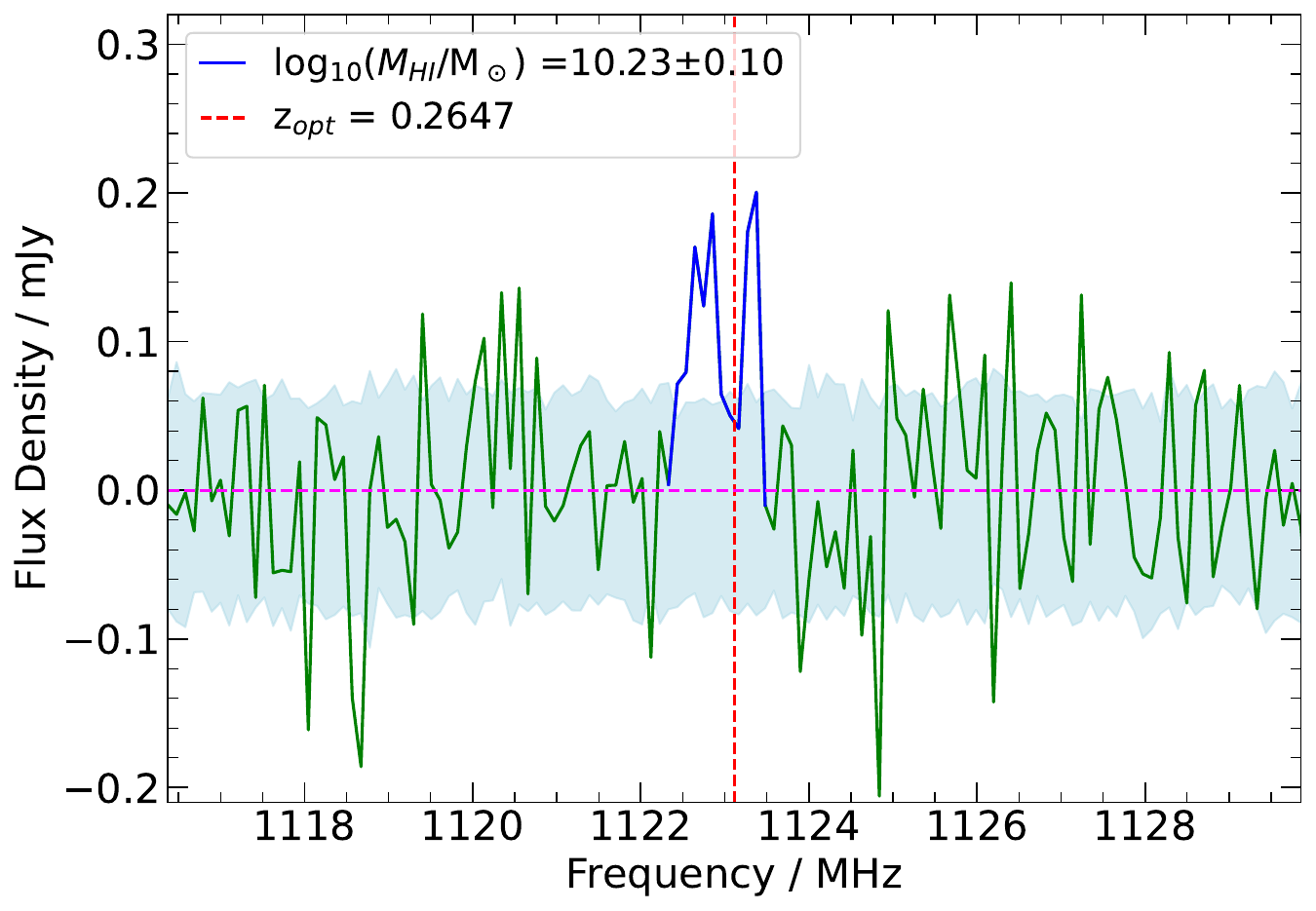}
        \includegraphics[width=0.95\columnwidth]{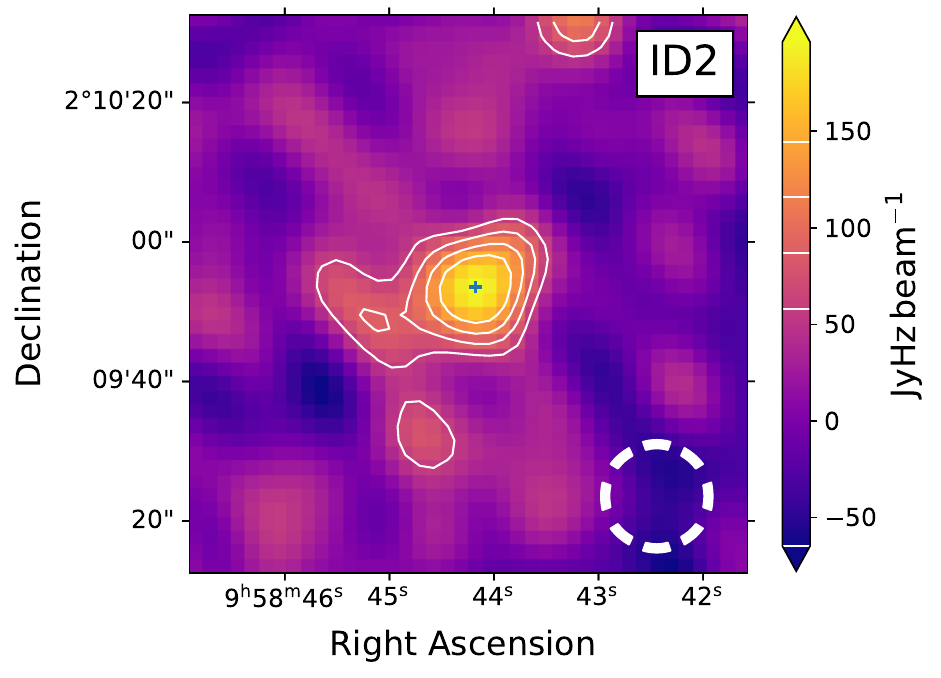}
        \includegraphics[width=0.95\columnwidth]{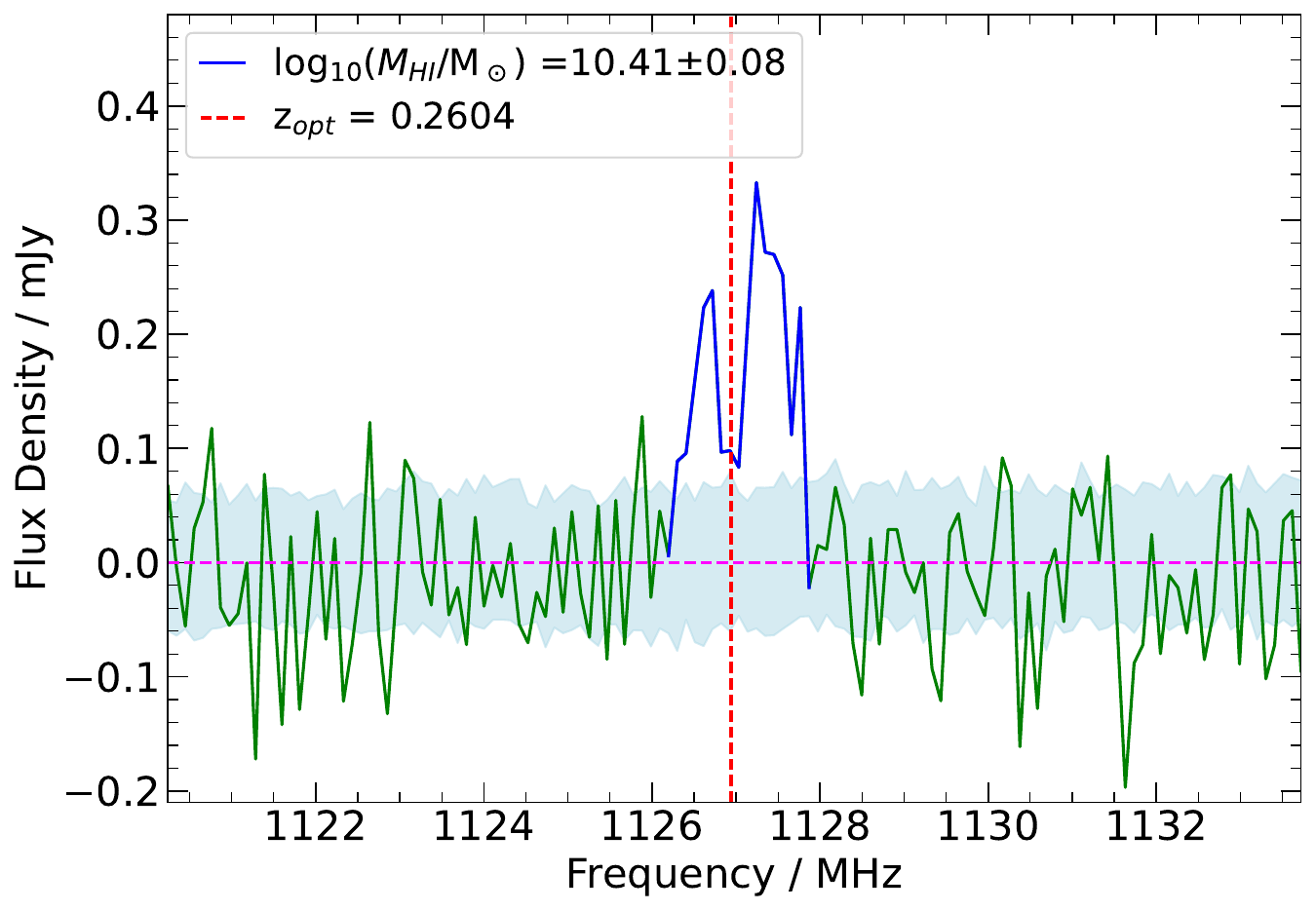}
        \includegraphics[width=0.95\columnwidth]{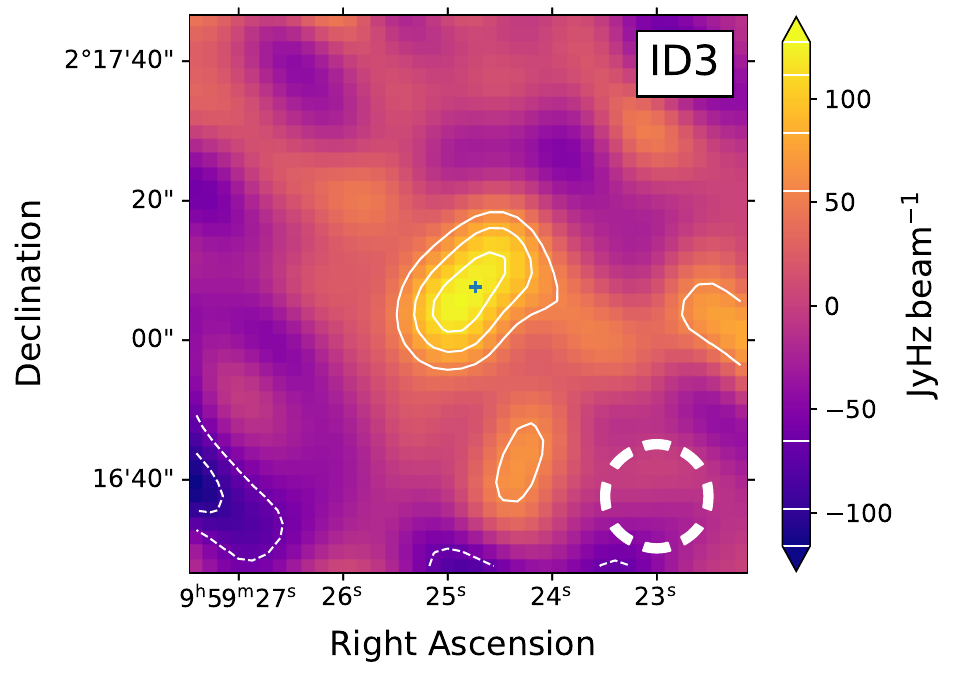}
        \includegraphics[width=0.95\columnwidth]{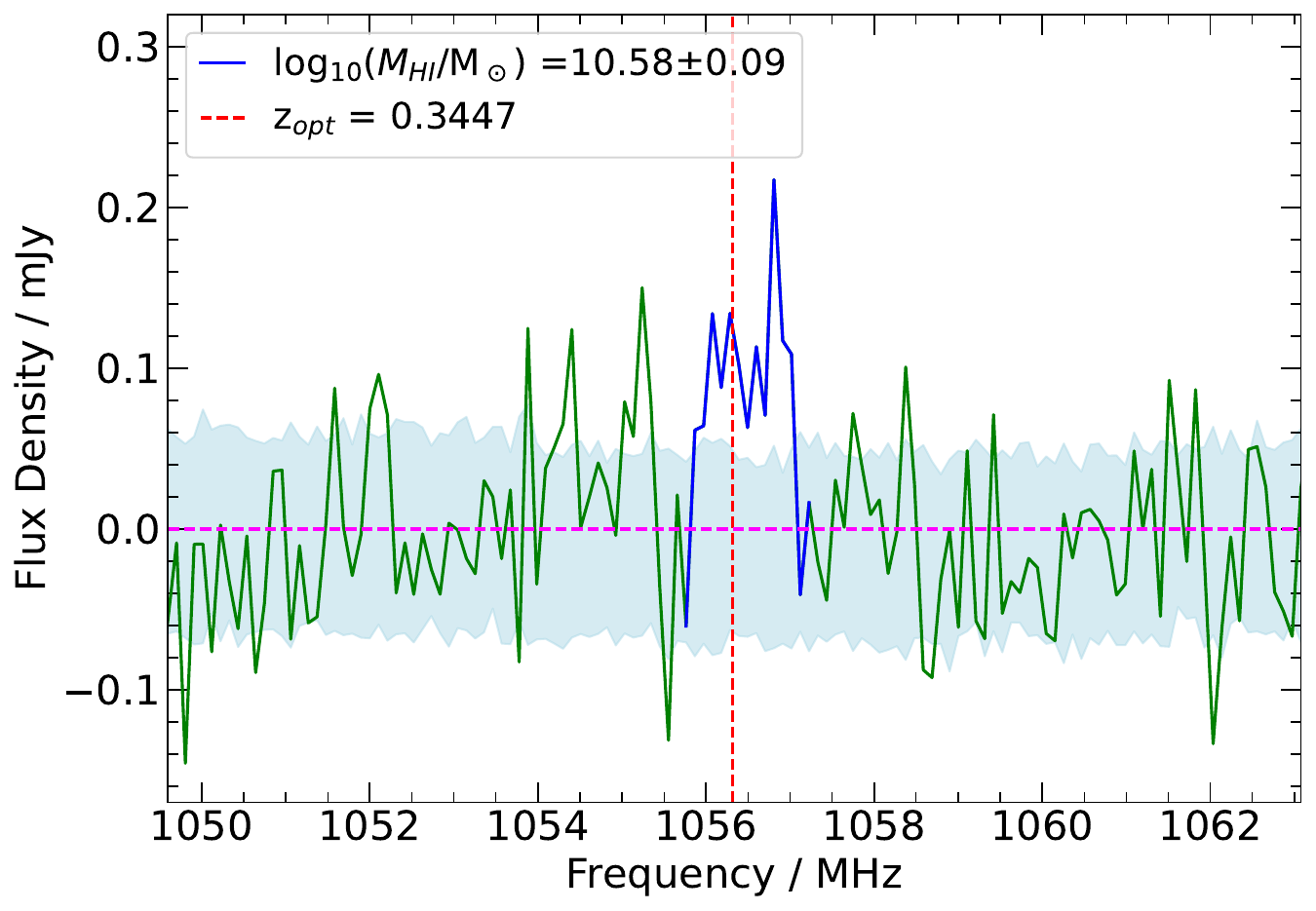}
        \includegraphics[width=0.95\columnwidth]{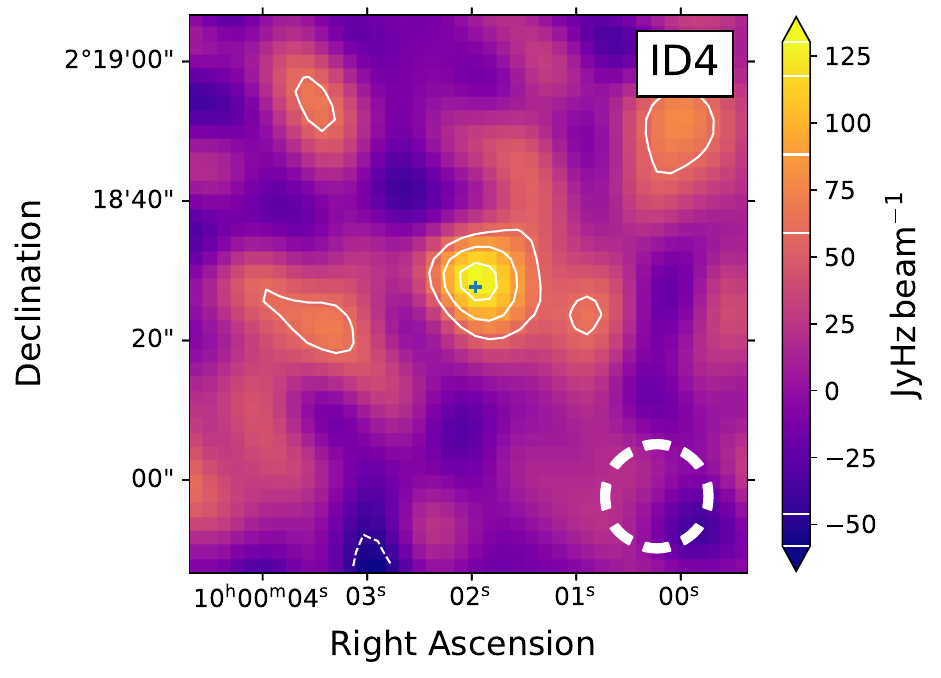}
        \includegraphics[width=0.95\columnwidth]{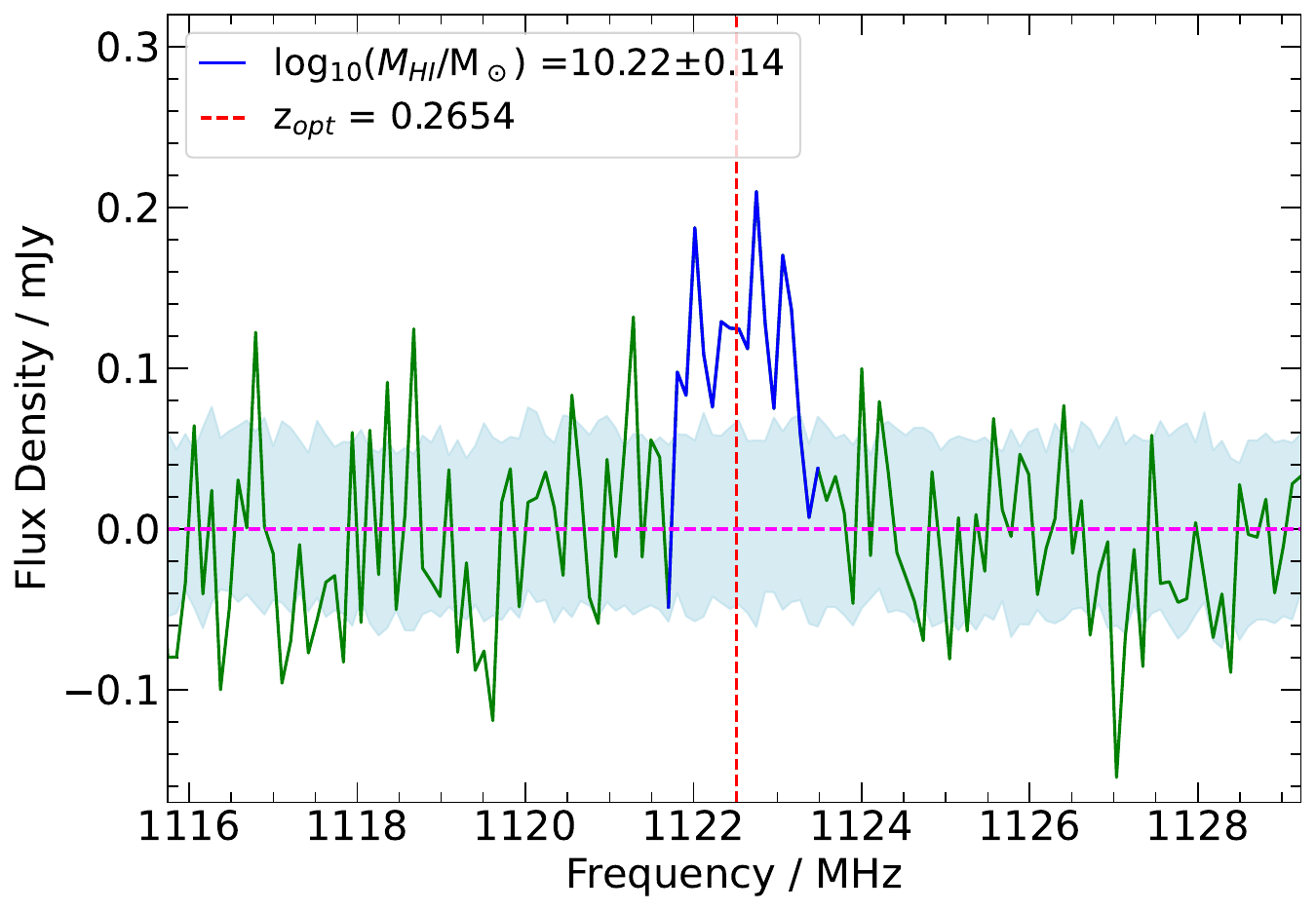}
\end{figure*}

\begin{figure*}
        \includegraphics[width=0.95\columnwidth]{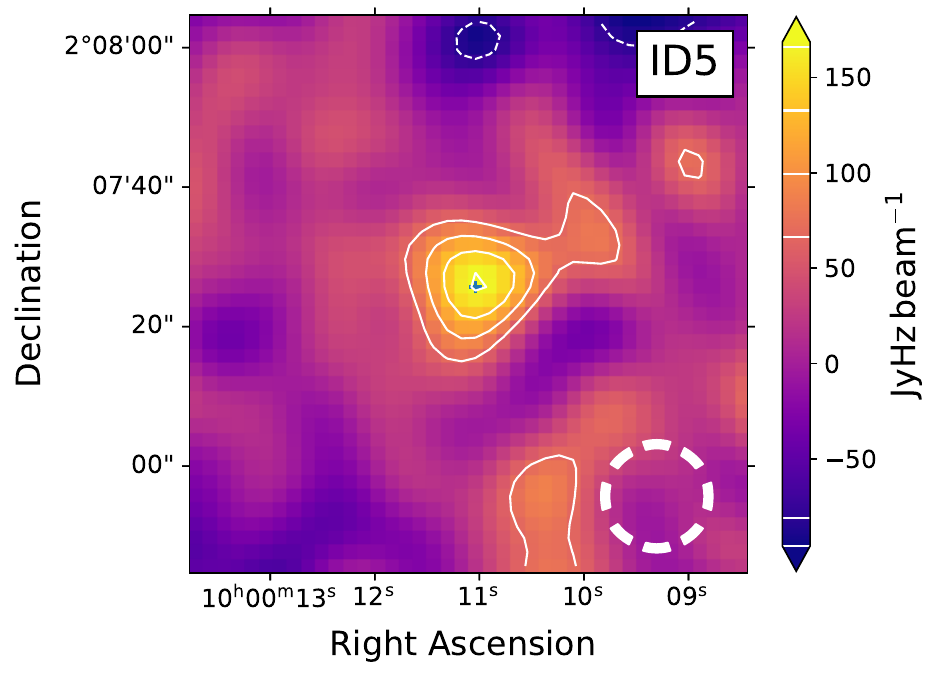}
        \includegraphics[width=0.95\columnwidth]{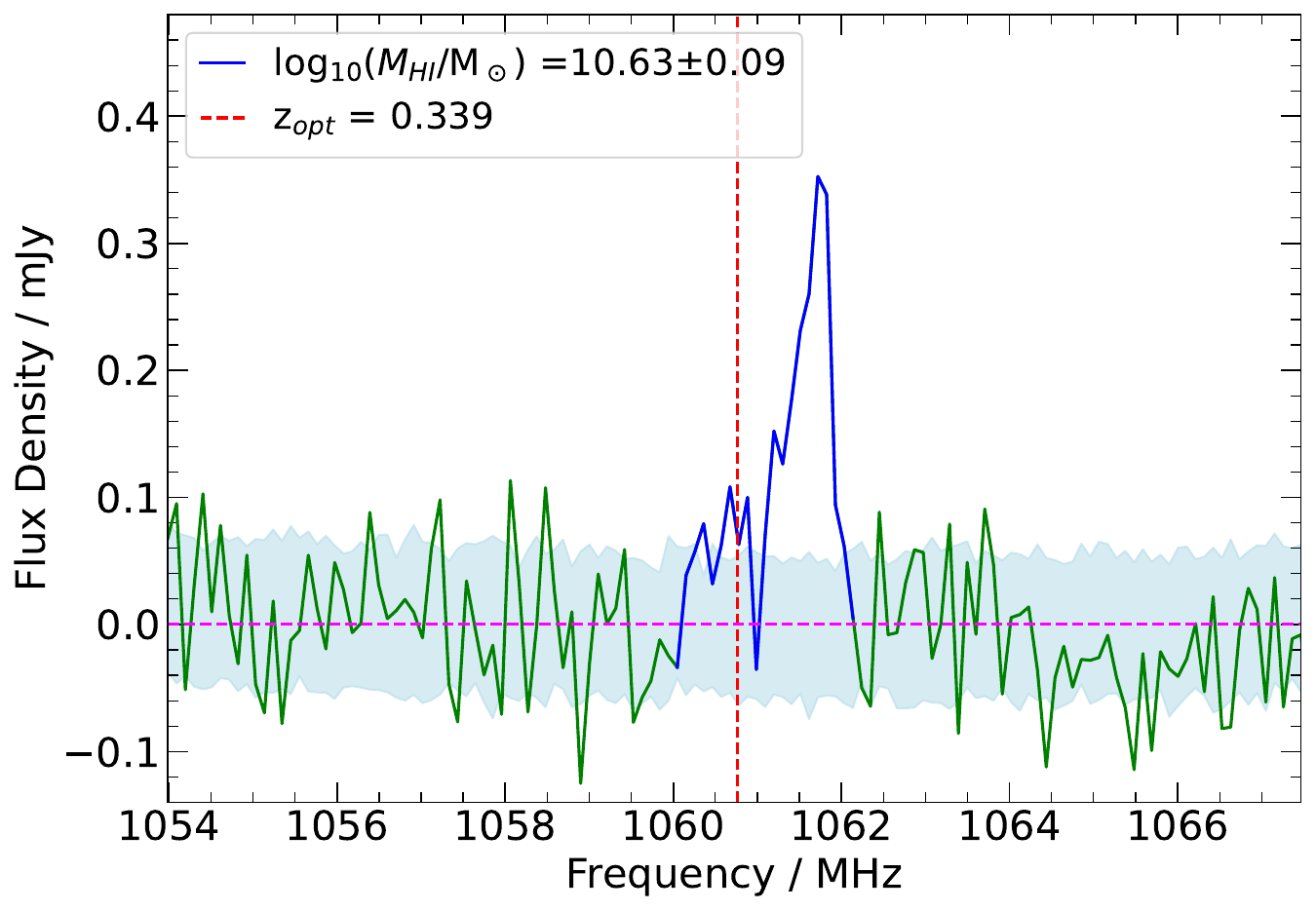}
        \includegraphics[width=0.95\columnwidth]{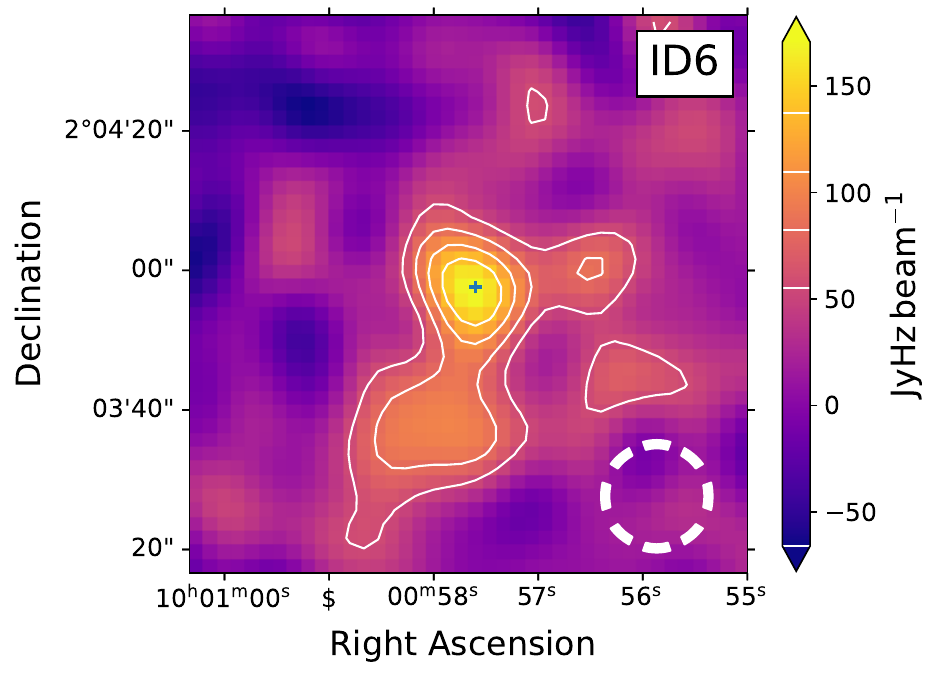}
        \includegraphics[width=0.95\columnwidth]{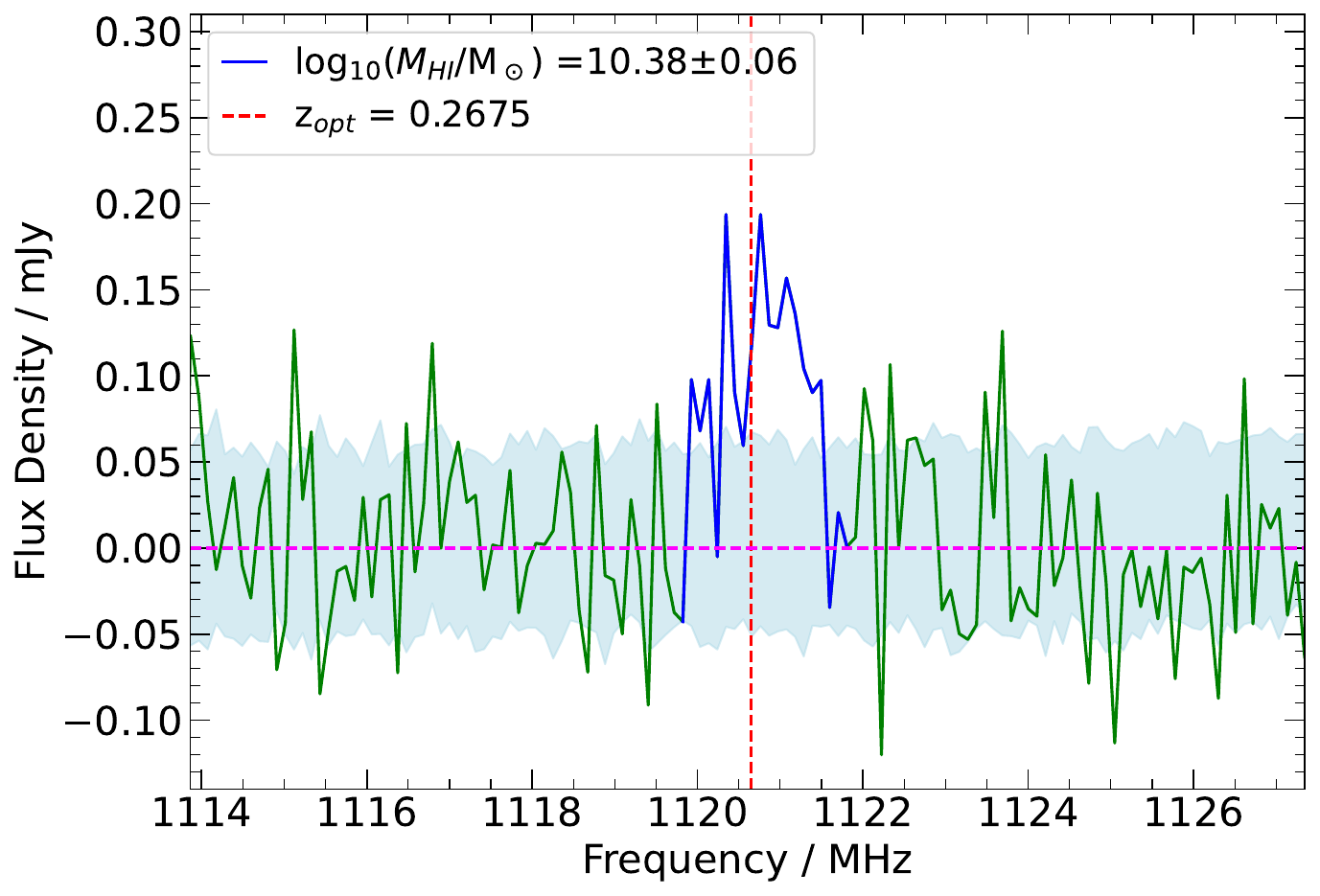}
        \includegraphics[width=0.95\columnwidth]{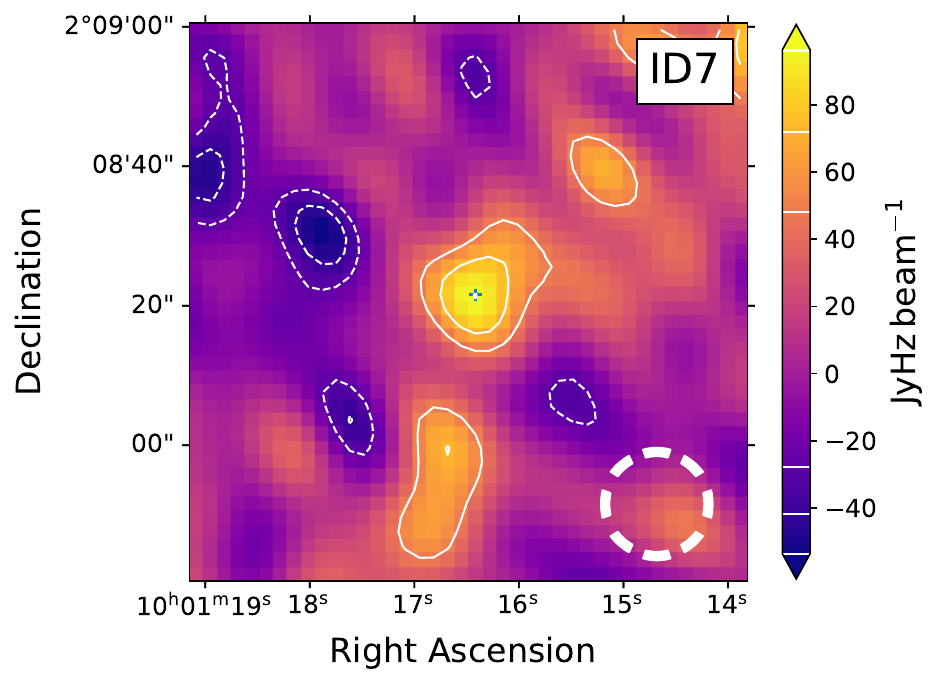}
        \includegraphics[width=0.95\columnwidth]{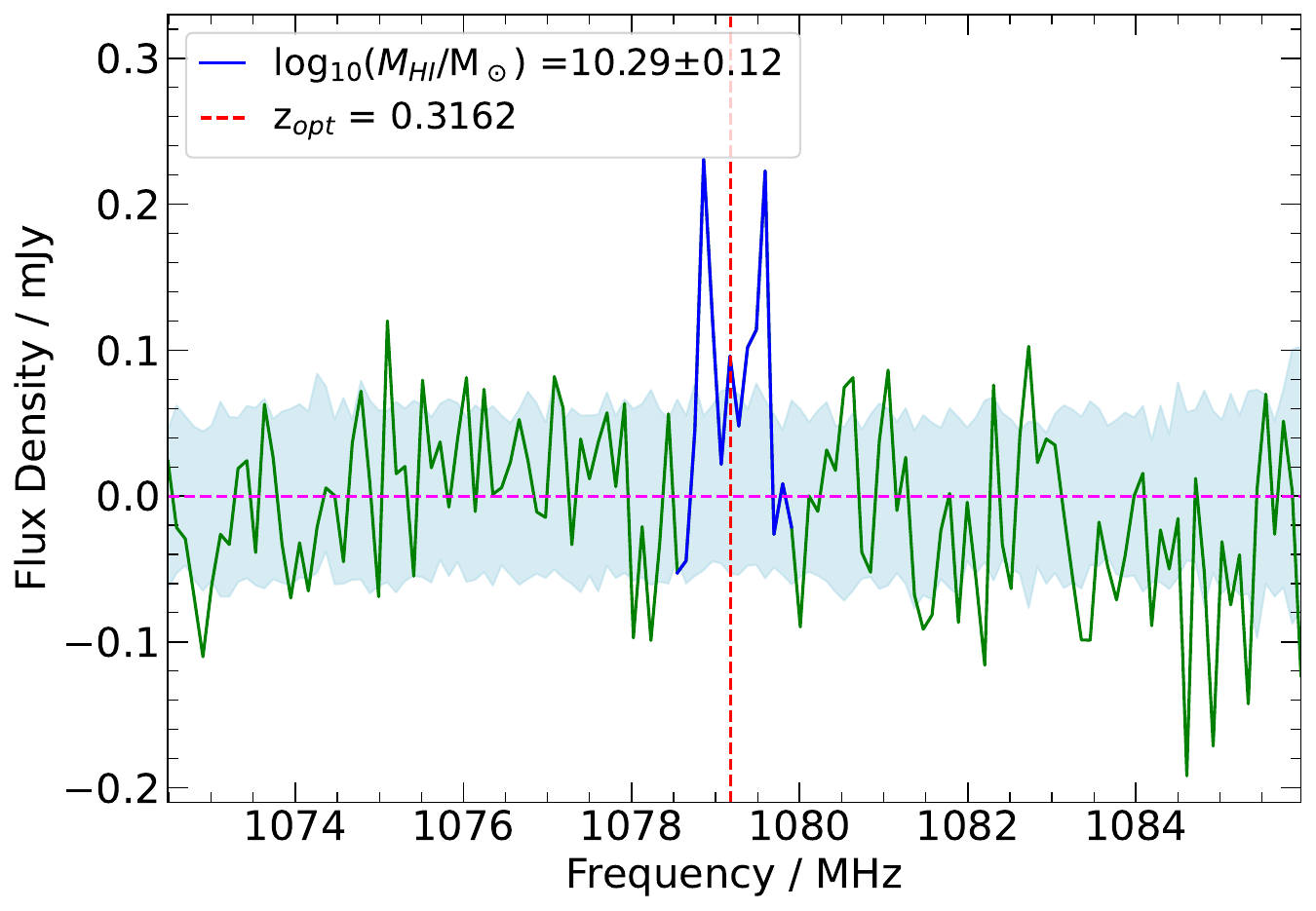}
        \includegraphics[width=0.95\columnwidth]{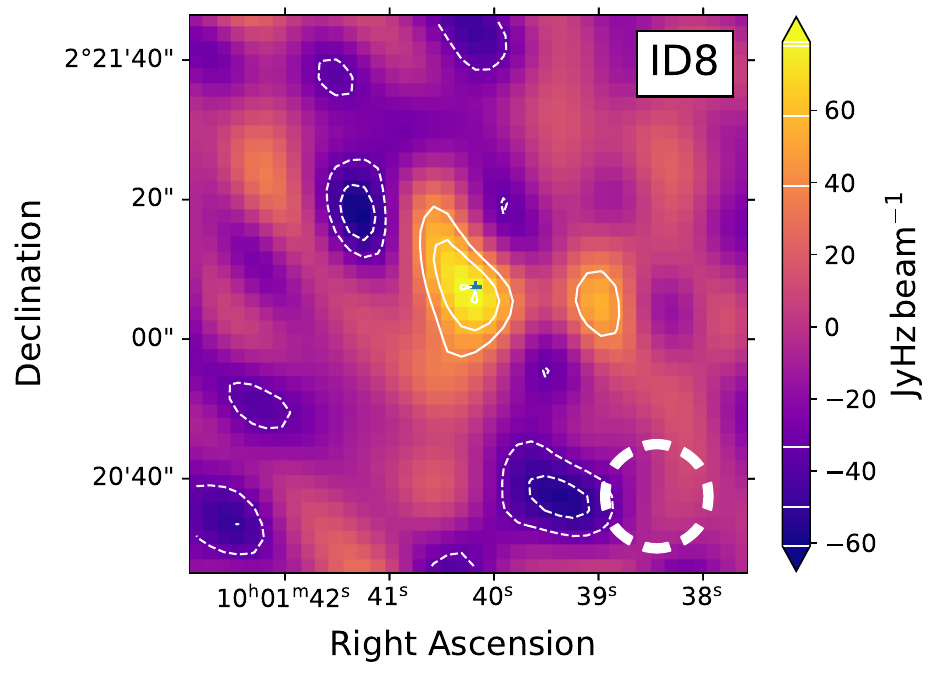}
        \includegraphics[width=0.95\columnwidth]{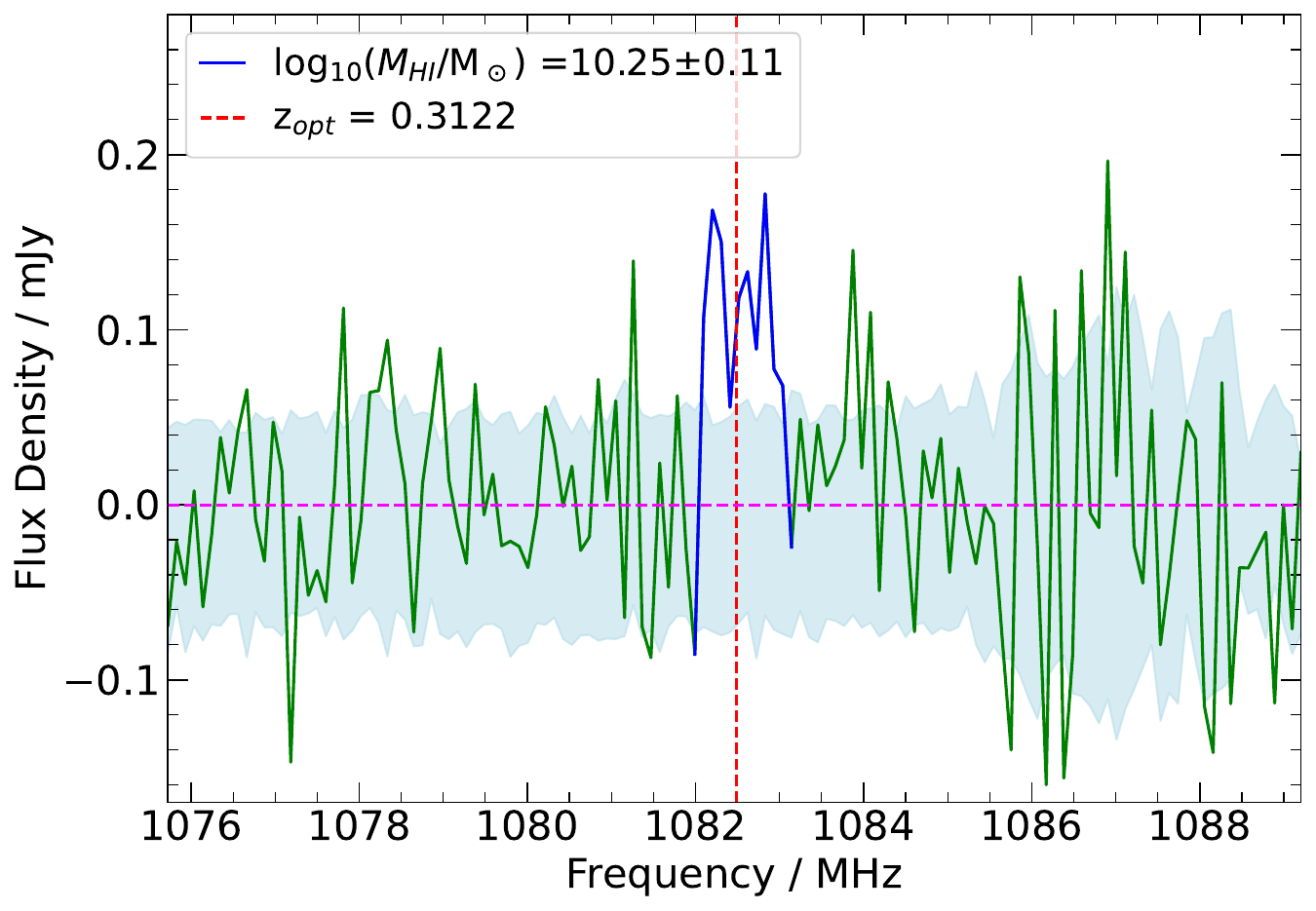}
\end{figure*}
      
\begin{figure*}
        \includegraphics[width=0.95\columnwidth]{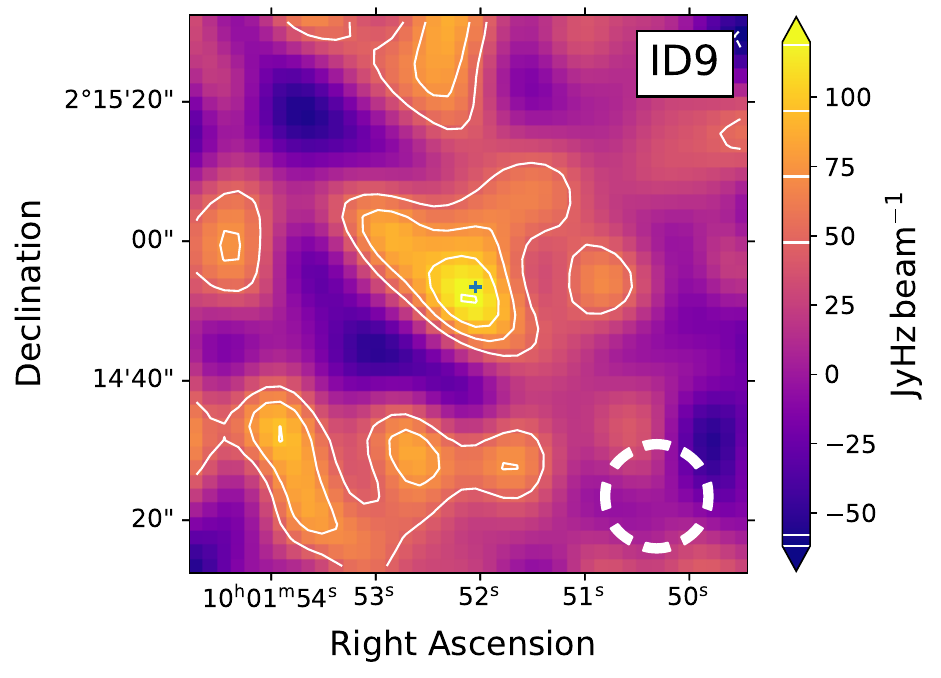}
        \includegraphics[width=0.95\columnwidth]{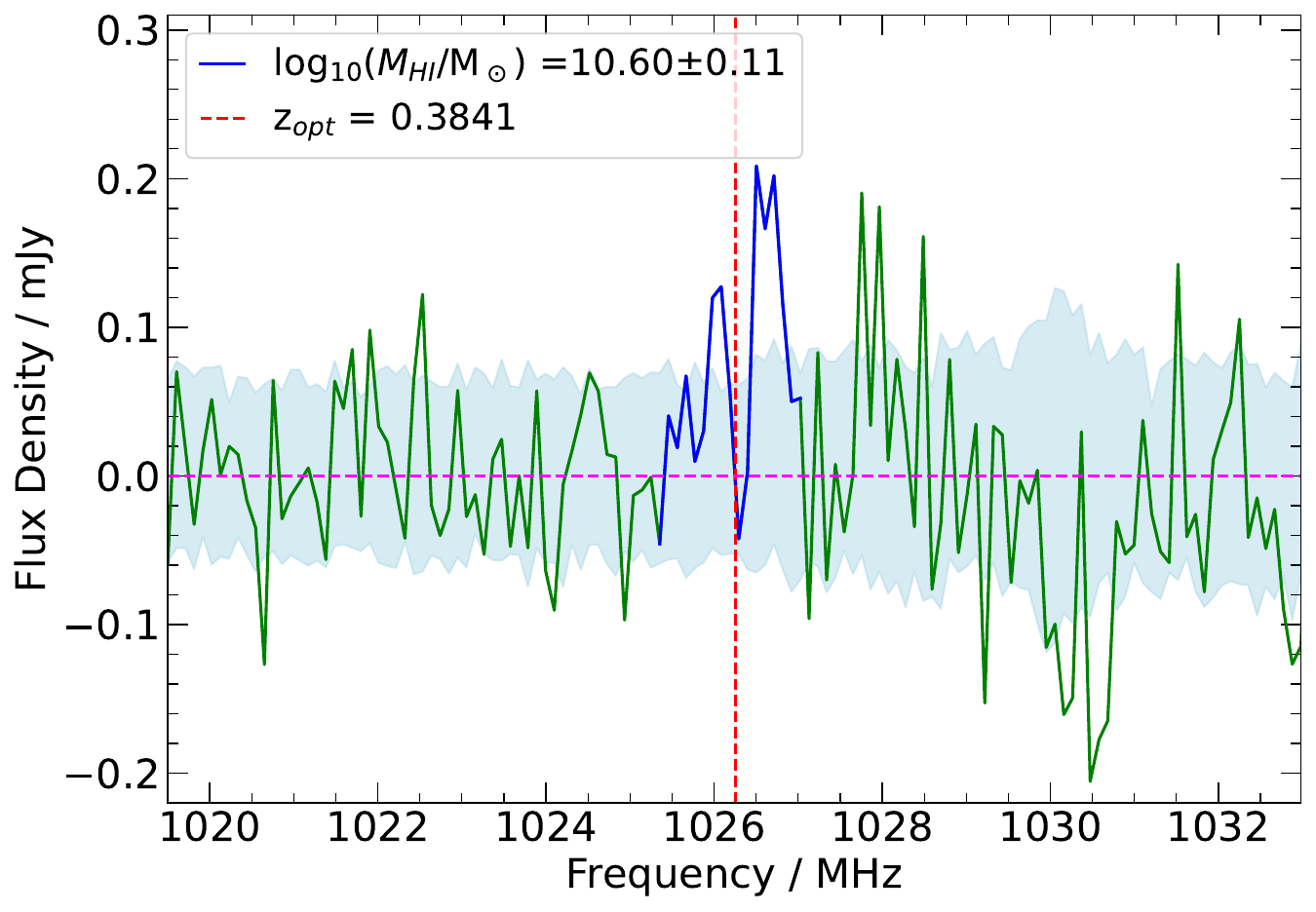}
        \includegraphics[width=0.95\columnwidth]{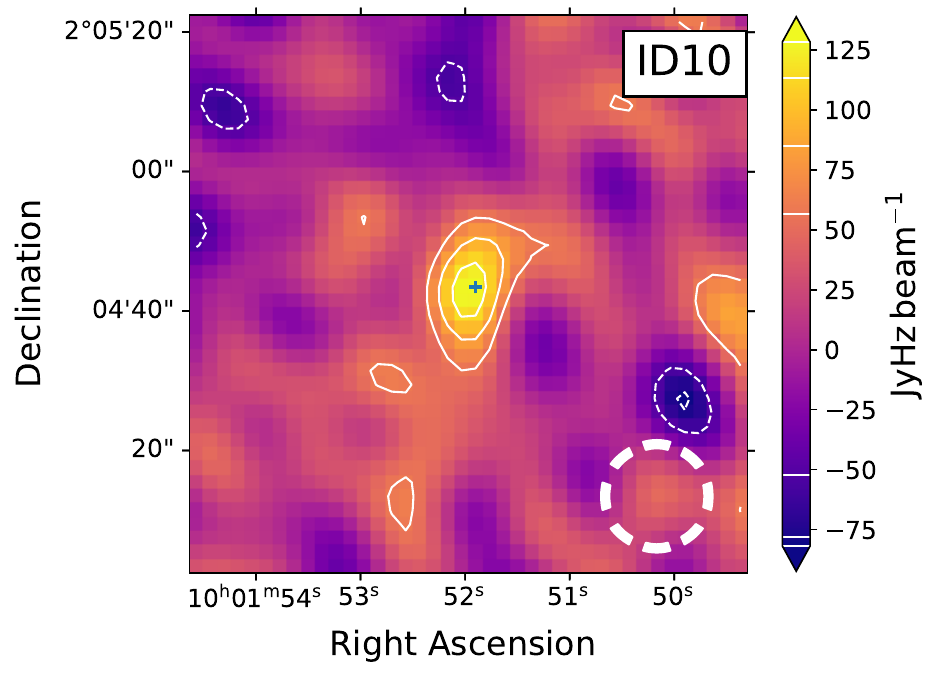}
        \includegraphics[width=0.95\columnwidth]{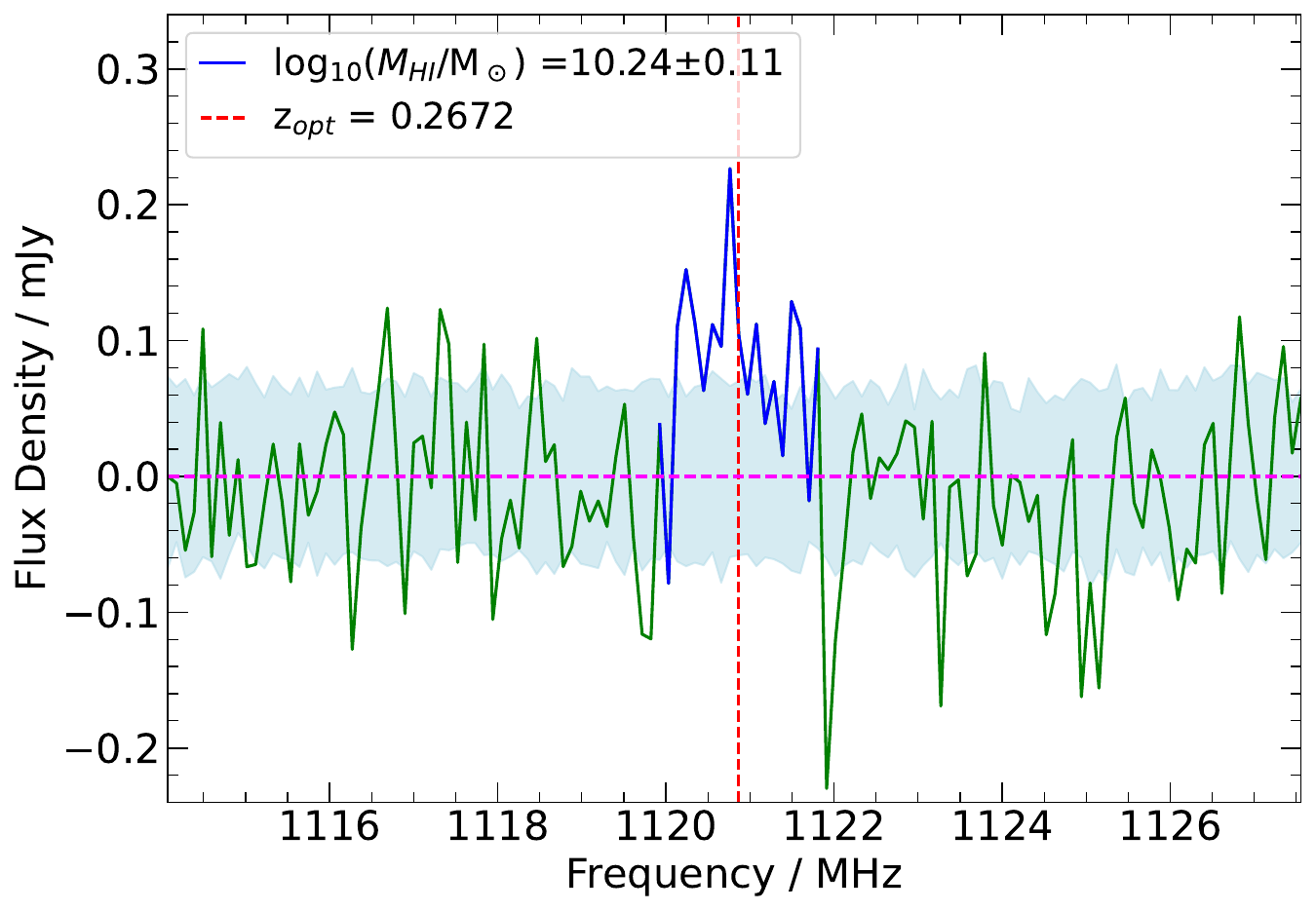}
        \includegraphics[width=0.95\columnwidth]{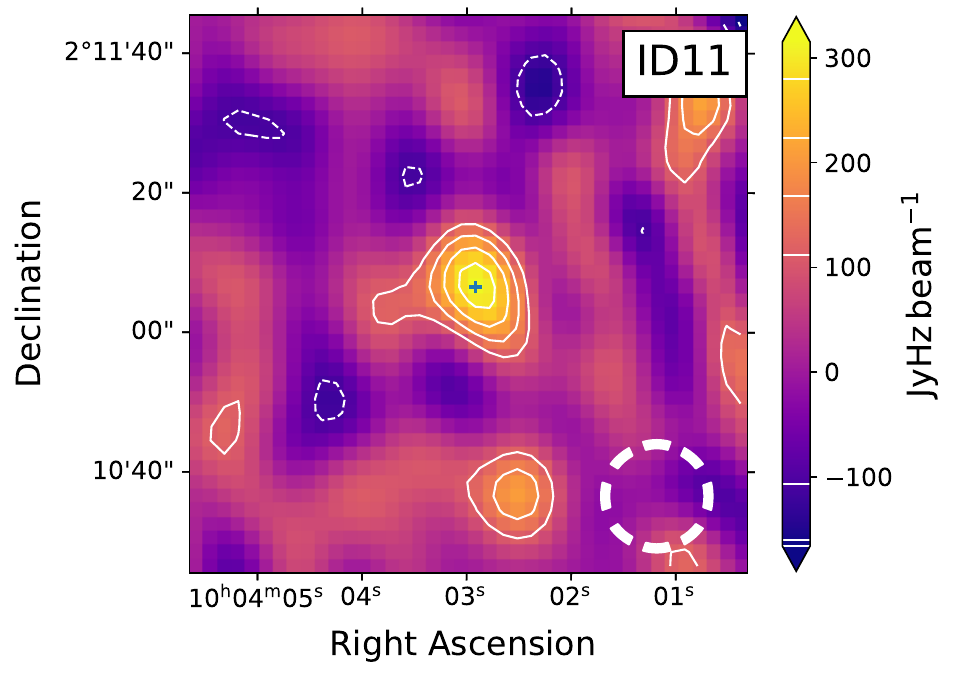}
        \includegraphics[width=0.95\columnwidth]{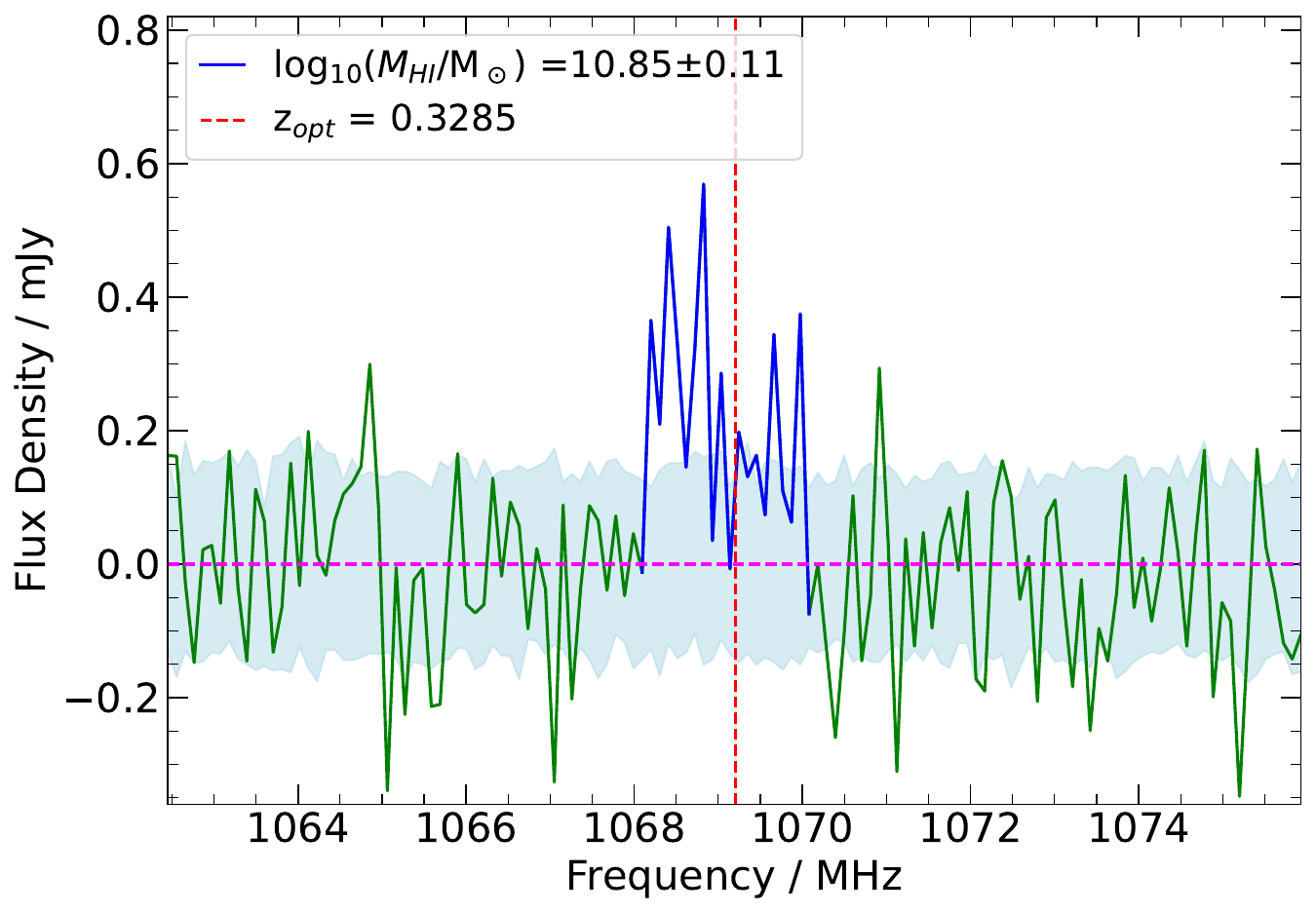}
        \caption{$80\times 80$\,arcsec$^2$ moment-0 maps ({\em left panels}) and the 1-D spectra ({\em right panels}) extracted around the optical redshift, denoted by the vertical dashed line. The solid contour levels shown in the moment-0 maps denote the 2, 3, 4, 5 and 6$\sigma$ levels and dashed contours are the negatives of these. The dashed circles represent the FWHM of the synthesised beam of the MIGHTEE L1 data.
        The part of the 1-D spectrum coloured dark blue is the spectral range used for measuring the SNR and also that used for creating the moment-0 map. The contour levels in the moment-0 maps and the uncertainty on the H{\sc i} masses given in the legend of the spectra are determined by measuring the standard deviation in $\sim 500$ spatial and spectral apertures, as described in Sec.~\ref{sec:search}. To demonstrate the stability of the noise properties of our data as a function of frequency, the shaded light blue regions in the 1-D spectra show the uncertainty measured in each channel from placing 500 apertures in the 3-D data cube and measuring 16th and 84th percentiles from the distribution of measured fluxes in each individual channel. However, these are not used to determine the integrated SNR of the emission line, which is determined as described in Section~\ref{sec:search}. 
        The dashed horizontal magenta line denotes the zero flux-density. We note the noise in ID11 is larger than the other objects due to this object lying towards the edge of the MIGHTEE COSMOS coverage.}
        \label{fig:M0Spec}
    \end{figure*}

\begin{figure}
  \includegraphics[width=\columnwidth]{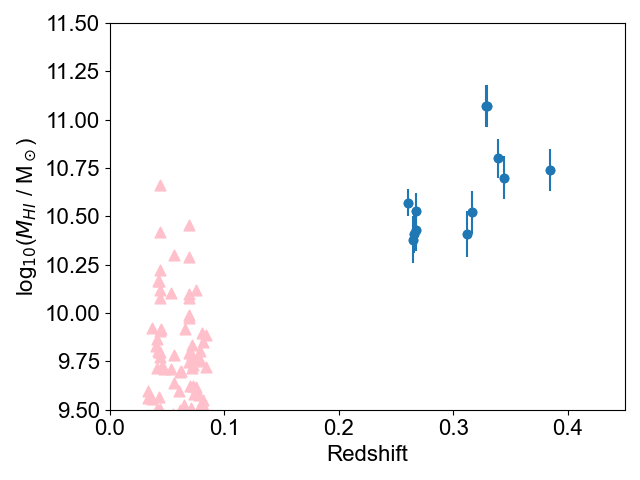}
   \caption{Redshift versus H{\sc i}-mass for the final sample of H{\sc i} detected galaxies presented in this paper (blue circles). Also shown are the H{\sc i}-detected galaxies from the MIGHTEE ES data release (pink triangles). The gap between $0.1<z<0.23$ corresponds to the spectral region with significant radio-frequency interference \citep[RFI; see][]{Heywood2024}.}
    \label{fig:HIzdist}
\end{figure}

\begin{figure*}
  \includegraphics[width=0.8\columnwidth]{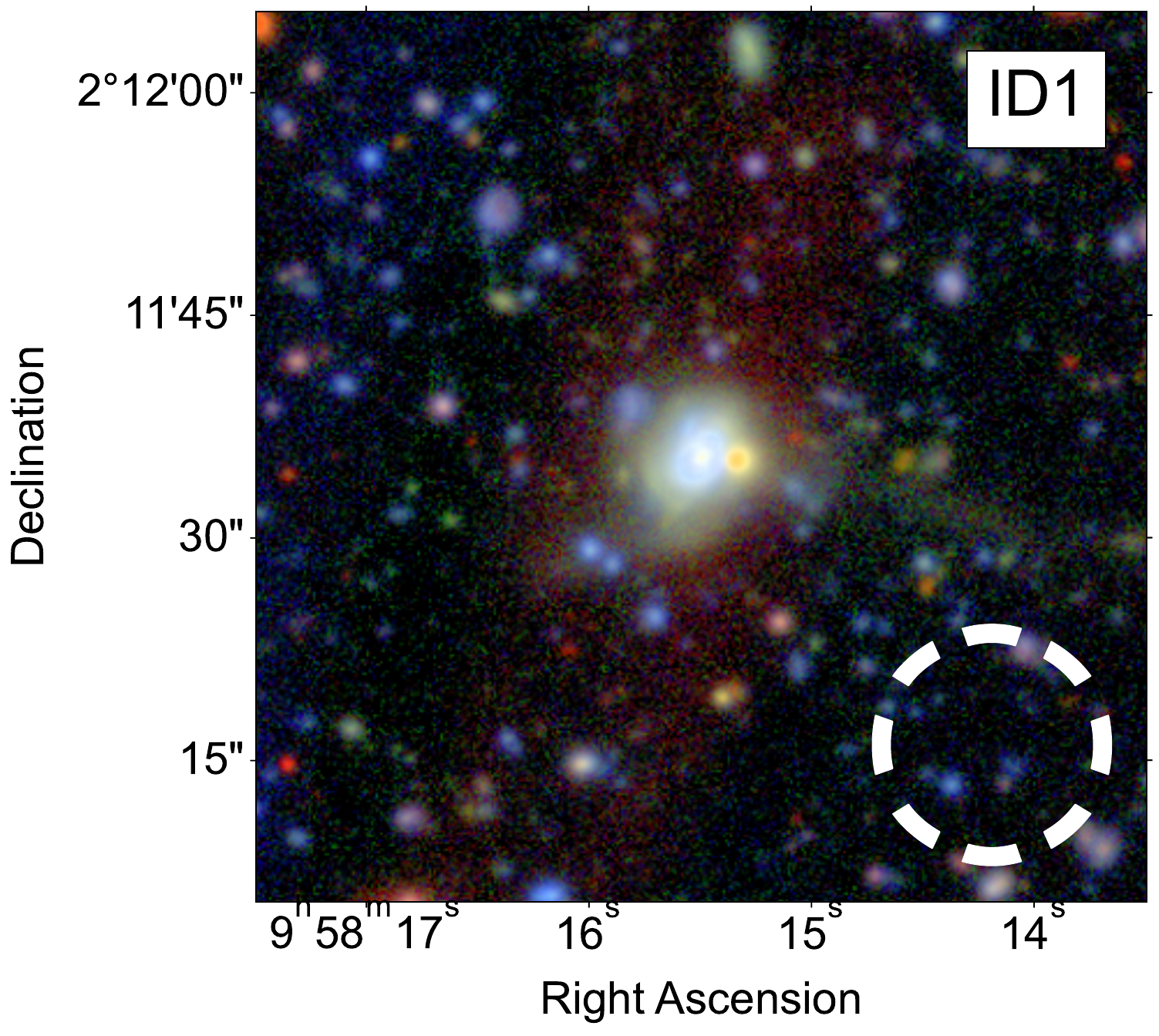}
  \includegraphics[width=1.\columnwidth]{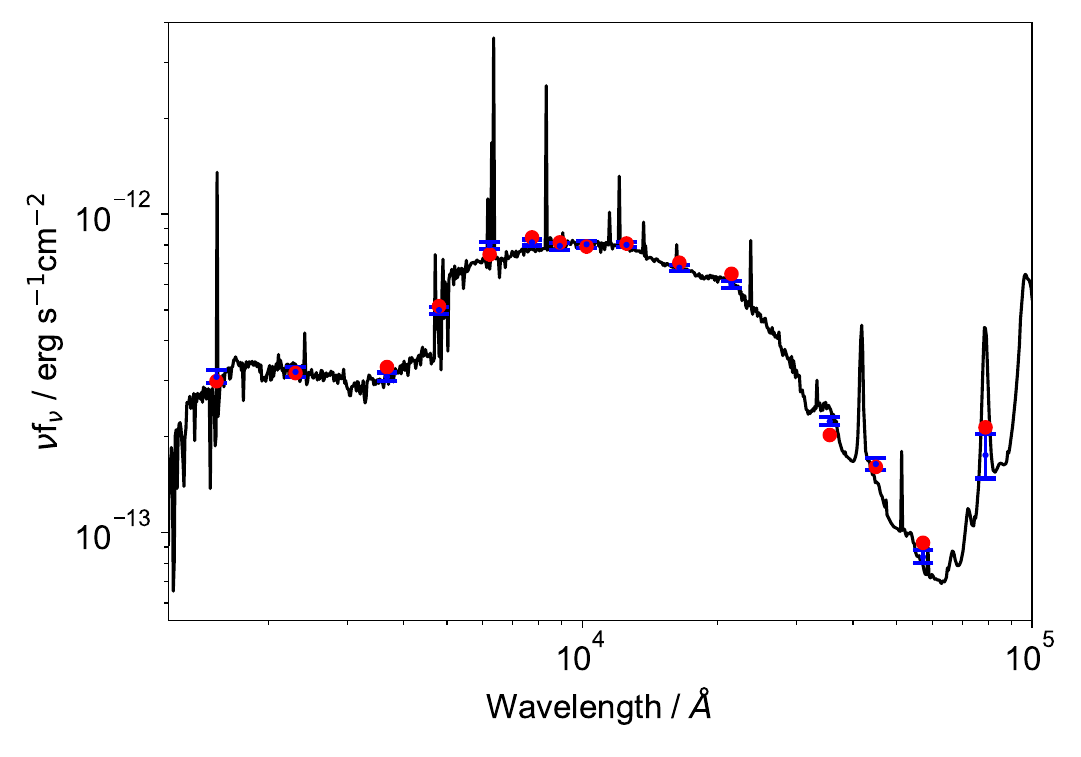}
  \includegraphics[width=0.8\columnwidth]{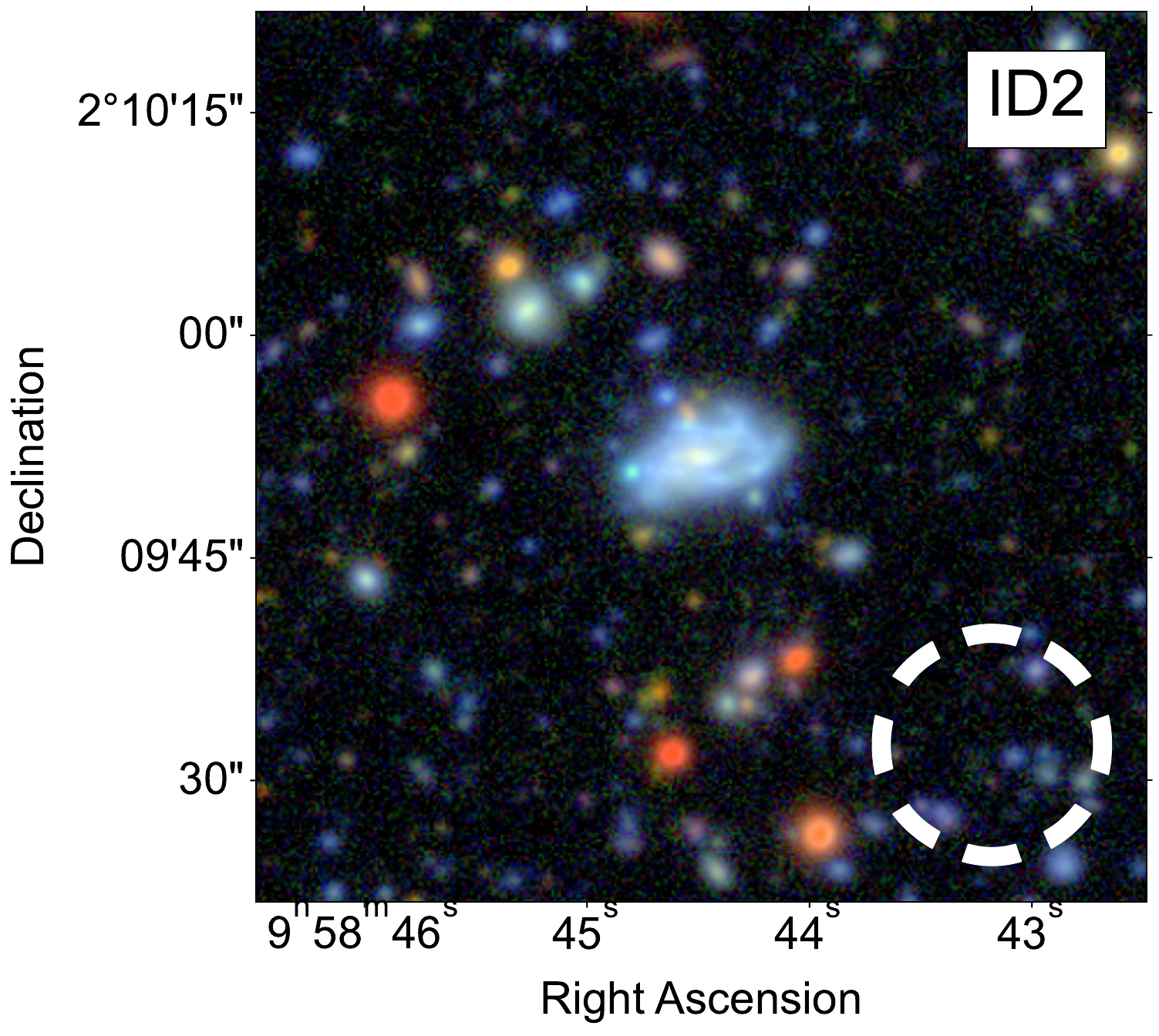}
  \includegraphics[width=1.\columnwidth]{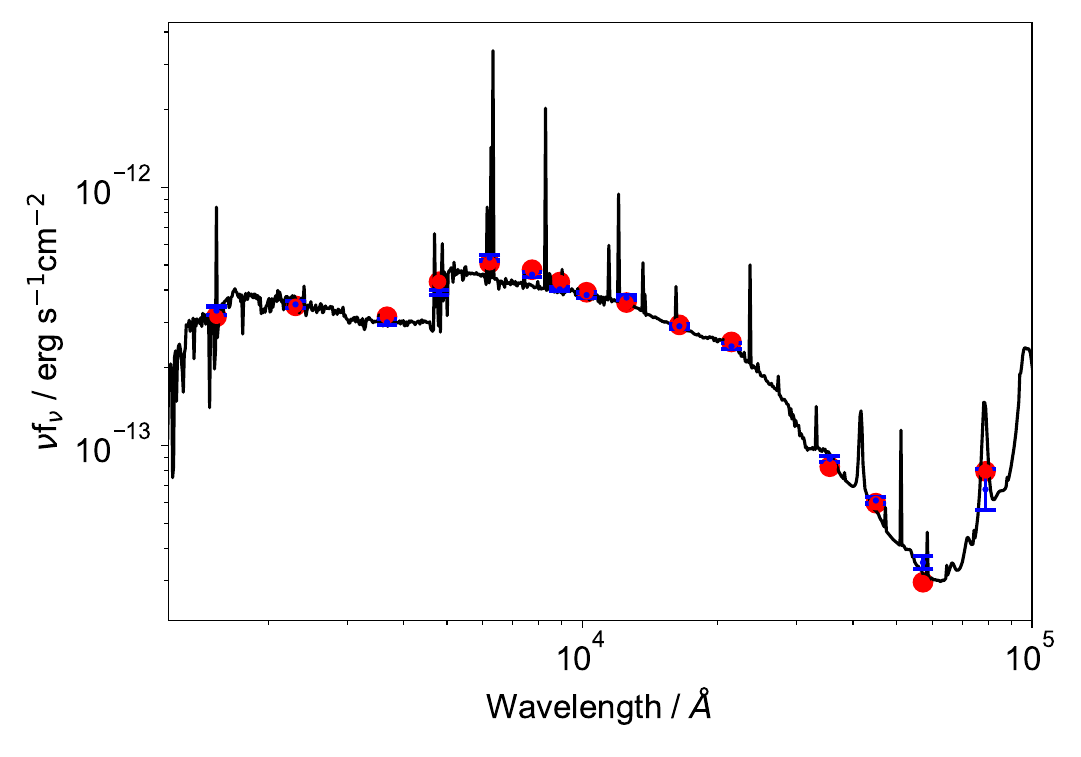}
\includegraphics[width=0.8\columnwidth]{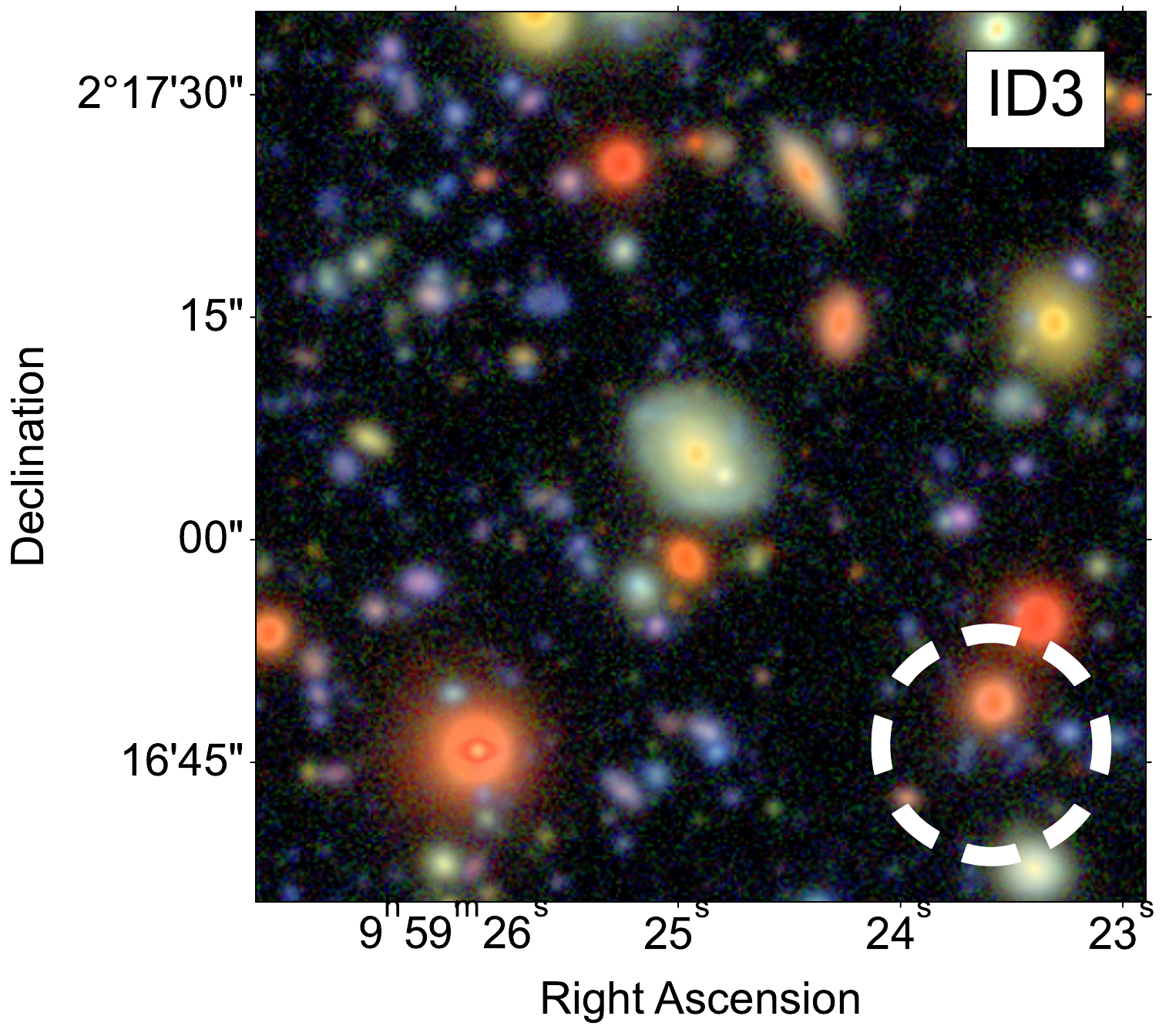}
  \includegraphics[width=1.\columnwidth]{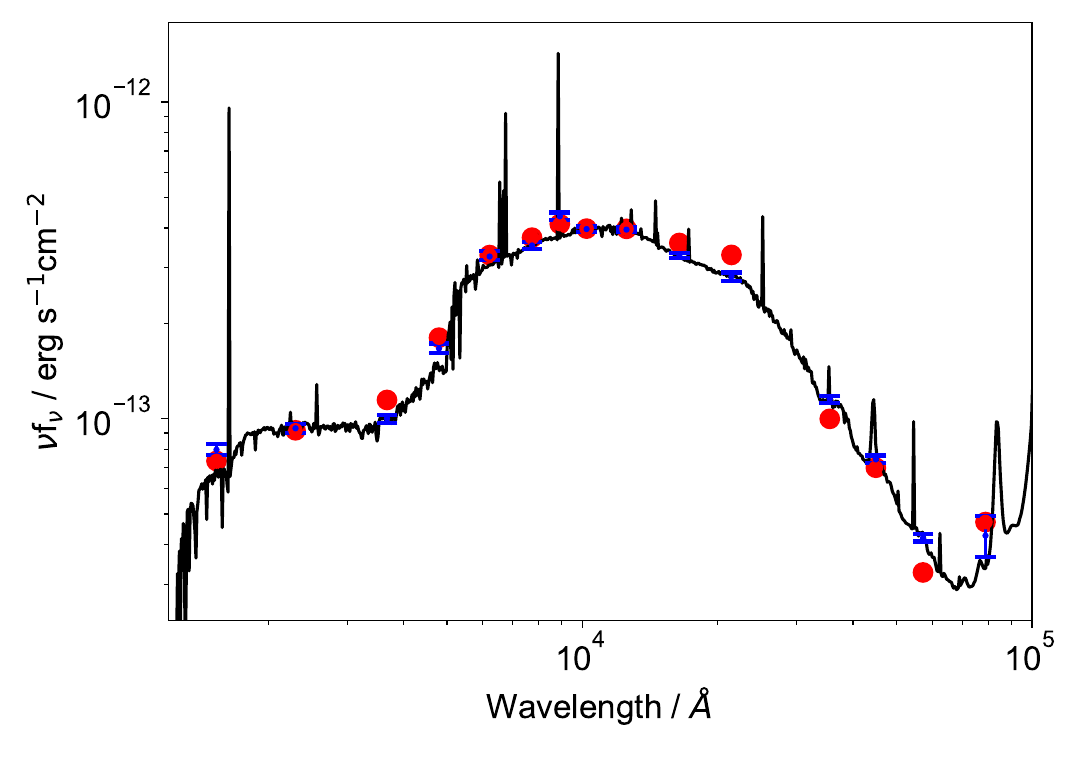}
\includegraphics[width=0.8\columnwidth]{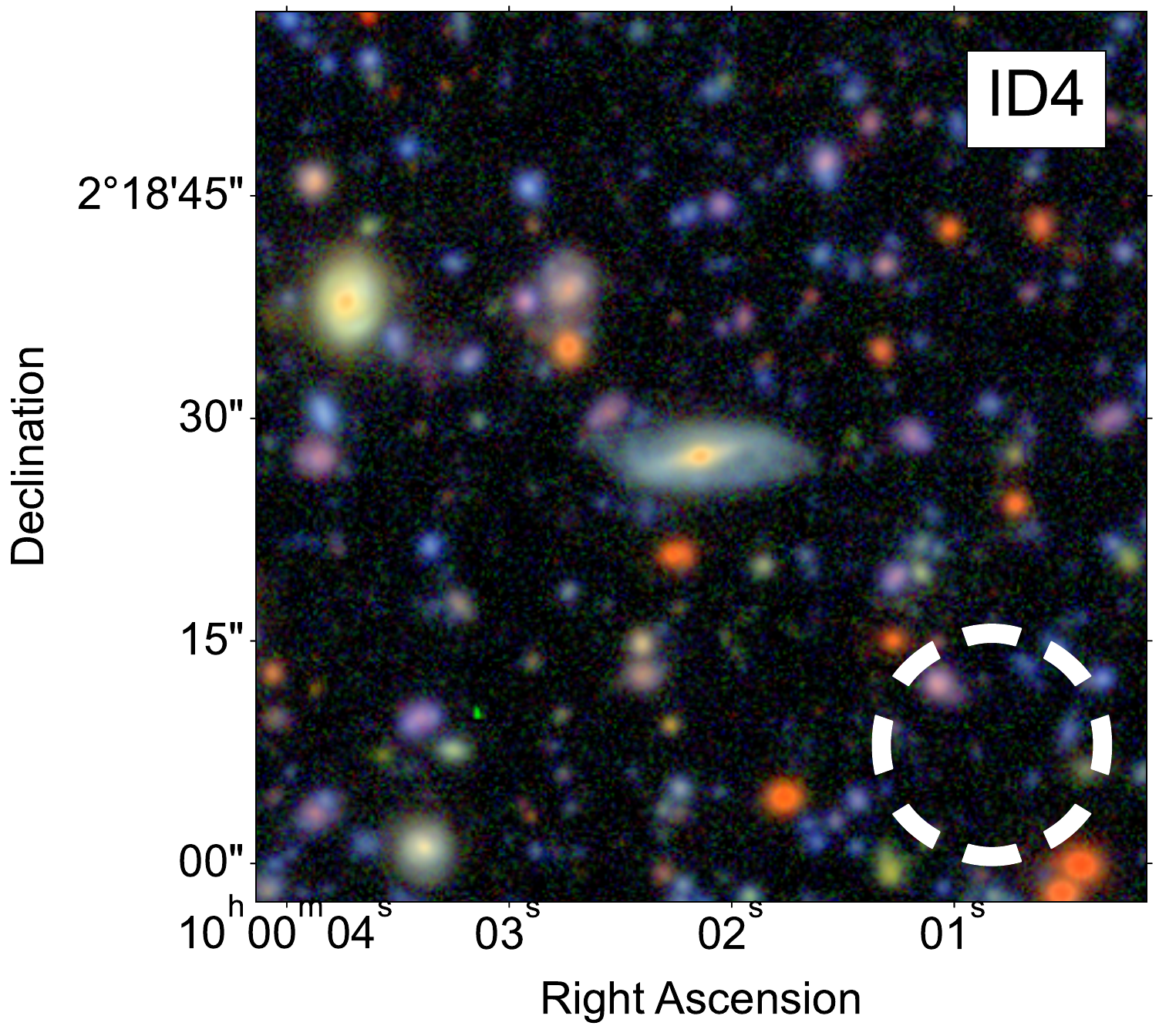}
\includegraphics[width=1.\columnwidth]{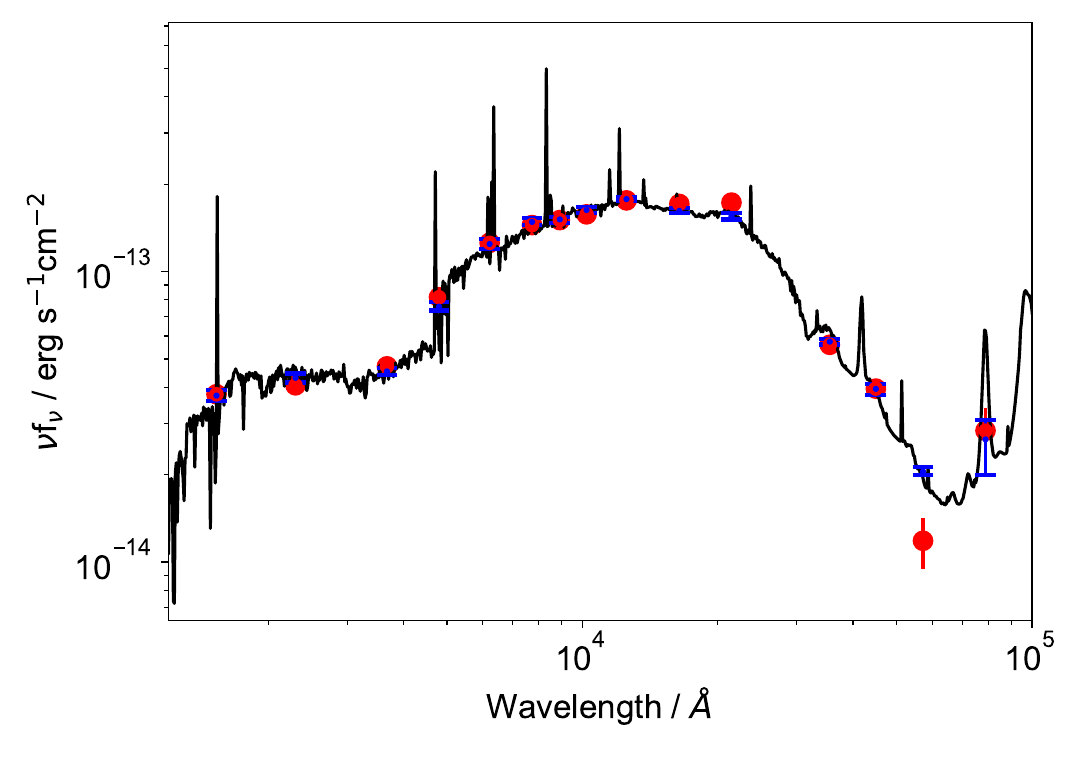}
\end{figure*}

\begin{figure*}
\includegraphics[width=0.8\columnwidth]{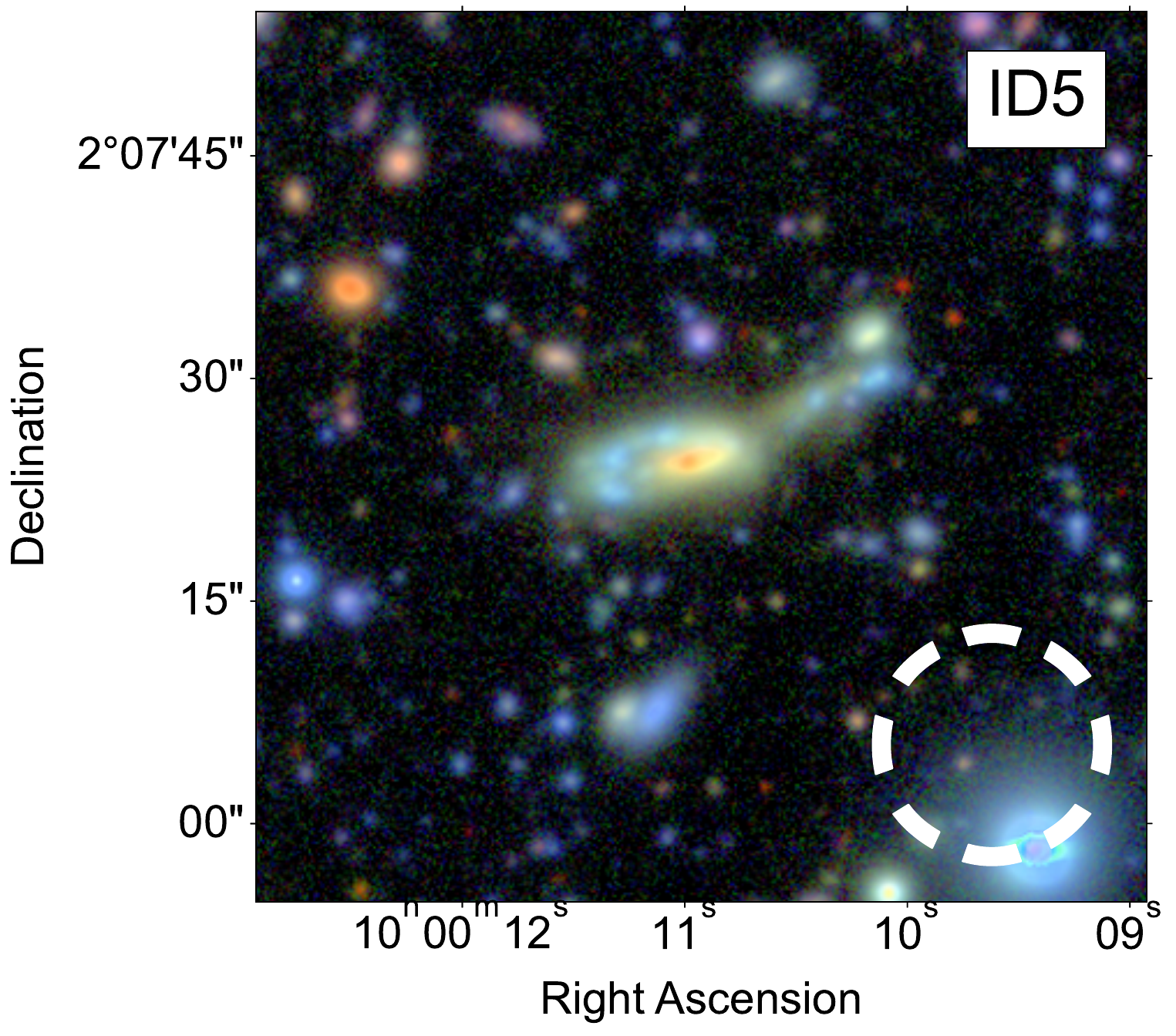}
\includegraphics[width=1.\columnwidth]{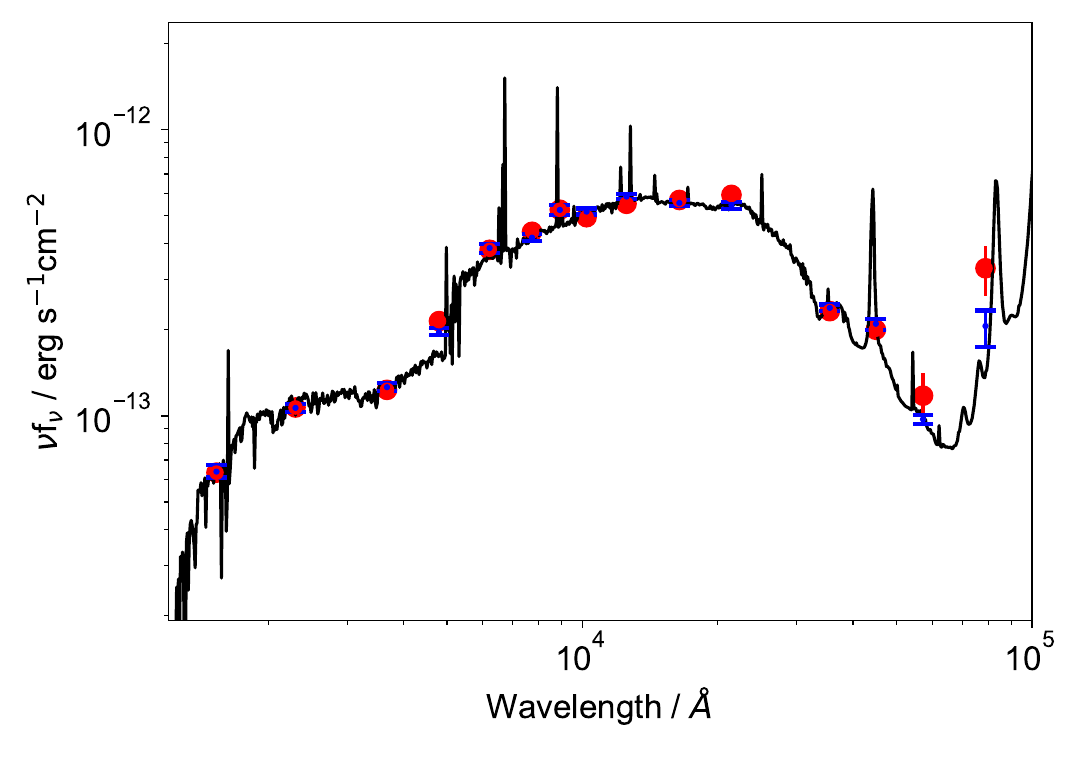}
\includegraphics[width=0.8\columnwidth]{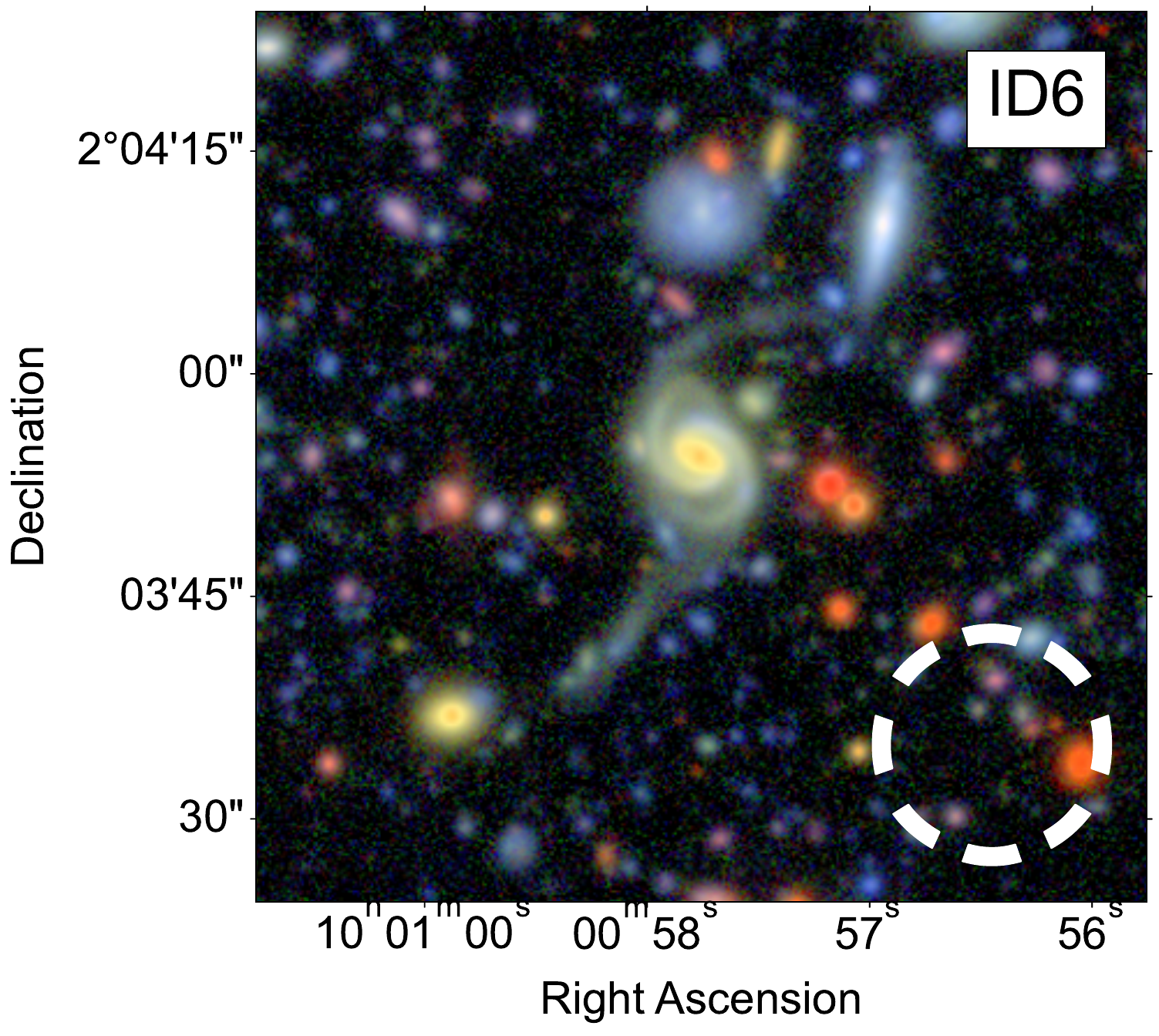}
\includegraphics[width=1.\columnwidth]{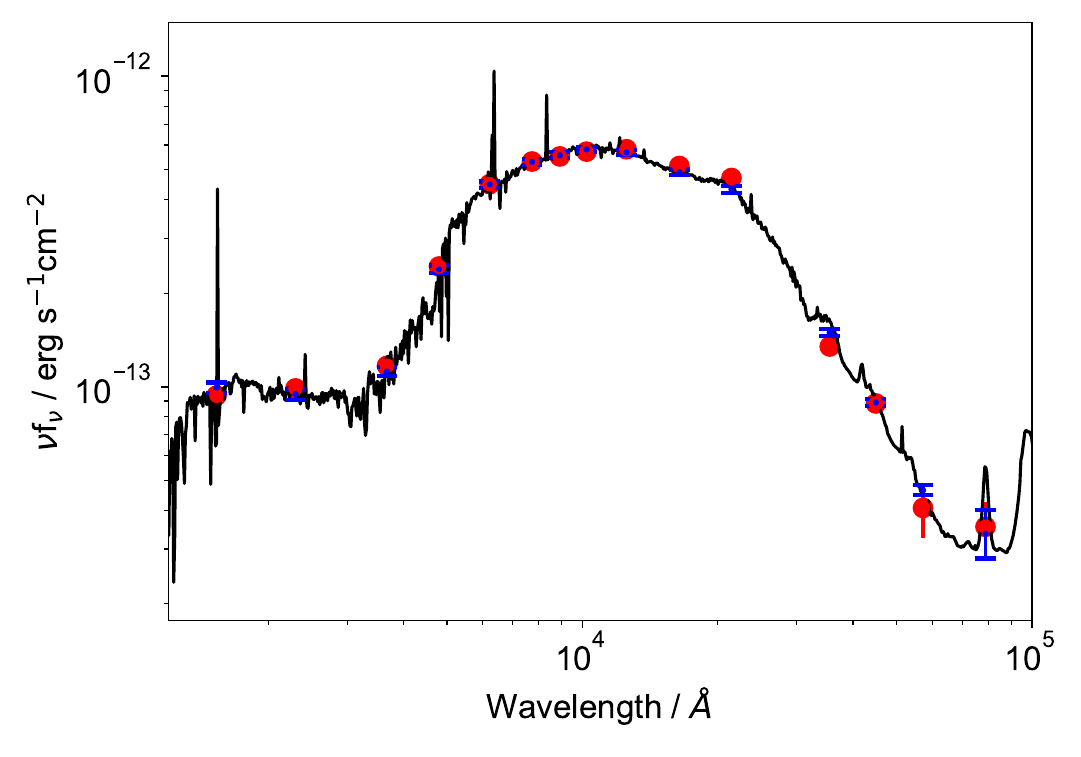}
\includegraphics[width=0.8\columnwidth]{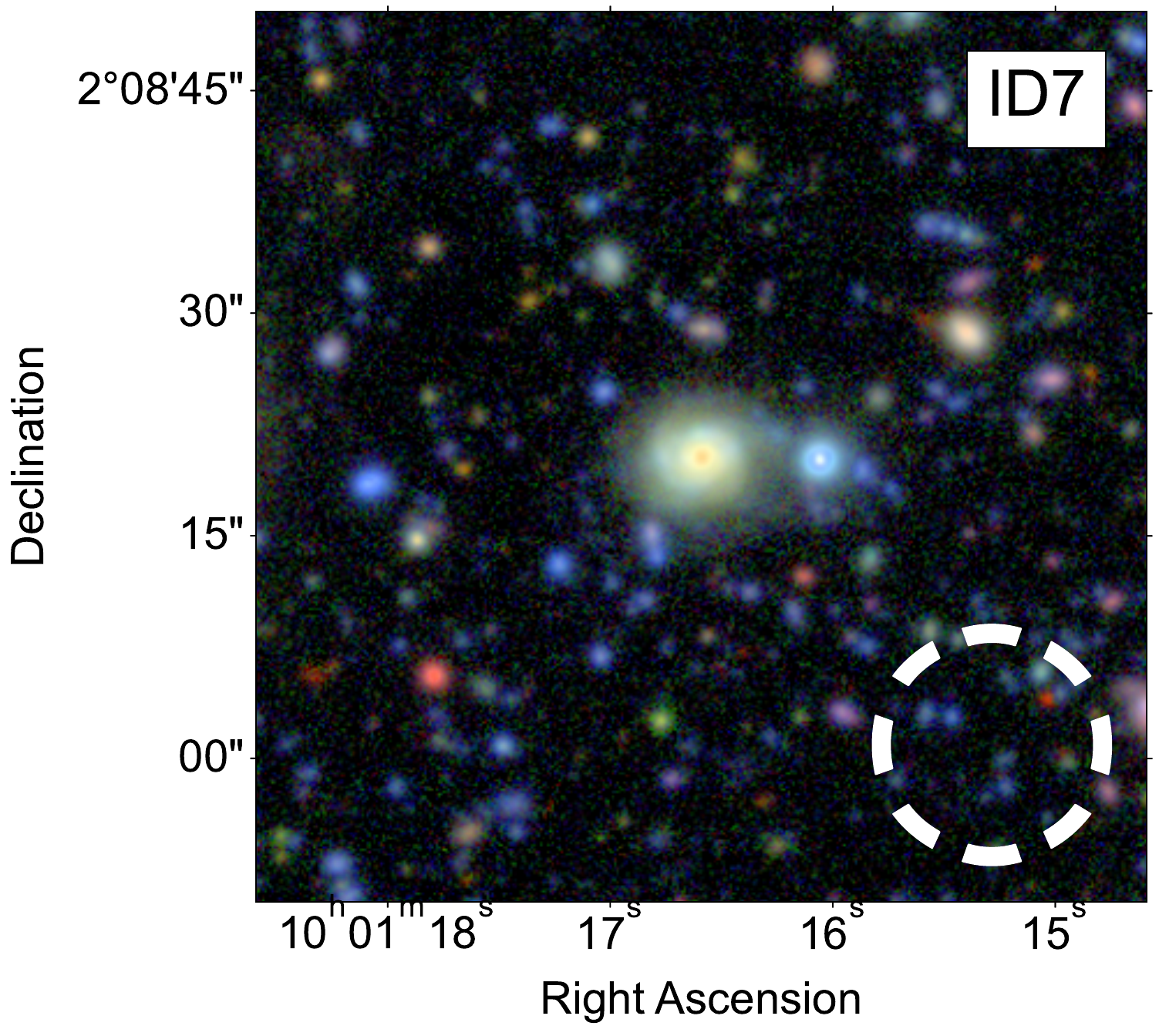}
\includegraphics[width=1.\columnwidth]{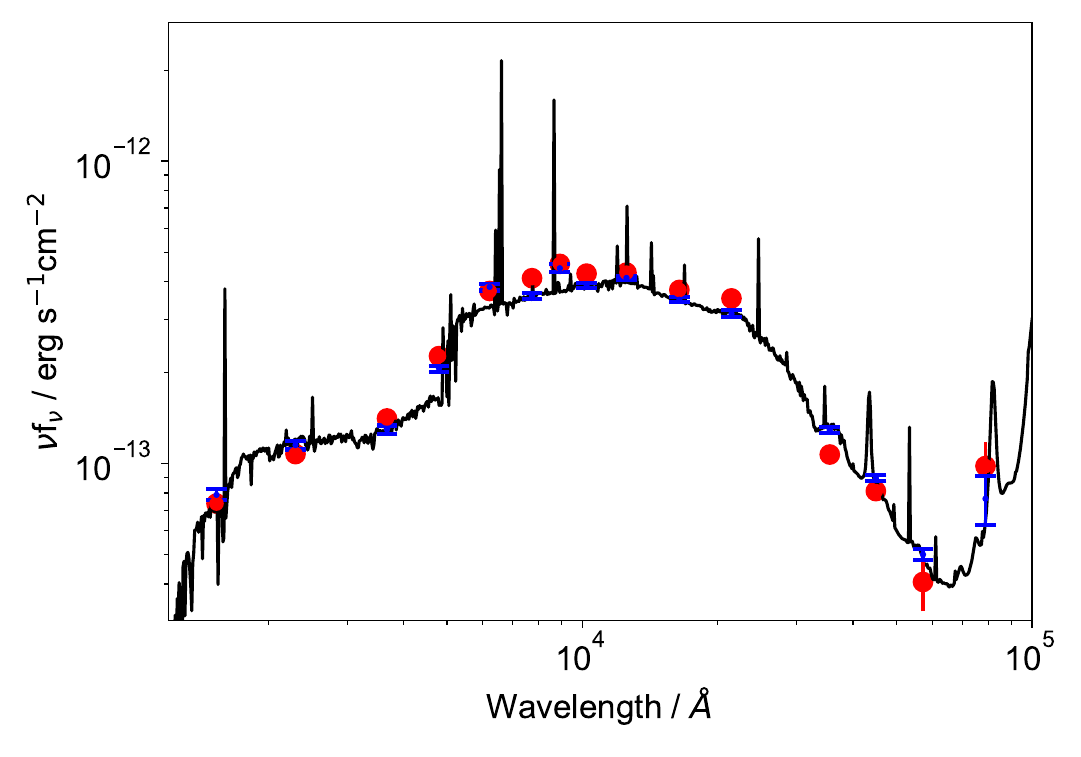}
\includegraphics[width=0.8\columnwidth]{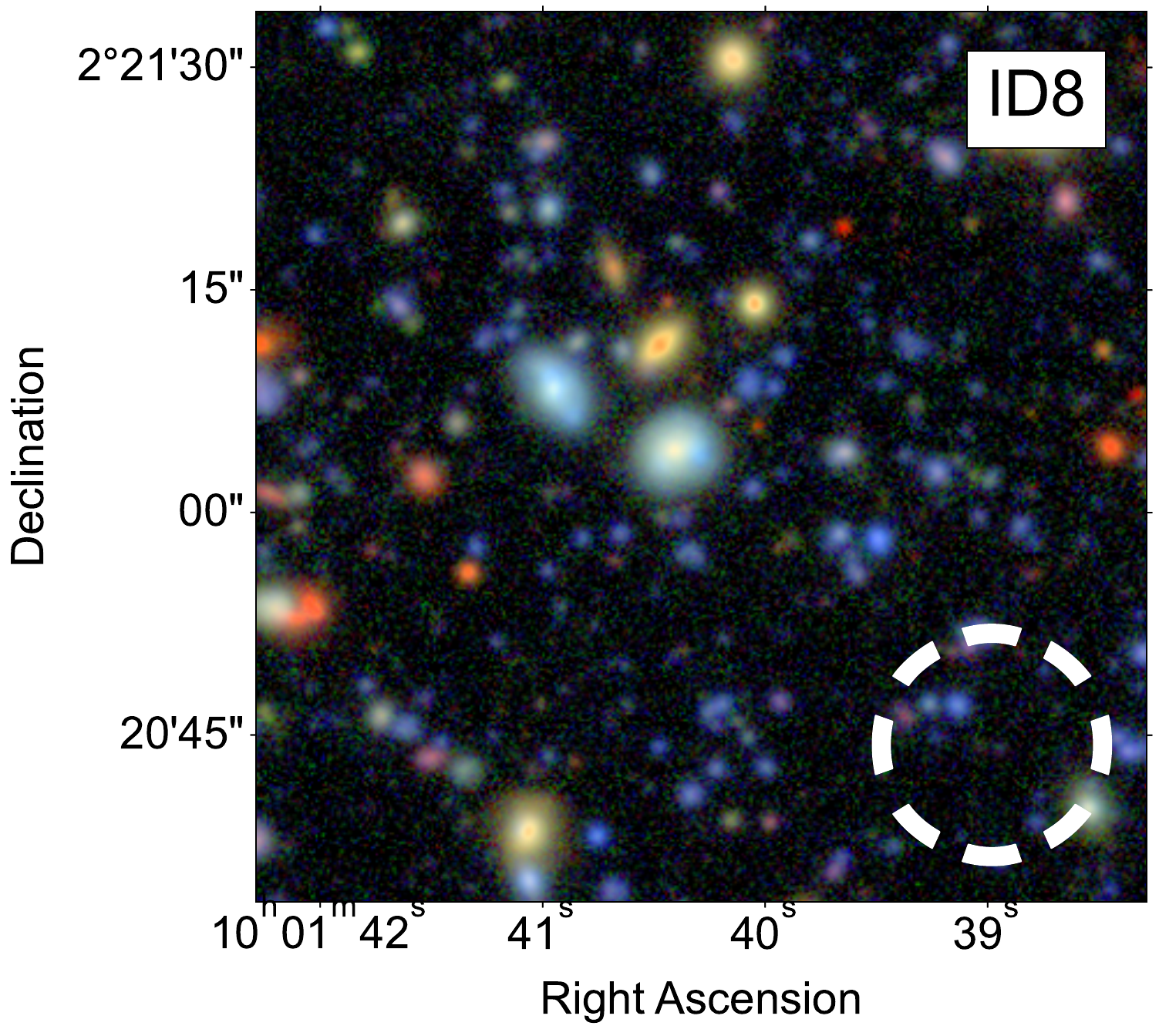}
\includegraphics[width=1.\columnwidth]{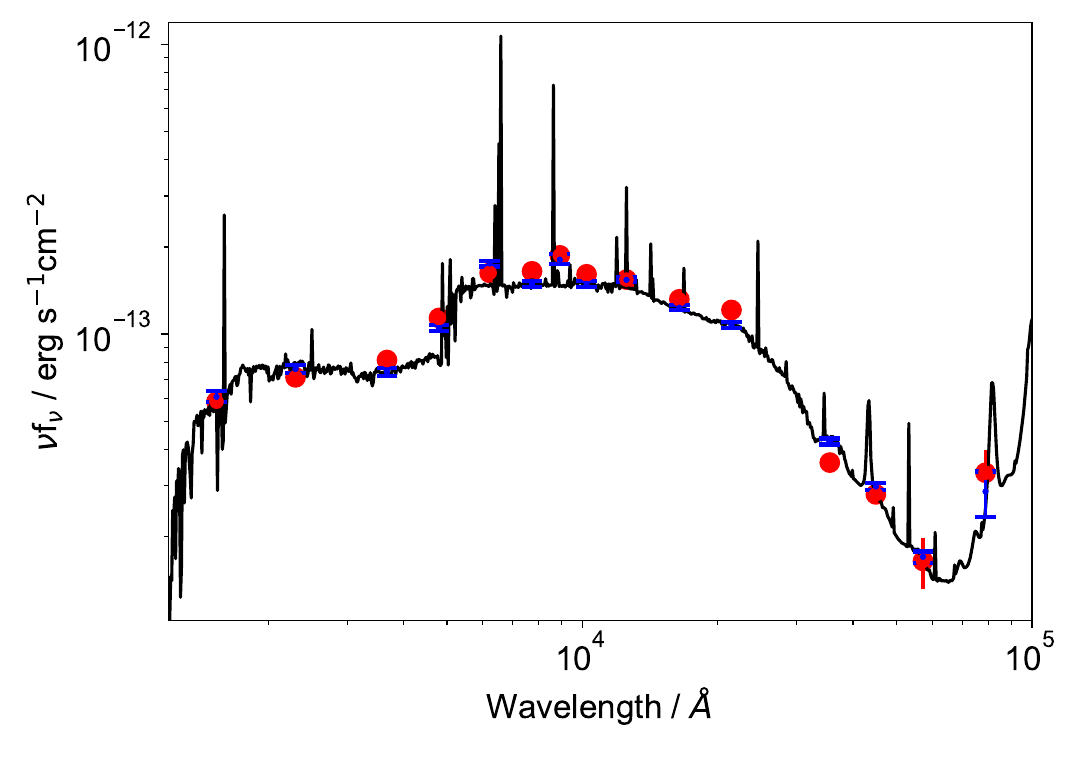}
\end{figure*}

\begin{figure*}
\includegraphics[width=0.8\columnwidth]{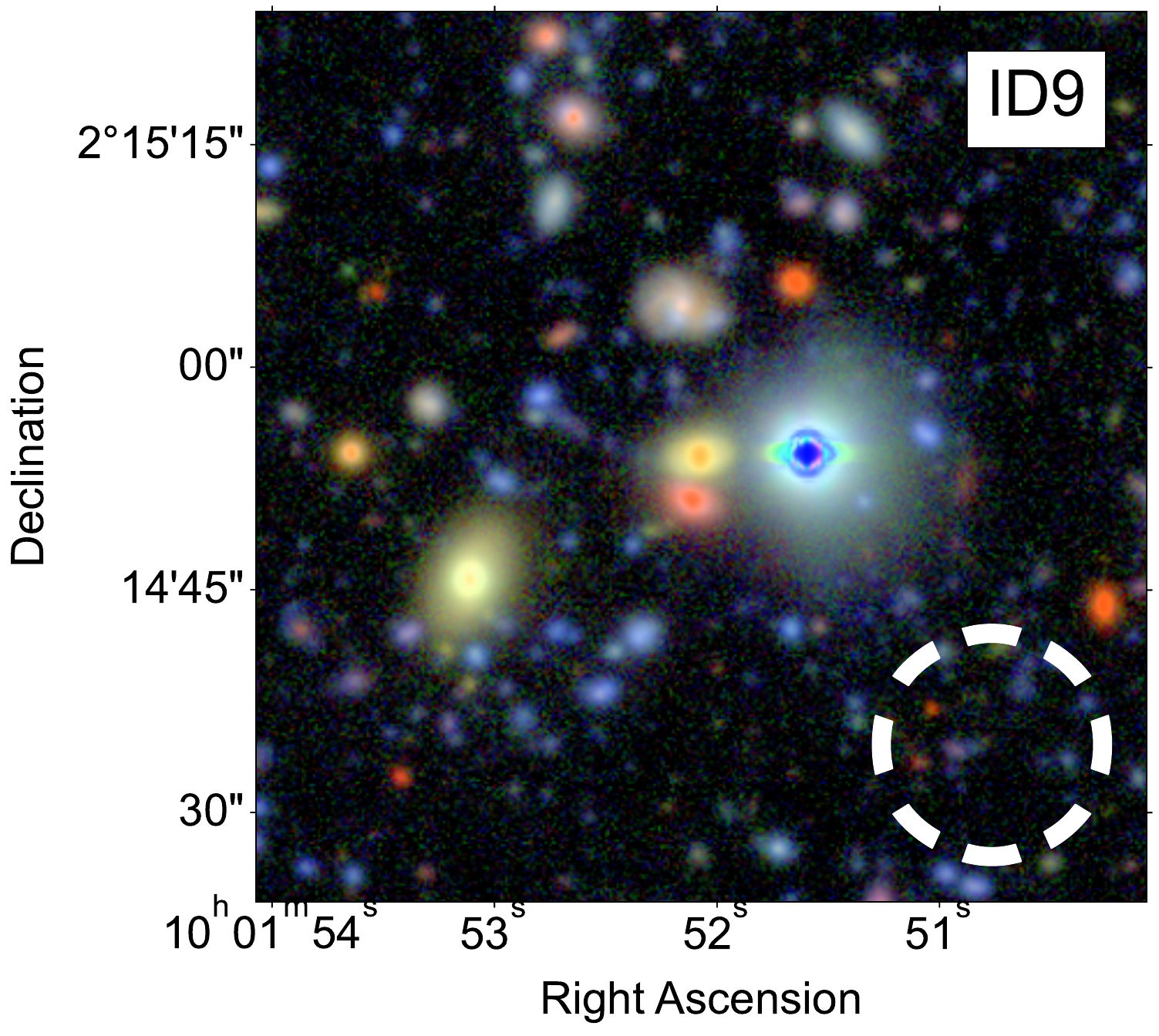}
\includegraphics[width=1.\columnwidth]{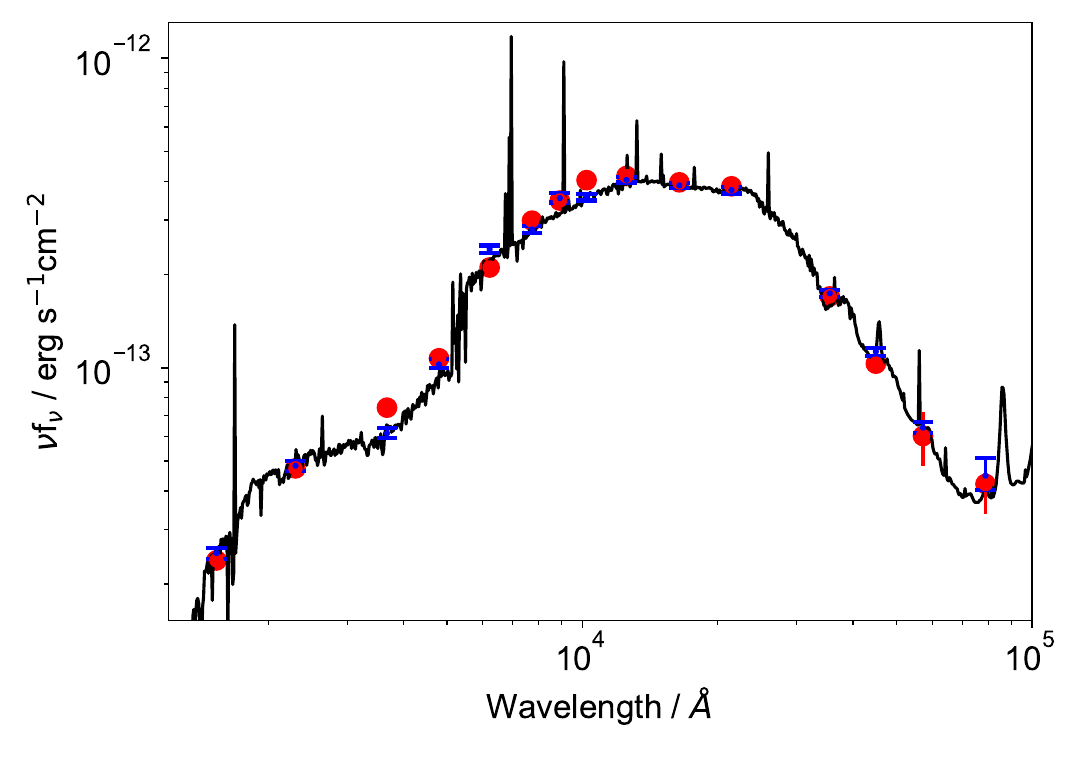}
\includegraphics[width=0.8\columnwidth]{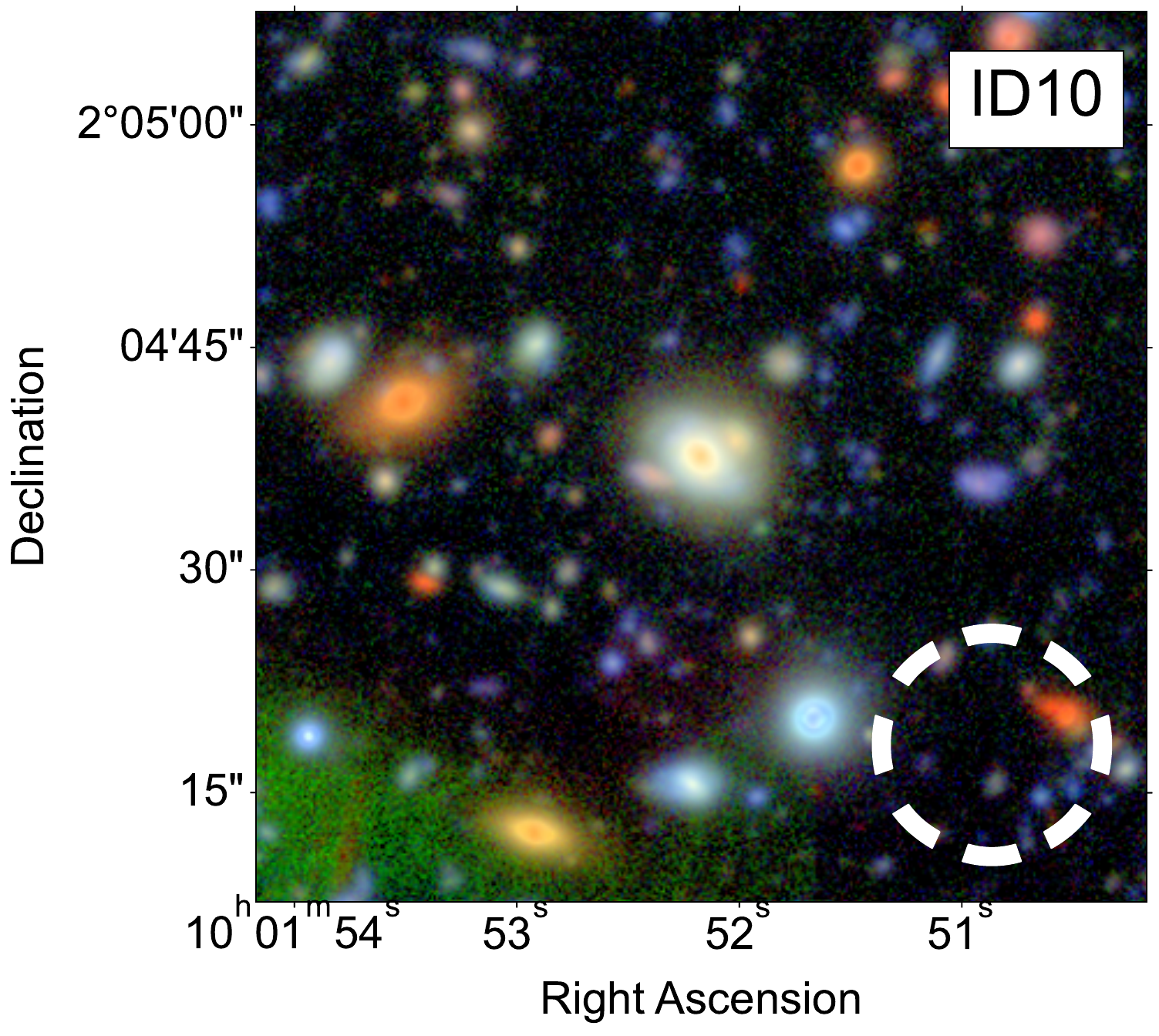}
\includegraphics[width=1.\columnwidth]{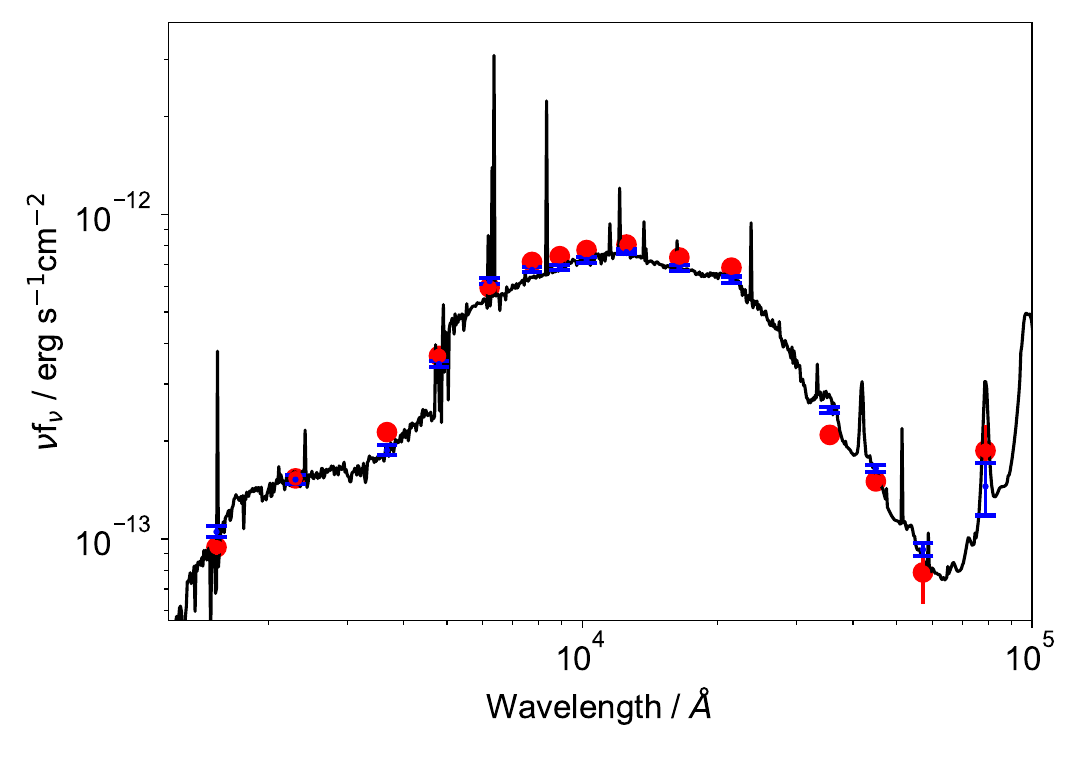}
\caption{HSC $gri$ images (left panels) and the best fit SEDs to the multi-wavelength photometry using {\sc bagpipes} for IDs 1--10 (right panels). The dashed circles in the images represent the FWHM of the synthesised beam of the MIGHTEE L1 data. The red circles in the SED panels denote the data used for fitting the SEDS (black line) and the blue points and error bars represent the 50th, 16th and 84th percentiles of the posterior distributions output from {\sc bagpipes}. We only show the SEDs to $10\mu$m for clarity.}\label{fig:RGB_SED}
\end{figure*}

\begin{figure}
\includegraphics[width=0.95\columnwidth]{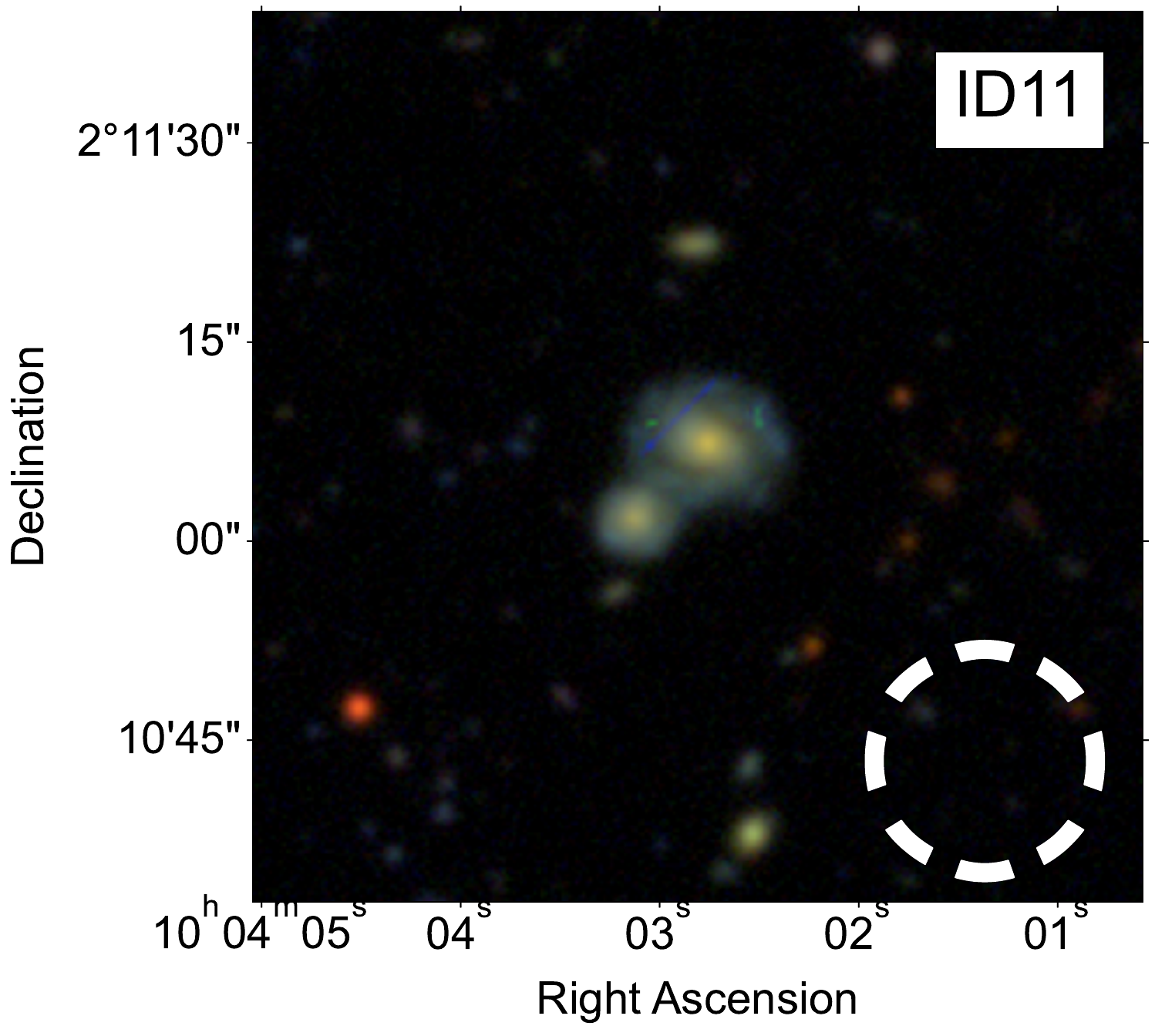}
   \caption{{\sc legacy} survey $gri$ image of the two potential galaxies associated with ID11. The large galaxy in the centre of the image is ID11a in Table~\ref{tab:sample} and the smaller galaxy to the south-east is ID11b. The dashed circle represents the FWHM of the synthesised beam of the MIGHTEE L1 data.}\label{fig:RGB11}
\end{figure}

\begin{table}
    \centering
    \caption{Priors on the parameters used for {\sc bagpipes} SED fitting. The parameters are (from top to bottom): the age of the galaxy, the SFR e-folding time $\tau$, the stellar mass of the galaxy $M_{\star}$, the metallicity $Z$ (in units of solar metallicity), the dust attenuation coefficient $A_V$, the PAH mass fraction $q_{\text{PAH}}$, the lower limit of starlight intensity distribution $u_{\text{min}}$, the fraction of stars at $u_{\text{min}}$ $\gamma$ and the ionisation parameter, $U$.}
    \label{table:fitting-priors}
\begin{tabular}{|cl|}
\hline
\textbf{Parameter} & \textbf{Prior distribution} \\
\hline 
Age & uniform $\in [0.1, 15.0]$ \\
$\tau$ & uniform $\in [0.3, 10.0]$\\
$\log{M_{\ast}}$ & uniform $\in [1.0, 15.0]$\\
$\log{Z}$ & uniform $\in [0.0, 2.5]$\\
A$_{V}$ & uniform $\in [0.0, 3.0]$ \\
q$_{\text{PAH}}$ & uniform $\in [0.1, 4.58]$ \\
u$_{\text{min}}$ & uniform $\in [0.1, 20.0]$ \\
$\gamma$ & uniform $\in [0.0, 0.5]$ \\
$\log_U$ & uniform $\in [-4.0, -1.0] $ \\
\hline 
\end{tabular}
\end{table}

For IDs 1--10, we use \textsc{Bagpipes}\footnote{\url{http://bagpipes.readthedocs.io}} \citep{Carnall_2018} to fit the SEDs.  For our sample of galaxies, given that we already have spectroscopic redshifts, we fix the redshift values and run \textsc{Bagpipes} with the \cite{BC03} stellar population models with a \cite{Chabrier2003} initial mass function to estimate the stellar mass and star-formation rates for our galaxies. We apply the \cite{Calzetti2000} attenuation law and adopt uniform priors for all parameters (shown in Table~\ref{table:fitting-priors}) with an exponentially delayed star-formation history, following \cite{Tudorache2024}. We use 22-band photometry from the far-ultraviolet through to the far-infrared for determining the stellar mass and star-formation rate from these SED fits. We adopt a minimum flux-density uncertainty of 5 per cent for the observations up to 4.5$\mu$m and 20 per cent for the observations longward of this, to account for both zero-point fluctuations, recognising that the synthetic templates are not perfect representations of SEDs of real galaxies.
However, for some of our galaxies the relatively low spatial resolution of the mid- and far-infrared data may result in some of this emission arising from neighbouring galaxies that lie within the point-spread function of these data. We therefore, also fit the SEDs using a restricted set of imaging data up to and including the 8$\mu$m.
The SED fits, using the 50th percentile of the posterior distribution, for these ten galaxies are shown in Figure~\ref{fig:RGB_SED} and the derived properties are given in Table~\ref{tab:seds}. The quoted uncertainties are the 16th and 84th percentiles of the posterior distributions over all parameters within the prior range, but do not account for variations that may arise by using different stellar population synthesis models or initial mass functions, which can result in systematic offsets of up to 0.1~dex in stellar mass and 0.3~dex in SFR \citep{Pacifici2023}.

We find that the stellar mass and SFRs are broadly consistent between the fits with and without the far-infrared data, although using the far-infrared data tends to reduce the stellar masses and  SFRs slightly due the far-infrared emission effectively limiting the dust emission, and thus reducing the SFRs. Thus, the true uncertainty on the stellar mass and SFRs are more accurately reflected in the differences between these fits, with and without long-wavelength data.
However, we use the stellar masses and SFRs derived from just using the data up to $8\mu$m in the SED fitting for the remainder of the paper, to mitigate against confusion at the longer wavelengths. {For completeness,} In Table~\ref{tab:seds} we also provide the SFRs determined using the radio continuum data, using the relation between SFR and 1.4\,GHz radio luminosity from \cite{Bell2003}. We adopt a spectral index of $\alpha = -0.7$ for the spectral slope ($S_{\nu} 
\propto \nu^{\alpha}$) to convert the radio flux densities from 1.28\,GHz to rest-frame 1.4\,GHz luminosity and using the observed scatter in this relation of 0.26\,dex to estimate the uncertainties. We do not use these values for the SFR, due to their large uncertainties compared to those derived from the SED-fitting, but their general consistency demonstrates that the galaxies are consistent with being normal star-forming galaxies, irrespective of how this is measured. 

\subsection{The star-forming galaxy main sequence}
\begin{figure}
  \includegraphics[width=\columnwidth]{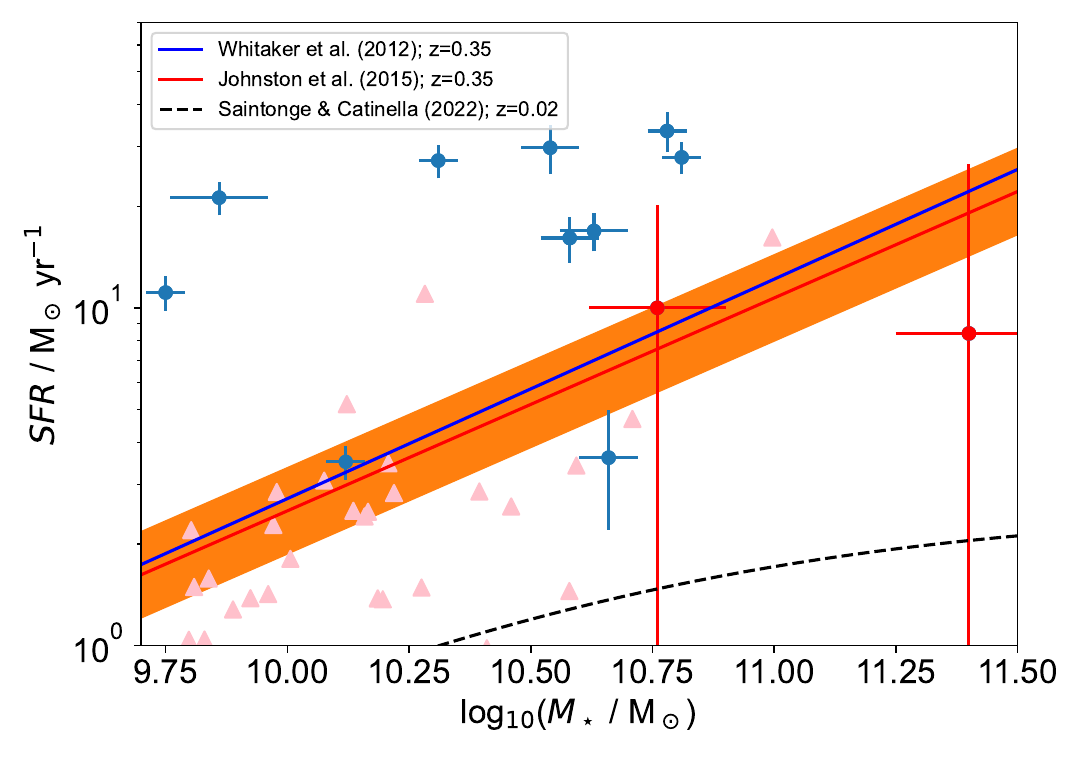}
   \caption{Star-formation main sequence for the H{\sc i}-detected galaxies 
    (filled blue circles), with the two red circles representing the two possible counterparts to ID11. Uncertainties on the SFRs for many of the galaxies are within the size of the symbols (Table~\ref{tab:sample}).
    The blue line denotes the SFMS line from \protect\cite{Whitaker_2014} and the red line is for the SFMS from \protect\cite{johnston2015}, alongside the 1$\sigma$ scatter around the main sequence (filled region), for galaxies at $z=0.35$. The pink triangles denote the low-redshift MIGHTEE 
    ES H{\sc i} detections at $z<0.08$ and the dashed black line shows the SFMS at $z \sim 0.02$ from \protect\cite{Saintonge_2022}.}\label{fig:SFMS}
\end{figure}

With these measurements of the host galaxy properties we are able to investigate whether the galaxies are similar to H{\sc i} detected galaxies in the local Universe. In Figure~\ref{fig:SFMS} we show where the galaxies lie on the $z\sim 0.35$ star-formation main sequence \citep[SFMS; e.g.][]{Noeske_2007,Whitaker2012,johnston2015}. As Figure~\ref{fig:SFMS} shows, the galaxies lie on and above the main sequence for star-forming galaxies at these redshifts, with just one galaxy lying below the main sequence (ID6), which is a spiral galaxy (Figure~\ref{fig:RGB_SED}) with a low star-formation rate (Table~\ref{tab:sample}). It is unsurprising that we preferentially select galaxies lying above the main sequence given that the parent DESI sample is targeted at emission line galaxies with relatively high SFRs.

Also shown in Figure~\ref{fig:SFMS} is the main sequence at $z \sim 0.02$ from \citet{Saintonge_2022} and data from \cite{Tudorache2024} using a similar analysis as presented here, but for the direct detections of H{\sc i} galaxies from the MIGHTEE Early Science (ES) data release \citep[see e.g.][]{Ponomareva2023} without requiring an optical spectroscopic redshift. The H{\sc i}-rich galaxies presented \cite{Tudorache2024} generally lie above the local star-forming main sequence (SMFS). This suggests that pure H{\sc i}-selection at $z\sim 0$ preferentially selects galaxies with significant amount of ongoing star formation compared to the more typical star-forming galaxies used in studies that define the main sequence, whereas the galaxies that we detect at $z\sim 0.35$ appear to have a much greater variability in their position with respect to the main sequence, although small number statistics prevent any stronger statements.

\subsection{The stellar-mass -- H{\sc i}-mass relation}

The stellar-mass -- H{\sc i}-mass scaling relationship with a sample selected on H{\sc i} emission is inherently biased to only probing the upper envelope of this relation \citep[see e.g. ][ for a discussion]{Pan_2023}. This stems from the fact that  we are only able to detect the highest H{\sc i} mass sources at these high redshifts, due to the flux-density limit of the survey. However, for fields with very deep optical and near-infrared data, as is the case here, we are not as limited in terms of the stellar mass that we can measure for such objects. This means that we can detect galaxies with stellar mass of $M_{\star} \geq 10^{8}$\,M$_{\odot}$ at $z\sim 0.5$ \citep[see e.g. Figure~3 in ][]{Adams2023}.

\begin{figure}
  \includegraphics[width=\columnwidth]{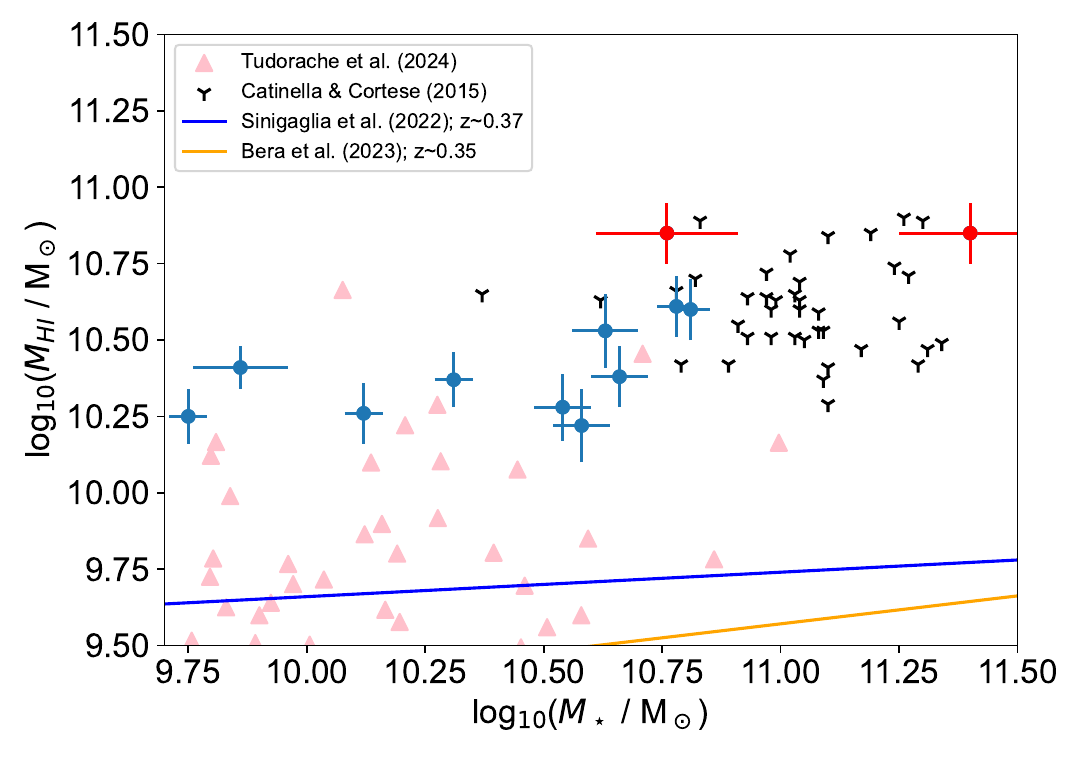}
   \caption{H{\sc i} mass versus the stellar mass for the H{\sc i}-detected galaxies (filled blue circles), with the two red circles representing the two possible counterparts to ID11.
  The pink triangles denote the low-redshift MIGHTEE ES H{\sc i} detections at $z<0.08$ from \protect\citet{Tudorache2024} and the black tri symbol represent the optically selected galaxies at $z>0.16$ with H{\sc i} detections from \protect\citet{CatinellaCortese2015}. The solid lines denote the scaling relations from stacking analyses of the H{\sc i} emission based on optical spectroscopic redshifts for SFGs at $z \sim 0.35$ \protect\citep{Sinigaglia2022,Bera2023}.}
    \label{fig:MHIMstar}
\end{figure}

Figure~\ref{fig:MHIMstar} shows the stellar-mass -- H{\sc i}-mass relation for our sample. One can immediately see that the objects presented in this paper lie significantly above the stacking-derived scaling relations at similar redshifts to our sample from \cite{Sinigaglia2022} and \cite{Bera2023}. For the reasons highlighted above, this is unsurprising for a sample with clear H{\sc i} detections. However, in Figure~\ref{fig:MHIMstar} we also show the stellar-mass -- H{\sc i}-mass for the low-redshift detections using the MIGHTEE ES release from \cite{Tudorache2024}. We see that all of our galaxies have much higher H{\sc i} masses than the low-redshift detections. This suggests that we are only detecting the galaxies at the bright end of the galaxy H{\sc i} mass function, which are rare in the local Universe and require large-area surveys to find them \citep[e.g.][]{Huang2014}, and are not present in the relatively small volume covered by the MIGHTEE ES data. If we use the H{\sc i} mass function from \cite{Ponomareva2023} and assume no evolution, then we would expect to detect of the order $\sim 50$ galaxies with $\log_{10}(M_{\rm HI}/$M$_{\odot} > 10.5$) over a volume commensurate with an effective area of $\sim $4~degree$^2$ over the redshift range $0.24<z<0.4$. Thus, the discovery of these 11 H{\sc i} galaxies within this field is expected, and more should be detected when a robust untargeted search is performed (Maksymowicz-Maciata et al. in prep).

\subsection{The SFR -- H{\sc i}-mass relation and gas depletion timescale}

\begin{figure}
  \includegraphics[width=\columnwidth]{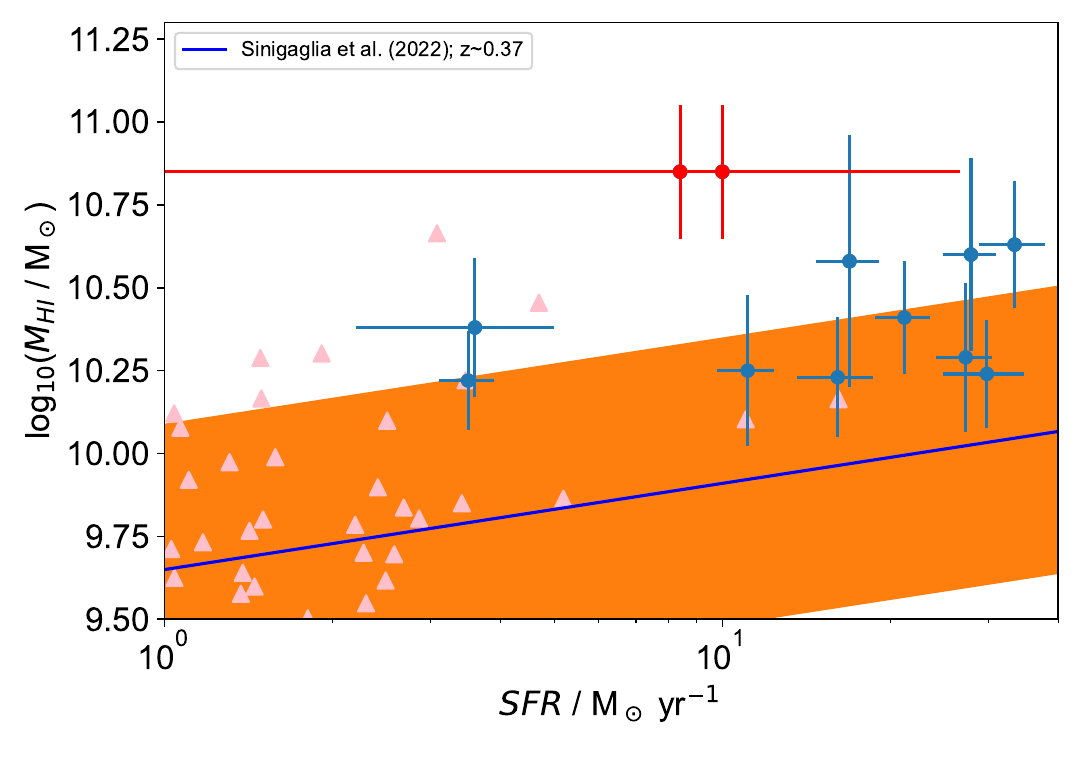}
   \caption{H{\sc i} mass versus the SFR for the H{\sc i}-detected galaxies (filled blue circles), with the two red circles representing the two possible counterparts to ID11. The pink triangles denote the low-redshift MIGHTEE ES H{\sc i} detections at $z<0.08$.   The  solid blue line and orange band shows the scaling relations from the stacking analyses of the H{\sc i} emission based on optical spectroscopic redshifts for SFGs at $z \sim 0.35$ from \protect\cite{Sinigaglia2022}.}
    \label{fig:MHISFR}
\end{figure}

In Figure~\ref{fig:MHISFR} we show the SFR against the H{\sc i} mass for our galaxies. We find that the galaxies at $z\sim 0.35$ generally lie above the scaling relation between SFR and H{\sc i} determined from the stacking analysis of the MIGHTEE ES data \citep{Sinigaglia2022}. This difference between the stacking results and our results here can again be explained by the fact that with direct H{\sc i} detections we are always biased towards galaxies containing the most H{\sc i}, otherwise they would fall below our detection threshold. This is reinforced by the fact that there is an overlap with the low-redshift H{\sc i}-selected sample presented in \cite{Tudorache2024}, although at these higher redshift the distribution is skewed towards higher SFRs.

\begin{figure}
  \includegraphics[width=\columnwidth]{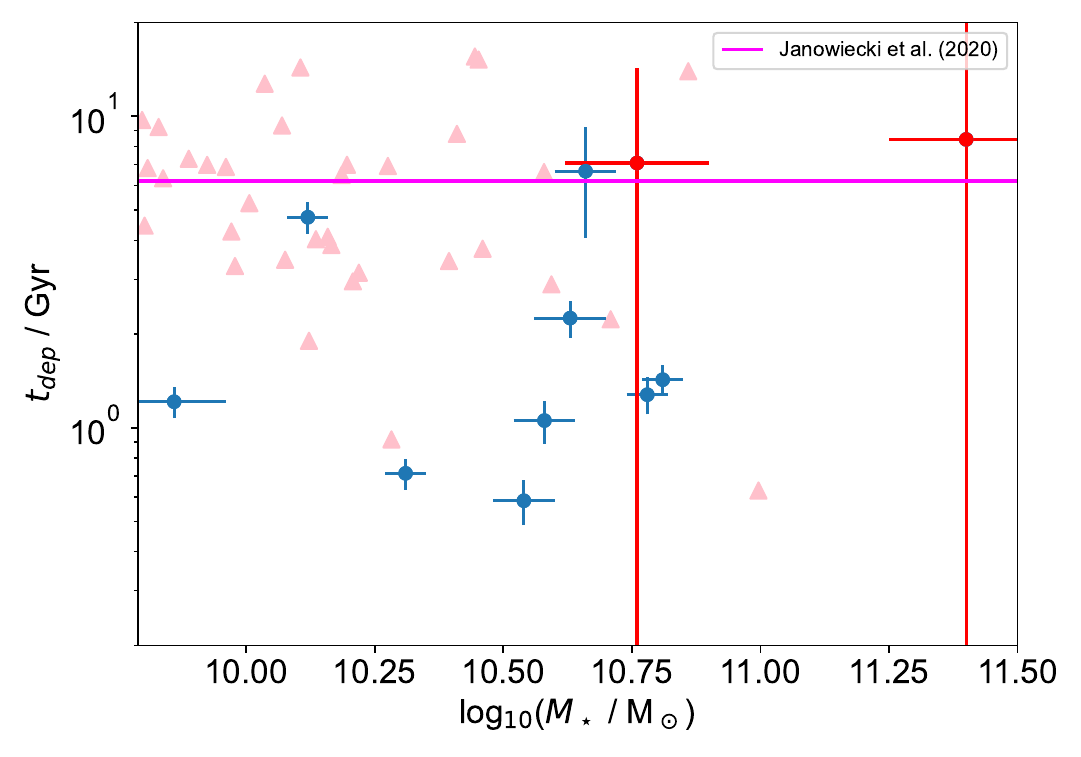}
   \caption{H{\sc i} mass versus the H{\sc i}-gas depletion time for the H{\sc i}-detected galaxies (filled blue circles) with the red circles denoting ID11. Also shown are the low-redshift H{\sc i} selected galaxies from \protect\cite{Tudorache2024} (pink triangles). The horizontal magenta line is the median depletion time from \citet{Janowiecki2020} for a stellar-mass selected H{\sc i} sample at $0.01<z<0.05$. }
    \label{fig:tdepMstar}
\end{figure}

In Figure~\ref{fig:tdepMstar} we show the gas depletion time, defined as $t_{\rm dep} = M_{\rm HI}/SFR$, against the stellar mass for our sample. We find that the gas depletion times for these H{\sc i} rich galaxies at relatively high redshift tend to be lower than those detected at low redshift from the MIGHTEE ES data, with a mean (median) depletion time of $3.1$\,Gyr (1.5\,Gyr), respectively, compared to 9.6\,Gyr  (9.5\,Gyr) for the MIGHTEE ES H{\sc i} sample from \cite{Tudorache2024} and also the stellar-mass selected xGASS sample \citep{Janowiecki2020}. Given the large scatter in the gas depletion time for our sample, we cannot make any strong statements with respect to any differences to the low-redshift H{\sc i} sample. Indeed, although the galaxies presented in this paper tend to have higher H{\sc i} masses than their low-redshift counterparts, this is counter balanced by their higher SFRs and underlines the importance of the galaxy stellar mass in driving the key scaling relations  \citep{Tudorache2024} that we observe in H{\sc i} selected samples.

\begin{table*}
    \centering
        \caption{Right Ascension (RA), Declination (Dec), redshift ($z$) and redshift of the optical galaxy, alongside the emission-line flux, the H{\sc i} mass, the measured $W_{50}$, inclination-corrected $W^{c}_{50}$ using the inclination $i$, from the galaxy ellipticity in the $g-$band. 
        SNR$^a$ corresponds to the SNR calculated from measuring the 16th and 84th percentiles of the aperture fluxes extracted at random spatial and spectral positions around the  galaxy candidate, SNR$^b$ corresponds to the SNR measured from random spatial positions around the  galaxy candidate but with spectral window fixed to that of the emission line and SNR$^c$ corresponds to the SNR measured using a single-pixel measurement of flux of the emission line with the noise measured using spatial positions around the galaxy candidate but with spectral window fixed to that of the emission line. The uncertainty on the H{\sc i} mass uses the SNR$^a$ case, and the uncertainty given in parentheses is an estimate of the systematic uncertainty derived from the difference between the flux measured in a single pixel at the position of the candidate galaxy and that obtained from the aperture flux used for the initial selection.
        ID11 has two possible counterparts and we provide positions, redshifts and inclinations for both. We note that the very large inclination-corrected $W_{50}$ for ID11b suggests that the measured $W_{50}$ for this object is either overestimated due to confusion between the two objects, or that the line emanates from ID11a.  }\label{tab:sample}
\scriptsize
\begin{tabular}{l|l|l|r|c|c|c|c|c|c|c|c}
\hline
\multicolumn{1}{|c|}{ID} &
  \multicolumn{1}{|c|}{RA} &
  \multicolumn{1}{c|}{DEC} &
  \multicolumn{1}{c|}{$z$} &
  \multicolumn{1}{c|}{Line flux} &
  \multicolumn{1}{c|}{$\log_{10}(M_{\rm HI}/{\rm M}_\odot)$} &
  \multicolumn{1}{c|}{SNR$^a$} &
   \multicolumn{1}{c|}{SNR$^b$} &
    \multicolumn{1}{c|}{SNR$^c$} &
  \multicolumn{1}{c|}{$W_{50}$} & 
   \multicolumn{1}{c|}{$W^{c}_{50}$} &
  \multicolumn{1}{c|}{$i$} \\
  \multicolumn{4}{|c|}{} &
    \multicolumn{1}{|c|}{(Jy\,Hz)} &
    \multicolumn{3}{|c|}{} &
    \multicolumn{1}{c|}{(km~s$^{-1}$)} &
    \multicolumn{1}{c|}{(km~s$^{-1}$)} &
    \multicolumn{1}{c|}{(degrees)} \\
\hline
1 & 09:58:15.49 & +02:11:35.4 & 0.2647 & $177\pm 41$ & 10.23 $\pm$  0.10 ($\pm 0.15$) & 4.1 & 4.4 & 4.0 & 252 $\pm$  25 & 347$\pm$46 & 46$\pm$ 5  \\
2 & 09:58:44.49 & +02:09:51.8 & 0.2604 & $278\pm52$  & 10.41 $\pm$  0.08 ($\pm 0.15$) & 6.4 & 5.7 & 7.1 & 373 $\pm$  19 & 432$\pm$38 & 60$\pm$ 5 \\
3 & 09:59:24.90 & +02:17:05.5 & 0.3447 & $216\pm45$ & 10.58 $\pm$  0.09 ($\pm 0.37$) & 4.1 & 4.4 & 1.5 & 315 $\pm$  19 & 379$\pm$39 & 56$\pm$ 5 \\
4 & 10:00:02.14 & +02:18:27.4 & 0.2654 & $172\pm56$ & 10.22 $\pm$  0.14 ($\pm 0.05$) & 4.7 & 4.7 & 5.3 & 337 $\pm$  20 & 346$\pm$29 & 77$\pm$ 5  \\
5 & 10:00:10.93 & +02:07:24.8 & 0.3390 & $252\pm53$ & 10.63 $\pm$  0.09 ($\pm 0.17$) & 4.6 & 4.7 & 5.4 & 455 $\pm$  19 & 506$\pm$36 & 64$\pm$ 5 \\
6 & 10:00:57.76 & +02:03:54.4 & 0.2675 & $244\pm34$ & 10.38 $\pm$  0.06 ($\pm 0.20$) & 5.1 & 7.0 & 3.7 & 307 $\pm$  34 & 438$\pm$53 & 44$\pm$ 5 \\
7 & 10:01:16.59 & +02:08:20.3 & 0.3162 & $109\pm31$ & 10.29 $\pm$  0.12 ($\pm 0.19$) & 4.9 & 4.0 & 3.6 & 226 $\pm$  29& 339$\pm$51 &  42$\pm$ 5 \\
8 & 10:01:40.28 & +02:21:03.8 & 0.3122 & $127\pm33$ & 10.25 $\pm$  0.11 ($\pm 0.20$)  & 4.5 & 4.2 & 5.7 & 246 $\pm$  18& 308$\pm$39 &  53$\pm$ 5 \\
9 & 10:01:52.07 & +02:14:54.0 & 0.3841 & $176\pm45$ & 10.60 $\pm$  0.11 ($\pm 0.27$) & 4.2 & 4.1 & 2.7 & 305 $\pm$  34&  435$\pm$52 & 45$\pm$ 5 \\
10 & 10:01:52.17 & +02:04:37.6 & 0.2672 & $177\pm45$ & 10.24 $\pm$  0.11 ($\pm 0.12$) & 4.7 & 4.1 & 7.1 & 387 $\pm$  77& 692$\pm$92 & 34$\pm$ 5 \\
11a$^\dagger$ & 10:04:02.76 & +02:11:07.3 & 0.3285 & $450\pm115$ & 10.85 $\pm$ 0.11 ($\pm 0.17$) & 5.7 & 4.0 & 4.5 & 490 $\pm$ 50 & 680$\pm$64 & 46$\pm$ 5 \\
11b$^\dagger$ & 10:04:03.12 & +02:11:01.7 & 0.3293 & " &  "  & " & " & " & " & 1032$\pm$140 & 28$\pm$ 5 \\
\hline\end{tabular}
\end{table*}

\begin{table*}
    \centering
        \caption{Host galaxy properties derived from the SED fitting with {\sc BAGPIPES} for fits both with and without data longward of $8$\,$\mu$m, to ensure that contributions to the low-spatial resolution long-wavelength data from nearby galaxies do not affect the derived properties. We provide information for both galaxies associated with ID11, using information provided in the DESI catalogue due to a lack of deep-field multi-wavelength coverage for this detection. Uncertainties on stellar mass and SFR from the SED fitting are the 16th and 84th percentiles of the posterior distributions from the prior range considered. We note that these do not account for any variations away from the \protect\cite{BC03} stellar population synthesis models assuming a \protect\citet{Chabrier2003} initial mass function.
        $S_{1.28}$ is the radio flux-density measured using the catalogue provided in \protect\cite{Hale2024} and we use the relation from \protect\citet{Bell2003} to estimate the SFR$_{\rm radio}$ based on the radio flux density. The uncertainty on SFR$_{\rm radio}$ is dominated by the intrinsic scatter in the \citet{Bell2003} relation of $0.26$\,dex. The galaxies flagged with $^\ast$ have their radio continuum flux densities determined from measuring the peak flux density directly from the imaging data as they do not meet the 5$\sigma$ catalogue threshold. We note that ID4 has a measured negative flux density at the galaxy position but we give the formal uncertainty around zero for the SFR.}\label{tab:seds}
\begin{tabular}{l|r|l|r|l|c|c}
\hline
\multicolumn{1}{|c|}{} &
   \multicolumn{2}{c|}{With FarIR} &
   \multicolumn{2}{c|}{No FarIR} &
   \multicolumn{2}{c|}{Radio} \\
\multicolumn{1}{|c|}{ID} &
   \multicolumn{1}{c|}{$\log_{10}(M_{\star} /$M$_{\odot})$} &
  \multicolumn{1}{c|}{$SFR$ / M$_{\odot}$~yr$^{-1}$} &
     \multicolumn{1}{c|}{$\log_{10}(M_{\star} /$M$_{\odot})$} &
  \multicolumn{1}{c|}{$SFR$ / M$_{\odot}$~yr$^{-1}$} &
    \multicolumn{1}{c|}{S$_{1.28}$ / $\mu$Jy} &
    \multicolumn{1}{c|}{$SFR$ / M$_{\odot}$~yr$^{-1}$}\\
\hline
1 &   $10.50 \pm 0.03$& 7.8 $\pm 0.5$ & $10.58\pm 0.06$& $16.1\pm 2.5$ &  290$\pm$7 & 32$\pm26$\\
2 &  $10.03 \pm 0.02$ & $6.2 \pm 0.5$& $9.86 \pm 0.10$ & $21.2\pm 2.4$ & 85$\pm$8 & 9$\pm$7\\
3$^\ast$ &  $10.45 \pm 0.04$& $3.3\pm 0.3$& $10.63 \pm 0.07$ & $16.9\pm2.2$ & 8$\pm$ 4 & $2\pm1$\\
4$^\ast$ & $9.93 \pm 0.03$& $1.0 \pm 0.1$& $10.12\pm 0.04$ & $3.5\pm 0.4 $ & $0\pm 4$ & $0\pm1$\\
5 &  $10.70 \pm 0.05$& $21.8 \pm 1.6$ & $10.78\pm 0.04$ & $33.4\pm 4.5$ & 420$\pm$5 & 81$\pm$66\\
6$^\ast$ &  $10.49 \pm 0.02$& $1.1 \pm 0.1$& $10.66\pm0.06$ & $3.6\pm 1.4$ & 9$\pm$4 &$1\pm1$\\
7 &  $10.39 \pm 0.02$ & $3.7\pm 0.3$& $10.31\pm 0.04$ & $27.3\pm 3.1$ & 39$\pm$5 & 6$\pm$5\\
8 &  $9.86 \pm 0.02$ & $2.7 \pm 0.2$ & $9.75\pm 0.04$ & $11.1\pm 1.3$ & 30$\pm$8 & 5$\pm$4\\
9 &  $10.75 \pm 0.02$& $4.7 \pm 0.4$& $10.81\pm 0.04$ & $27.9\pm 3.0$ & 19$\pm$4 & 5$\pm$4\\
10 &  $10.56\pm 0.02$ & $4.1 \pm 0.3$ & $10.54\pm 0.06$ & $29.8\pm 4.9$ & 82$\pm$6 & 9$\pm$8\\
11a$^\dagger$ & & & $11.40\pm0.15$ & $8.4\pm 18.3$&  71$\pm$8 & 12$\pm$10\\
11b$^\dagger$ & & & $10.76\pm0.14$ & $10.0 \pm 10.2$  & 76$\pm$8 & 14$\pm$11\\
\hline\end{tabular}
\end{table*}

\subsection{The $z\sim 0.35$ baryonic Tully-Fisher relation}
The Tully-Fisher relation \citep{TullyFisher} relates the dynamical mass of rotation-dominated galaxies through measurements of the velocity rotation curve, with the amount of light or observable mass present in the galaxy.
It has been used as a redshift independent method for determining the distance to galaxies and as such is one of the key elements in studies to understand the bulk flow of galaxies at relatively low redshifts \citep{Courtois2012, Tully2013,Tully2019}. However, the rotation curves of the galaxies are the result of the mass distribution from all components within the galaxy (stars, gas and dark matter), which may not be wholly captured by just tracing the stellar emission at a given waveband, although some wavebands are more suited to this than others \citep[see e.g.][]{Ponomareva2017}. In particular, low-stellar-mass gas-rich systems require the mass in the gaseous component to be properly accounted for, and this gave rise to the baryonic Tully-Fisher relation (bTFr), a tight relation that spans $\sim 5$ orders of magnitude in baryonic mass  \citep{McGaugh2012,Lelli2016,Lelli2019,Ponomareva2021}.

Cosmological hydrodynamic simulations, such as the {\sc simba} simulation \citep{Dave2019} predict evolution in the bTFr \citep{GlowackiDave2021}, which can be explained by the merger histories of galaxies. However, an accurate measurement of the bTFr beyond the local Universe is difficult and only a few studies have attempted such a measurement \citep[][]{CatinellaCortese2015,Gogate2023},
with the faintness of the 21-cm line inhibiting detailed studies of the evolution of the bTFr with redshift. Other emission lines have been used such as CO \citep{Topal2018} and ionised gas tracers such as H$\alpha$ \citep{diTeodoro2016,Tiley2016, Tiley2019}. However, both CO and H$\alpha$ tend to trace the central mass distribution of galaxies as the gas tracer does not extend beyond the optical disk, unlike the H{\sc i} \citep{Catinella2023}. Furthermore, it is often difficult to reconcile different results due to the varying quality of the observational data \citep{Tiley2019}. Therefore, it is important to compare like with like across all redshifts to mitigate differences in the mass distributions being traced by the dynamics, coupled with controlling the systematics between different observational strategies.

Although many studies of the bTFr rely on resolved rotation curves, much of the information is actually contained within the velocity width of the integrated H{\sc i} line, potentially removing the need to resolve galaxies in order to place them on the bTFr \citep{Lelli2019, Ponomareva2021,Yasin2025}. Simulations of H{\sc i} galaxies have also examined the impact of using different integrated H{\sc i} linewidths on the bTFr \citep{GlowackiDave2021}. This is particularly pertinent as we move to higher redshift, where resolving the H{\sc i} line with current interferometers remains problematic, due to the need to balance surface brightness sensitivity (requiring large single-dish telescopes or short baselines in an interferometer) with high spatial resolution ($<5$~arcsec), requiring high sensitivity on baselines extending to at least 12~km, and is one of the cornerstone science applications for the Square Kilometre Array.
Therefore, for our sample of unresolved galaxies we use the velocity width at 50 per cent of the peak flux density ($W_{50}$). In the cases where there appears to be a double-horned profile, we take the maximum flux density from each horn individually to estimate the lower and upper frequencies from which to determine the velocity width. We vary the position of the peak and zero-flux density baseline within the 1$\sigma$ noise of the data to estimate the measured uncertainty on $W_{50}$. 
We then correct the line width for instrumental broadening and random motions following \cite{VerheijenSancisi2001}, using the channel width of 104.5\,kHz (corresponding to an observed velocity width of $v_{\rm ob} \sim 16$\,km~s$^{-1}$ at $z=0.35$) and we adopt 5~km~s$^{-1}$ as an additional source of line broadening due to turbulent motions \cite[see e.g.][]{Ponomareva2016}. We note that the dominant contribution to the uncertainties on our measured line widths is the finite channel width of our data and does not depend strongly on the SNR of the detected line. An example of this is ID7, where although the integrated line flux has a SNR=4.5, the width of the line is very well defined by the two peaks of the line and the very steep emission line edges. Other cases are not so clear cut, but we note that varying the height of the peak of the line by the measured 1$\sigma$ noise does not significantly impact the measured line width and, as detailed below, the uncertainty on the inclination dominates over all these uncertainties.

The observed velocity widths then need to be corrected due to the inclination of the galaxy. We follow the standard procedure of using the galaxy elliptical measurement from the $g-$band imaging data, and correct for the inclination ($i$) to an edge-on disk using $\cos i = b / a$, where $a$ and $b$ are the major and minor axes of the galaxy. We assign an intrinsic uncertainty to the inclination of 5\,degrees for all our galaxies and this is carried through to the overall uncertainty on the inclination corrected measurement of $W_{50}$. We note that the uncertainty due to the finite channel width and the inclination correction dominate the uncertainties presented in Table~\ref{tab:sample}, and adopting a thickness for the disk does not lead to different results.

The baryonic mass ($M_{\rm bar}$) is determined from both the measured stellar mass and the H{\sc i} mass given in Table~\ref{tab:sample}. Following the literature we make a correction to the gas mass, multiplying by the universal value of a factor 1.4 \citep[see e.g.][]{McGaugh2012, Ponomareva2021}, to account for the primordial abundance of helium and metals. The contribution from the molecular gas has been found to make a negligible contribution to the statistical properties of the bTFr at least at low masses \citep{McGaugh2012, Ponomareva2018}. However, there is a potential higher contribution from molecular gas at higher galaxy masses. For example, \citet{Catinella2018} and \citet{Saintonge_2017} show that the molecular gas content may have a similar mass to that of the H{\sc i} for galaxies with stellar mass $M_{\star} > 10^{10}$\,M$_{\odot}$. Without a direct measurement of the molecular gas content of these galaxies it is impossible to accurately account for this in determining the baryonic mass of a galaxy, however we return to this later.

\begin{figure}
  \includegraphics[width=\columnwidth]{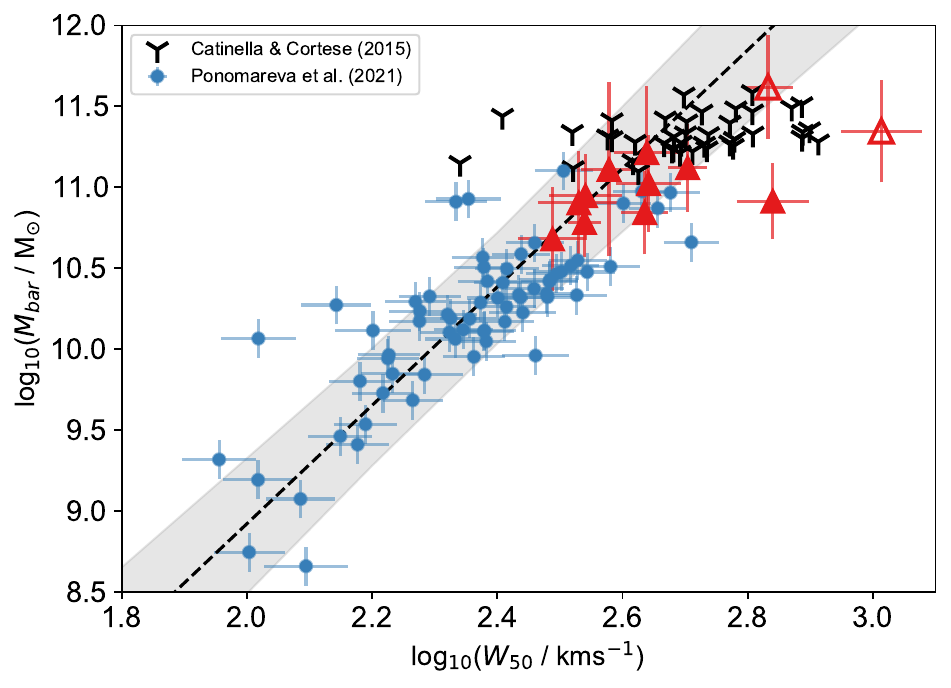}
   \caption{H{\sc i} derived baryonic Tully-Fisher relation for our sample at $z\sim 0.35$ (red triangles) using the inclination corrected measurements for $W_{50}$. The open red triangles represent the two plausible counterparts for ID11, both of which have stellar masses from the DESI catalogue. Note the filled and open triangles that are significantly offset from the local relation are IDs 10 and 11b, which have the lowest inclinations for our sample. The $W_{50}$ for these galaxies should be treated with caution given previous studies of the bTFr only use galaxies with inclinations $>30$\,degrees to minimise the uncertainty in the projection to a edge-on disk.
   The blue points and dashed line denotes the data and best-fit $0<z<0.08$ baryonic Tully-Fisher relation of \protect\cite{Ponomareva2021} based on the MIGHTEE ES Data Release, with the shaded region showing the observed scatter for this relation. The black tri symbols show the data for the optically-selected H{\sc i} detections in \protect\citet{CatinellaCortese2015}.}
    \label{fig:bTFr}
\end{figure}

In Figure~\ref{fig:bTFr} we show the position of our galaxies on the low-redshift Baryonic Tully-Fisher relation of \cite{Ponomareva2021}. We find that all our galaxies lie within the scatter of the local bTFr but probe much higher velocity widths and baryonic masses. The fact that we do not detect any significant evolution in the bTFr from the local Universe to $z\sim 0.35$ with this sample, is not altogether surprising given the result of simulations, which suggest relatively weak evolution \citep{GlowackiDave2021}. However, our high baryonic mass objects all tend to lie below the local relation, indicative of a flattening of the gradient at these high masses or high-redshifts. This would be in line with the expected evolution based on the simulations presented in \cite{GlowackiDave2021}. An alternative explanation for the flattening of the bTFr in the high-mass regime is that $W_{50}$ tends to trace the maximum rotational velocity (V$_{\rm max}$) rather than the velocity at the flat part of the rotation curve (V$_{\rm flat}$). In intermediate-mass galaxies with flat rotation curves, V$_{\rm max}$ and V$_{\rm flat}$ are consistent. However, high-mass systems often have rotation curves that decline beyond the turnover radius, making V$_{\rm flat}$ systematically lower than the circular velocity derived from the corrected width of the global H{\sc i} profile \citep{casertano1991, Ponomareva2016}. Therefore, measuring the rotational velocity from resolved rotation curves in these objects might place them back onto the straight line relation \citep{diteodoro2021}, but higher resolution data would be required to make these measurements.

\begin{figure}
  \includegraphics[width=\columnwidth]{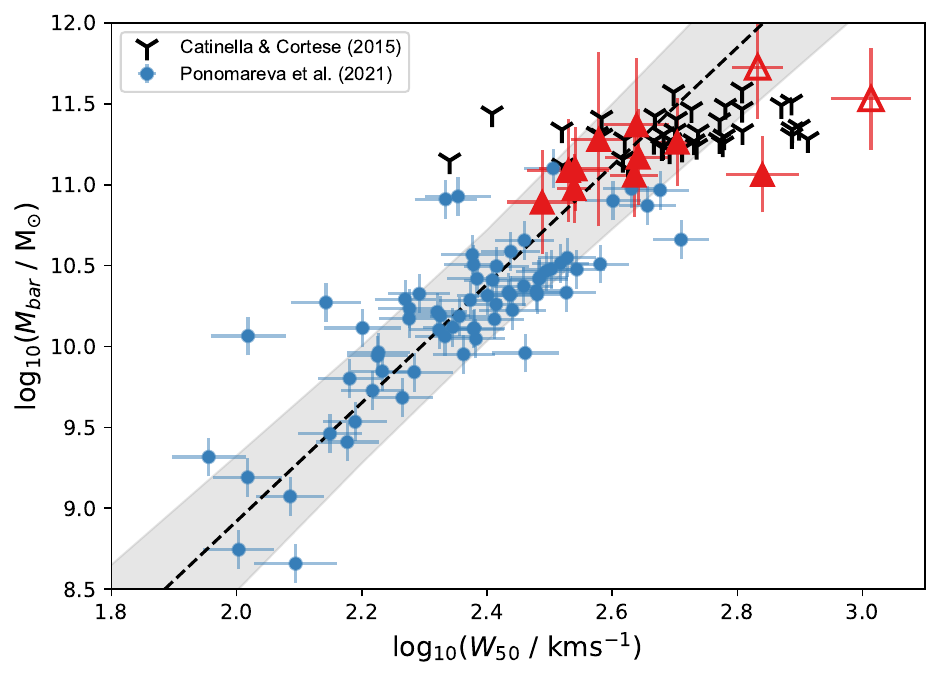}
   \caption{As Figure~\ref{fig:bTFr} with the baryonic mass estimate including the contribution from  the molecular gas mass, assuming the same mass as is present in H{\sc i} for the galaxies presented in both this paper and  \protect\citet{CatinellaCortese2015}.  }
    \label{fig:bTFr_mol}
\end{figure}

As mentioned previously, another possible explanation for the apparent flattening of the bTFr at high masses and large rotation velocities may also arise from our incomplete knowledge of the molecular gas mass in these galaxies. To illustrate this, in Figure~\ref{fig:bTFr_mol} we show the bTFr when we assume that the molecular gas mass is equal to that of the measured H{\sc i} mass \citep[see e.g. Figure~4 in ][]{Saintonge_2022}. We now find that these higher redshift galaxies are more aligned with the local bTFr defined by lower-mass galaxies. This result would need to be confirmed with direct measurements of the molecular gas mass to determine whether the observed trend can be explained without any evolution in the bTFr with redshift.\footnote{We note that using the scaling relations from the large compilation of galaxies at all redshifts from \cite{Tacconi2020} results in total baryonic masses $\sim 0.1$\,dex lower. The scatter in the total molecular gas mass as a function of galaxy mass is $\sim 0.5$\,dex and therefore this is well within the level of uncertainty. }

Although the data are inconclusive with respect to showing strong evidence for evolution in the bTFr, this study does highlight the potential of using H{\sc i} detections with the MeerKAT telescope (using both MIGHTEE and LADUMA) for studies of the evolution of the bTFr to these redshifts and potentially higher with lower-frequency observations and by utilizing stacking techniques \citep[e.g.][]{meyer2016,pan2021}.

\section{Conclusions}\label{sec:conclusions}

Using the Data Release 1 of the spectral line cube of the MIGHTEE survey we have searched for H{\sc i} emission at the position and redshift of emission-line galaxies with spectroscopic redshifts selected from the DESI survey. We identify eleven high-confidence ($>4\sigma$) H{\sc i} detections at $0.25<z<0.4$. These add to the six high-redshift H{\sc i} galaxies at $0.38<z<0.5$, recently discovered using data from the single-dish FAST deep survey \citep{Xi2024} demonstrating that we are now entering an era where direct detections of 21-cm H{\sc i} emission are possible over 5 billion years of cosmic time. 
By design, our H{\sc i}-rich galaxies have a wealth of ancillary data available, allowing us to measure the host galaxy properties. Using full spectral energy distribution fitting we derive the stellar mass and star-formation rates for the H{\sc i} galaxies. They generally lie on and above the galaxy star formation main sequence, with just a single galaxy lying significantly below the main sequence, and have similar stellar properties to their low-redshift H{\sc i} selected counterparts. However, we find much larger H{\sc i} gas reservoirs than their low-redshift ($z<0.08$) counterparts, but their discovery does not require any evolution in the local H{\sc i} mass function due to the much larger comoving volume accessible over the higher redshift range. Indeed, we expect many more galaxies to be detectable at these redshifts through untargeted searches.

Using the $W_{50}$ parameter as a proxy for the rotation velocity of the galaxies we are able, for the first time, to investigate whether $z>0.25$ H{\sc i} galaxies lie on the local baryonic Tully-Fisher relation, overcoming issues around different tracers of the gravitational potential of galaxies using different emission lines. We find that although the galaxies lie at the very high baryonic mass and high rotational velocity, they are consistent with the low-redshift relation and we do not find any evidence for evolution, as expected based on hydrodynamic simulations. However, we find tentative evidence for a flattening in the bTFr at these redshifts, that could be attributed to real evolution in the relationship between the baryonic mass and gravitational mass.  
On the other hand, if the molecular mass fraction in the most massive galaxies is higher than for their low-mass counterparts \citep[e.g.][]{Saintonge_2022}, and does become a significant component of the baryonic mass then including this component may be sufficient to resolve the apparent flattening at high masses/redshifts.
We also note that using the integrated line width of the H{\sc i} emission as a proxy for the flat part of the rotation curve at these high baryonic masses, which may lead to an overestimate of the true rotational velocity.

This study paves the way for future studies of H{\sc i} beyond the local Universe using searches targeted at known objects and  using pure-H{\sc i} selection.
The data used in this study are derived from approximately a fifth of the overall areal coverage of the MIGHTEE and LADUMA surveys, and we can therefore expect many more detections of H{\sc i}-rich galaxies at $z>0.25$ using both the method adopted here utilising the wealth of spectroscopic redshifts in these fields, but also with blind searches.  
By combining MIGHTEE data with the deeper observations from the LADUMA survey, we should be able to constrain the evolution of the H{\sc i} mass function and potentially the bTFr out to $z\sim 0.5$, before the Square Kilometre Array comes online towards the end of the decade.

\section*{Acknowledgements}

We thank the anonymous referee for helping clarify the initial selection of candidates in the paper, and for also providing general constructive feedback.
MJJ, MNT, IH, AV and HP acknowledge the support of a UKRI Frontiers Research Grant [EP/X026639/1], which was selected by the European Research Council, and MJJ, IH and AAP the STFC consolidated grants [ST/S000488/1] and [ST/W000903/1].
 MJJ, MNT and CLH also acknowledge support from the Oxford Hintze Centre for Astrophysical Surveys which is funded through generous support from the Hintze Family Charitable Foundation.  IH acknowledges support from the South African Radio Astronomy Observatory which is a facility of the National Research Foundation (NRF), an agency of the Department of Science and Innovation. 
MB acknowledges the financial support of the Flemish Fund for Scientific Research (FWO-Vlaanderen, Bilateral Scientific Cooperation Grant G0G0420N) and the Belgian Science Policy Office (BELSPO, Networking Grant BL/02/SA12). KS acknowledges support from the Natural Sciences and Engineering Research Council of Canada. 
A.B. acknowledges support from INAF under the Large Grant 2022 funding scheme (project "MeerKAT and LOFAR Team up: a Unique Radio Window on Galaxy/AGN co-Evolution”).
A.B. acknowledges financial support from the South African Department of Science and Innovation's National Research Foundation under the ISARP RADIOMAP Joint Research Scheme (DSI-NRF Grant Number 150551), and from the Italian Ministry of Foreign Affairs and International Cooperation under the “Progetti di Grande Rilevanza” scheme (project RADIOMAP, grant number ZA23GR03).
PEMP acknowledges the support from the Dutch Research Council (NWO) through the Veni grant VI.Veni.222.364.
Parts of this research were supported by the Australian Research Council Centre of Excellence for All Sky Astrophysics in 3 Dimensions (ASTRO 3D), through project number CE170100013. JD acknowledges support from an Africa-Oxford Catalyst Collaboration Grant (AfOx-290).

The MeerKAT telescope is operated by the South African Radio Astronomy Observatory, which is a facility of the National Research Foundation, an agency of the Department of Science and Innovation. We acknowledge the use of the ilifu cloud computing facility – www.ilifu.ac.za, a partnership between the University of Cape Town, the University of the Western Cape, Stellenbosch University, Sol Plaatje University and the Cape Peninsula University of Technology. The ilifu facility is supported by contributions from the Inter-University Institute for Data Intensive Astronomy (IDIA – a partnership between the University of Cape Town, the University of Pretoria and the University of the Western Cape), the Computational Biology division at UCT and the Data Intensive Research Initiative of South Africa (DIRISA). 

This research made use of Astropy,\footnote{\url{https://www.astropy.org/}} a community-developed core Python package for Astronomy \citep{astropy1, astropy2}; \texttt{TOPCAT} \citep{topcat1, topcat2}; \texttt{matplotlib} \citep{matplotlib}; \texttt{NumPy} \citep{numpy1,numpy2}; and \texttt{SciPy} \citep{scipy}. This work has made use of the Cube Analysis and Rendering Tool for Astronomy \citep[CARTA;][]{comrie2021}. This research has made use of NASA's Astrophysics Data System. 

The Legacy Surveys consist of three individual and complementary projects: the Dark Energy Camera Legacy Survey (DECaLS; Proposal ID \#2014B-0404; PIs: David Schlegel and Arjun Dey), the Beijing-Arizona Sky Survey (BASS; NOAO Prop. ID \#2015A-0801; PIs: Zhou Xu and Xiaohui Fan), and the Mayall z-band Legacy Survey (MzLS; Prop. ID \#2016A-0453; PI: Arjun Dey). DECaLS, BASS and MzLS together include data obtained, respectively, at the Blanco telescope, Cerro Tololo Inter-American Observatory, NSF’s NOIRLab; the Bok telescope, Steward Observatory, University of Arizona; and the Mayall telescope, Kitt Peak National Observatory, NOIRLab. Pipeline processing and analyses of the data were supported by NOIRLab and the Lawrence Berkeley National Laboratory (LBNL). The Legacy Surveys project is honored to be permitted to conduct astronomical research on Iolkam Du’ag (Kitt Peak), a mountain with particular significance to the Tohono O’odham Nation.

NOIRLab is operated by the Association of Universities for Research in Astronomy (AURA) under a cooperative agreement with the National Science Foundation. LBNL is managed by the Regents of the University of California under contract to the U.S. Department of Energy.

This project used data obtained with the Dark Energy Camera (DECam), which was constructed by the Dark Energy Survey (DES) collaboration. Funding for the DES Projects has been provided by the U.S. Department of Energy, the U.S. National Science Foundation, the Ministry of Science and Education of Spain, the Science and Technology Facilities Council of the United Kingdom, the Higher Education Funding Council for England, the National Center for Supercomputing Applications at the University of Illinois at Urbana-Champaign, the Kavli Institute of Cosmological Physics at the University of Chicago, Center for Cosmology and Astro-Particle Physics at the Ohio State University, the Mitchell Institute for Fundamental Physics and Astronomy at Texas A\&M University, Financiadora de Estudos e Projetos, Fundacao Carlos Chagas Filho de Amparo, Financiadora de Estudos e Projetos, Fundacao Carlos Chagas Filho de Amparo a Pesquisa do Estado do Rio de Janeiro, Conselho Nacional de Desenvolvimento Cientifico e Tecnologico and the Ministerio da Ciencia, Tecnologia e Inovacao, the Deutsche Forschungsgemeinschaft and the Collaborating Institutions in the Dark Energy Survey. The Collaborating Institutions are Argonne National Laboratory, the University of California at Santa Cruz, the University of Cambridge, Centro de Investigaciones Energeticas, Medioambientales y Tecnologicas-Madrid, the University of Chicago, University College London, the DES-Brazil Consortium, the University of Edinburgh, the Eidgenossische Technische Hochschule (ETH) Zurich, Fermi National Accelerator Laboratory, the University of Illinois at Urbana-Champaign, the Institut de Ciencies de l’Espai (IEEC/CSIC), the Institut de Fisica d’Altes Energies, Lawrence Berkeley National Laboratory, the Ludwig Maximilians Universitat Munchen and the associated Excellence Cluster Universe, the University of Michigan, NSF’s NOIRLab, the University of Nottingham, the Ohio State University, the University of Pennsylvania, the University of Portsmouth, SLAC National Accelerator Laboratory, Stanford University, the University of Sussex, and Texas A\&M University.

BASS is a key project of the Telescope Access Program (TAP), which has been funded by the National Astronomical Observatories of China, the Chinese Academy of Sciences (the Strategic Priority Research Program “The Emergence of Cosmological Structures” Grant \# XDB09000000), and the Special Fund for Astronomy from the Ministry of Finance. The BASS is also supported by the External Cooperation Program of Chinese Academy of Sciences (Grant \# 114A11KYSB20160057), and Chinese National Natural Science Foundation (Grant \#12120101003, \# 11433005).

The Legacy Survey team makes use of data products from the Near-Earth Object Wide-field Infrared Survey Explorer (NEOWISE), which is a project of the Jet Propulsion Laboratory/California Institute of Technology. NEOWISE is funded by the National Aeronautics and Space Administration.

The Legacy Surveys imaging of the DESI footprint is supported by the Director, Office of Science, Office of High Energy Physics of the U.S. Department of Energy under Contract No. DE-AC02-05CH1123, by the National Energy Research Scientific Computing Center, a DOE Office of Science User Facility under the same contract; and by the U.S. National Science Foundation, Division of Astronomical Sciences under Contract No. AST-0950945 to NOAO.
 
 For the purpose of Open Access, the author has applied a CC BY public copyright licence to any Author Accepted Manuscript (AAM) version arising from this submission.

\section*{Data Availability}

The raw visibility data are available from the SARAO archive by
searching for the capture block IDs listed in Table 1 of \cite{Heywood2024}. The image
products described in this article are available at {\url{https://doi.org/10.48479/jkc0-g916}}. The spectroscopic catalogue used in this paper is available from the DESI data release page {\url{https://data.desi.lbl.gov/doc/}}.



\bibliographystyle{mnras}
\bibliography{example} 


\bsp	
\label{lastpage}
\end{document}